\documentclass[a4paper,fleqn,usenatbib]{mnras}

\usepackage[T1]{fontenc}
\usepackage{ae,aecompl}

\usepackage{savesym}
\savesymbol{tablenum}
\usepackage{siunitx}
\restoresymbol{SIX}{tablenum}

\usepackage[maxfloats=40]{morefloats}

\usepackage[innercaption]{sidecap}
\sidecaptionvpos{figure}{m}

\title[The cloudy shape of hot Jupiter thermal phase curves]{The cloudy shape of hot Jupiter thermal phase curves}

\usepackage{natbib}
\usepackage{subfigure,wrapfig}
\usepackage[]{sidecap}
\usepackage[T1]{fontenc}
\usepackage{mathptmx}
\usepackage{hyperref}

\hypersetup{
   colorlinks=true,
   citecolor=blue,
   linkcolor=blue,
   urlcolor=black
   }
   
   \usepackage[figuresright]{rotating}

\date{Accepted 2020 October 6. Received 2020 September 28; in original form 2020 July 3.}

\pubyear{2020}

\author[Parmentier, V. al.]{
Vivien Parmentier,$^{1}$\thanks{E-mail: vivien.parmentier@physics.ox.ac.uk}
Adam P. Showman,$^{2,*}$
and Jonathan J. Fortney$^{3}$
\\
$^{1}$University of Oxford\\
$^{2}$Department of Planetary Sciences and Lunar and Planetary Laboratory, The University of Arizona, Tucson, AZ 85721, USA\\
$^{3}$Department of Astronomy and Astrophysics, University of California, Santa Cruz, CA 95064
$^{*}$Deceased 16 March 2020
}

\begin{document}
\label{firstpage}
\pagerange{\pageref{firstpage}--\pageref{lastpage}}
\maketitle

\begin{abstract}
Hot Jupiters have been predicted to have a strong day/night temperature contrast and a hot spot shifted eastward of the substellar point. This was confirmed by numerous phase curve observations probing the longitudinal brightness variation of the atmosphere. Global circulation models, however, systematically underestimate the phase curve amplitude and overestimate the shift of its maximum. 
We use a global circulation model including non-grey radiative transfer and realistic gas and cloud opacities to systematically investigate how the atmospheric circulation of hot Jupiters varies with equilibrium temperature from 1000 to 2200K.  We show that the heat transport is very efficient for cloudless planets cooler than 1600K and becomes less efficient at higher temperatures. When nightside clouds are present, the day-to-night heat transport becomes extremely inefficient, leading to a good match to the observed low nightside temperatures. The constancy of this low temperature is, however, due to the strong dependence of the radiative timescale with temperature. We further show that nightside clouds increase the phase curve amplitude and decreases the phase curve offset at the same time. This change is very sensitive to the cloud chemical composition and particle size, meaning that the diversity in observed phase curves can be explained by a diversity of nightside cloud properties. Finally, we show that phase curve parameters do not necessarily track the day/night contrast nor the shift of the hot spot on isobars, and propose solutions to to recover the true hot-spot shift and day/night contrast.
 
\end{abstract}

\begin{keywords}
planets and satellites: atmospheres, planets and satellites: gaseous planets
\end{keywords}

%
\section{Introduction}
Among the known exoplanets, planets in short { period} orbits are the easiest to characterize. Close-in, giant planets are believed to be tidally locked and are thus slow rotators~\citep{Lubow1997,Guillot2002}, with rotation periods from less than a day to several days. Their slow rotation, together with the large day/night contrast in the irradiation they receive makes their atmospheres intrinsically three-dimensional objects~\citep{Showman2002}. The atmospheric circulation transports heat from dayside to the nightside and the efficiency of this transport determines the nightside temperature of the planet. 

 In this study we focus on hot Jupiters, which we define as Jupiter mass planets on close-in orbits with equilibrium temperatures between 1000 and 2200K. Planets cooler than 1000K often called warm Jupiters, are unlikely to be tidally locked whereas in planets hotter than 2200K, often called ultra hot Jupiters~\citep{Parmentier2018,Lothringer2018}, heat transport through H$_2$ dissociation/recombination~\citep{Bell2018,Tan2019} and magnetically coupled atmospheric circulation~\citep{Batygin2013,Rogers2014,Rogers2014a} likely complicates the atmospheric dynamics. Nonetheless, from equilibrium temperatures of 1000 to 2200K, the irradiation received by hot Jupiters can vary by a factor 20 and a variety of behaviour is expected. { For instance}, over such a range of temperatures, brown dwarfs have been classified into 16 different subclasses, from L1 to T6~\citep{Kirkpatrick2005}. 

Hot Jupiters can be observed during a variety of orbital phases. More than 80 have been observed during secondary eclipse~\citep{Garhart2020,Baxter2020} when the mean dayside flux can be observed. A sub-sample of a dozen planets have been monitored during their whole orbit~\citep{Parmentier2018}. These phase curves provide valuable information on the longitudinal aspect of the planet.  When observed in thermal emission, information on the thermal structure and the chemical composition is gathered. When reflected light is observed, information on the longitudinal variation of the albedo can be obtained. 

The hundreds of known hot Jupiters and the dozen of planets with observed phase curves provide an opportunity to understand how the atmospheric dynamics scales with parameters such as their equilibrium temperature. Systematic studies have been carried out to understand how the atmospheric dynamics of hot Jupiters is expected to scale with their equilibrium temperature.~\citet{Showman2002} first postulated that the day/night temperature contrast in hot tidally locked planets should be determined by a competition between radiative cooling and day/night heat transport through winds and gravity waves. Later on~\citet{Perna2012}, \citet{Perez-Becker2013a}, \citet{Komacek2017}, ~\citet{Komacek2018}, ~\citet{Zhang2017} combined models of varying complexity, from analytical to shallow-water and semi-grey 3D Global Circulation Models (GCM) to predict how the wind speed, wave speed and strength of radiative cooling scale differently with equilibrium temperature, leading to an expected increase of the day/night temperature contrast with equilibrium temperature.

The first goal of this study is to { pursue this systematic approach to understand} how the atmospheric dynamics of { cloudless} hot Jupiter atmospheres scales with equilibrium temperature { when non-grey radiative transfer and realistic gaseous opacities are used in the global circulation model}. Planets at different temperatures are expected to have different opacities because of the the pressure and temperature broadening of the lines, the change of chemistry, and the fact that the emission of hotter planets peaks towards shorter wavelengths, where the opacities are often smaller. Because the opacity variations are correlated with temperature, we expect to see a quantitative difference in the trends between atmospheric circulation and equilibrium temperature { compared to the ones seen in the previous systematic studies using grey or semi-grey opacities. Additionally, the flux received at different wavelengths originates from different atmospheric layers~\citep{Knutson2009} and these levels can vary with latitude and longitude~\citep{Dobbs-dixon2017,Drummond2018a}}. Our study allows for the first time a direct comparison of the numerous phase curve measurements taken in different bandpasses for different planets with a set of non-grey global circulation simulations derived from the same model.

{ Although non-grey global circulation models have been used to directly interpret observations of specific planets~\citep[e.g.][among others]{Showman2009,Dobbs-Dixon2013,Lewis2014,Mayne2014,Kataria2015,Charnay2015,Charnay2015a,Lee2016,Lee2017,Lewis2017,Lines2018,Drummond2018a,Lines2019,Steinrueck2019,Parmentier2018,Kreidberg2018,Venot2020} they often overestimate the shift of the maximum of the phase curve and underestimate the phase curve }amplitude~\citep[e.g.][]{Kataria2015,Lewis2017,Zhang2018,Parmentier2018a}, leading to modelled nightsides that are hotter than observed~\citep{Keating2019,Beatty2018}. Several explanations have been put forward to explain the larger-than-expected phase curve amplitude and lower-than-expected phase curve offset. An increased metallicity~\citep{Showman2009,Kataria2015,Drummond2018} would increase the day/night contrast at the photosphere and reduce the hot spot offset by lowering the photospheric pressure and reducing the radiative timescale. However the metallicities that hot Jupiter are expected to reach fall short at explaining this trend~\citep{Kataria2015,Drummond2018}. { The quenching of CO on the planet nightside has been suggested by~\citet{Knutson2012} to explain the large phase curve amplitude at $4.5\mu m$. However, this hypothesis was disproved by~\citet{Drummond2018a},~\citet{Mendonca2018b} and ~\citet{Steinrueck2019}. Disequilibrium chemistry leading to a CO2 absorption was also proposed by~\citet{Mendonca2018a} but later disproved by~\citet{Mendonca2018b} and~\citet{Drummond2018b} using more sophisticated chemical models. The presence of drag limiting the wind speed has also been proposed~\citep{Perna2010,Rauscher2012a,Showman2013,Komacek2017}. Although magnetic drag is a promising explanation for the ultra-hot Jupiters, no good drag mechanism is known to sufficiently slow the winds for planets cooler than $\approx 1600-1800\,\rm K$~\citep{Batygin2013,Rogers2014,Rogers2014a,Koll2018}. }

The last explanation, which we explore in more detail in this paper, is that a thick layer of clouds impedes the infrared radiation to emerge from the nightside~\citep{Showman2009,Kataria2015,Oreshenko2016,Beatty2018,Mendonca2018a,Keating2019,Lines2019}. Indeed the presence of clouds and hazes in exoplanet atmospheres seems ubiquitous~\citep{Pont2013,Sing2016,Gao2020a}. They shape the transmission spectrum of most observed planets, leading to a major source of uncertainties when retrieving molecular abundances~\citep{Greene2016}. Evidence for partially cloudy atmosphere has been found, both at the limb through transmission spectroscopy~\citep{Pinhas2019} and on the dayside through the observations of optical phase curves~\citep{Demory2013,Shporer2015}. Interestingly, the albedo maps and temperature maps of hot Jupiters appear to be anti-correlated: whereas the hottest hemisphere of the planet is shifted east of the substellar point, the most reflective region is shifted west of the substellar point{ ~\citep{Webber2015,Munoz2015}}. ~\citet{Showman2002} first pointed out the possibility of nightside clouds: as the superrotating equatorial jet transports dayside air to the nightside, the temperature drops, causing condensation and the formation of nightside clouds; the clouds then dissipate as the air travels back to the dayside where temperatures rise and cloud particles sublimate.  This can lead to a configuration with a dayside that is predominantly cloud free and a nightside that is predominantly cloudy.~\citet{Parmentier2016} carried the argument much farther, presenting detailed SPARC/MITgcm calculations with post-processed clouds to determine the cloud distributions and lightcurves for a wide range of planetary conditions and cloud compositions. These models showed that clouds commonly persist on the western edge of the dayside, simply because the eastward-flowing air that reaches the western terminator from the nightside is still relatively cool, and takes time to heat up.  The coolest regions on the dayside therefore tend to be toward its western edge.  Thus a common configuration on hot Jupiters comprises thermal hotspots shifted to the east but cloudy regions shifted to the west of the substellar point. { Other works involving global circulation models~\citep{Oreshenko2016,Lee2016,Roman2019,Lines2019} and/or microphysical cloud modelling~\citep[e.g.][]{Lee2015,Lee2016,Helling2016,Powell2018,Helling2019,Lines2018} confirmed this general picture: clouds tend to form on the cool nightside and evaporate on the hot dayside.} Understanding the distribution of clouds on hot Jupiter can constrain their physical and chemical properties, leading to important insights into atmospheric mixing processes, microphysics and deep atmospheric cold traps.

The second goal of this paper is to investigate how the presence of nightside clouds on hot Jupiters affect their phase curves from the optical to the far infrared. Nightside clouds are expected to change both the phase curve amplitude and the phase curve offset through two different mechanisms. First, by changing the albedo and the greenhouse effect of the atmosphere they can alter the total incoming flux and the day-to-night heat redistribution, effectively modifying the thermal structure of the planet~\citep{Oreshenko2016,Roman2017,Lines2018,Lines2019,Mendonca2018a,Roman2019,Lines2019}. Second, by changing the opacities, they can change the photospheric levels that are observed and introduce sharp variations of the photospheric levels probed by the observations~\citep{Dobbs-dixon2017}. As an example, when sharp brightness gradients are present, such as produced by nightside clouds, the brightest hemisphere is not necessarily centred around the hottest point of the atmosphere. That is because if the hottest point of the atmosphere is shifted away from the substellar point, the hemispherically averaged brightness when facing the hot spot is the combination of a part of the dayside flux and a part of the nightside flux. If the nightside flux is reduced by the presence of clouds, then the planets would not look at bright as expected at this orbital phase. { Here we reproduce qualitatively the findings of ~\citet{Mendonca2018a}, ~\citet{Roman2019} and~\citet{Lines2019} but focus our work on a more thorough exploration of the non-grey effects of nightside clouds over a range of equilibrium temperatures.}

 The last goal of our study is to provide specific predictions that can be used to better prepare the coming observations. In the coming decade, the amount of 3D information we have about hot Jupiters is going to increase by orders of magnitudes. Both the number of targets with precise 3D observations and the accessible wavelength range will be multiplied by one order of magnitude. JWST will provide phase curves of a few planets with a wavelength coverage ranging from 0.6$\mu m$ to $20\mu m$~\citet{Stevenson2016a,Bean2018}. Particularly, the phase curve of WASP-43b with the MIRI instrument will be one of the first observation that should be carried out~\citep{Venot2020}. Ariel~\citep{Tinetti2018} plans to dedicate 10$\%$ of its science time to phase curves, corresponding to more than 30 planets being observed from 2 to 8$\mu m$. Currently TESS~\citep{Ricker2014} is observing phase curves of many ultra-hot Jupiters and PLATO 2.0~\citep{Rauer2013} should have the precision needed to observe the optical phase curves of cooler planets. Missions smaller in scale, such as EXCITE~\citep{Nagler2019}, a ballon based spectrometer, will observe the phase curve of approximately 10 objects from 1 to 4$\mu m$ during each flight. Finally, ground based measurements at high resolution are now able to resolve the transit of a planet~\citep{Ehrenreich2020} and should be able to obtain wind measurements separately on both limbs of the planet. With increased precision will come a higher sensitivity to the 3D structure of the planet. Several studies have looked at the biases that can be driven by the non-homogeneous 3D structure. Atmospheric variation along and across the limb can mimic high mean molecular weight atmosphere~\citep{Line2016,Kempton2017}, lead to biased temperature estimates~\citep{Caldas2019,MacDonald2020} or biased abundance estimates~\citep{Pluriel2020}. Non-homogeneous thermal structure can, when interpreting a hemispherically averaged spectrum, lead to the spurious detection of molecules~\citet{Feng2016,Taylor2020,Feng2020}. Although mitigation strategies have been proposed by parametrising the inhomogeneities~\citet{Line2016,Blecic2017,Pinhas2019,Irwin2019,MacDonald2020,Taylor2020,Changeat2019,Changeat2020}, forward global circulation models are needed to anchor the priors of their parameters. 

We begin by describing the data and the models we use in section~\ref{sec::Methods}. Then in section~\ref{sec::GCMResults} we discuss how the efficiency of heat redistribution varies with equilibrium temperature in our non-grey simulations, then evaluate the changes induced by the  presence of nightside clouds. In section~\ref{sec::ComparingNS} we show how the nightside clouds can explain the low measured nightside brightness temperatures of hot Jupiters. In section~\ref{sec::Effect} we investigate the effect of nightside clouds on the spectrum and phase curve of hot Jupiters and challenge the idea that the phase curve offset should track the atmospheric hot spot shift. In section~\ref{sec::PhaseCurvesobs} we show that the diversity of phase curve offset and amplitude seen in hot Jupiter phase curves can be explained by a diversity in the nightside cloud properties. Section~\ref{sec::Consequences} provides a few thoughts on { how to marginalise over} the clouds and retrieve useful atmospheric properties. Finally section~\ref{sec::Conclusion} summarises our conclusions.

\section{Methods}
\label{sec::Methods}
\subsection{Data}
Phase curves observed with either the Hubble Space Telescope or the Spitzer Space Telescope have been published for 14 planets and a few more planets possess a phase curve observed with the Kepler spacecraft. We use the phase curve amplitude and phase curve offsets summarised in ~\citet{Parmentier2018a}. Amplitude and offsets of phase curves published after~\citet{Parmentier2018a} were calculated based on the parameters of~\citet{Beatty2018} for KELT-1b,~\citet{Zhang2018} for HD149026b and WASP-33b and~\citet{Dang2018} for CoRoT-2b. A missing minus sign was noticed in~\citet{Parmentier2018a} and the offset of HAT-P-7b was corrected to $6.8\pm7.5^{\circ}$ at $3.6\mu m$ and $-4.1\pm7.5^{\circ}$ at $4.5\mu m$ based on~\citet{Wong2016}. For the specific case of the Spitzer phase curves of WASP-43b we use the most recent re-analyses from~\citet{Morello2019} but also show the reductions of ~\citet{Mendonca2018a} and ~\citet{Stevenson2017} for completeness. Dayside and nightside brightness temperatures are taken directly from~\citet{Beatty2018} apart for WASP-14b where we used the value from the original paper~\citet{Wong2015}, { for Qatar-1b for which we use the values in~\citet{Keating2020}} and for WASP-43b where { we used the values from~\citet{May2020} for the 4.5$\mu$m case and the one from~\citet{Morello2019} for the 3.6$\mu$m case. }

\subsection{Dynamics}
We use the SPARC/MITgcm global circulation model to model the atmospheric circulation of Jupiter size planets in a tidally locked orbit around a sun-like star. The model solves the primitive equations on a cubed-sphere grid. It has been successfully applied to a wide range of hot Jupiters~\citep{Showman2009,Showman2015,Kataria2015,Kataria2016,Lewis2017,Steinrueck2019,Parmentier2013,Parmentier2016,Parmentier2018}.

The simulations presented here have a similar setup to the ones in~\citet{Parmentier2016}. We performed a set of simulations for planets with equilibrium temperatures (defined with no albedo and full redistribution) ranging from $1000\, \rm K$ to $2200\,\rm K$ with intervals of $100\,\rm K$. The gravity of the planet was assumed to be $10\rm m/s^2$, the specific heat capacity of the atmosphere $C{\rm p}=1.3\times 10^4\rm Jkg^{-1}K^{-1}$, the specific heat capacity ratio $\gamma=1+\frac{2}{7}$, and the mean molecular weight $\mu=2.3m_{\rm H}$, valid for a $\rm H_2-He$ dominated atmosphere.

 { All simulations were run for $1000$ days and all quantities have been averaged over the last 100 days of the simulation. We do not integrate the model for the several thousands of days that would be necessary to reach a fully converged state~\citet{Mendonca2018a,Wang2020,Mendonca2020}. This has several consequences. First, the integration time is too small for our choice of deep boundary condition to be able to affect the results. Second, the integration time is too small for the presence of a deep circulation to emerge and significantly affect the photospheric level~\citep{Mayne2017,Sainsbury-Martinez2019,Carone2020,Wang2020}. A pseudo-steady state, is, however, reached at the photospheric levels~\citep{Showman2009}. This steady-state is correct to within the assumption that the deep flow does not significantly change the photospheric levels equilibrium.}
 
{ We do not add any explicit Rayleigh drag in our simulations, as was the case in~\citet{Showman2009}. The main damping is performed by the fourth order Shapiro filter applied to the temperature and the velocity fields.  The simulation therefore relies on numerical dissipation to equilibrate kinetic energy~\citep{Koll2018}. The Shapiro filter dissipates kinetic energy by smoothing horizontal gradients at the grid level and should therefore not significantly affect the atmospheric circulation that is driven by large, planetary scale wave flow interactions~\citep{Showman2011,Hammond2018}. The lack of explicit Rayleigh drag should not affect the general features of the simulations as they have been found to be robust agains the choice of dissipation mechanism~\citep{Heng2011a,Liu2013,Mayne2014,Koll2018}. Quantitatively, however, the wind speeds at the photosphere would likely be smaller if additional Rayleigh drag were implemented. }

Our pressure levels range from 200 bars to 2$\mu\rm bar$ with 53 levels so that we have a resolution of almost three levels per scale height. We use a horizontal resolution of C32, equivalent to an approximate resolution of 128 cells in longitude and 64 in latitude and a time-step of 25 s.

We stop our grid of models at an equilibrium temperature of $2200\,\rm K$ in order to avoid the complications raised by molecular dissociation~\citep{Parmentier2018}, H$_2$ binding latent heat transport~\citep{Bell2018,Tan2019} and the presence of strong atomic iron opacities~\citep{Lothringer2018}. On the lower bound we do not model planets with an equilibrium temperature lower than 1000 K as these planet are in long enough orbits so that the tidal locking assumption starts to be questionable~\citep[see e.g. fig. 2 of][]{Parmentier2014,Showman2015}.

\subsection{Radiative transfer}
Radiative transfer is handled both in the 3D simulations and during the spectral calculations using the plane-parallel radiative transfer code of~\citet{Marley1999}. The code was first developed for Titan’s atmosphere~\citep{McKay1989} and since then has been extensively used for the study of giant planets~\citep{Marley1996}, brown dwarfs~\citep{Marley2002}, hot Jupiters~\citep{Fortney2005a, Fortney2008} and ultra-hot Jupiters~\citep{Parmentier2018}. 

Molecular and atomic abundances are calculated using a modified version of the NASA CEA Gibbs minimisation code (see Gordon \& McBride 1994) as part of a model grid previously used to explore gas and condensate equilibrium chemistry in substellar objects  over a wide range of atmospheric conditions~\citep{Moses2013,Skemer2016,Kataria2015,Wakeford2017,Burningham2017,Marley2017DPS,Parmentier2018}. Here we assume solar elemental abundances and local chemical equilibrium with rainout of condensate material, meaning that the gas phase does not interact with the solid phase. { We note that chemical might not be the best assumption for hot Jupiter atmospheres where chemical quenching is expected to smooth out the large day/night chemical gradients present in the atmosphere~\citep{Cooper2006,Drummond2018b,Mendonca2018b,Steinrueck2019}. Although the effect of disequilibrium chemistry on the energy balance of the atmosphere is small, its consequence in specific observational bandpasses can be large, particularly in the Spitzer $3.6\mu m$ bandpass. This is further discussed in section~\ref{sec::PCA}.}

We model the non-grey atmospheric opacities through correlated-k distributions, allowing the information of several tens of thousands of individual wavelengths to be compressed into 8 k-coefficients. Molecular opacities were calculated following~\citet{Freedman2008} including more recent updates described in~\citet{Freedman2014}.  { Although more recent line lists have been published, the ones of water and CO, which are the main radiatively active molecules in this study have been rather unchanged in the past years. Among the updated linelists we do not use here are the methane~\citet{Yurchenko2017} and alkali~\citet{Allard2019}. We do not expect the trends seen in this paper to be qualitatively affected by a change of linelist, nor do we expect it to change our conclusion on the role of clouds. For specific planets and specific bandpasses, however, the quantitative values might be affected.} The absorption and asymmetric scattering of the clouds in both the thermal and stellar components are calculated with Mie theory using refractive indexes specific for each species. We assume that the particle size follow a gaussian distribution of width 1.05 in order to wash out the Mie scattering oscillation present for a particle with a unique size. 

Our radiative transfer calculations can be done with two different spectral resolutions. When coupled to the GCM, we use 11 frequency bins that have been carefully chosen to maximise the accuracy and the speed of the calculation~\citep{Kataria2013}. We then use the thermal structure outputted from the global circulation model and post-process it with a higher-resolution version of our radiative transfer model (196 frequency bins ranging from 0.26 to 300$\mu m$). For this, we solve the two-stream radiative transfer equations along the line of sight for each atmospheric column and for each planetary phase considering absorption, emission, and scattering. This method naturally takes into account geometrical effects such as limb darkening. The stellar flux is assumed to be a collimated flux propagating in each atmospheric column with an angle equal to the angle between the local vertical and the direction of the star. More details about the numerical methods, similar to ~\citet{Fortney2006a}, can be found in \citet{Parmentier2016} section 2.2.
\begin{figure*}
\includegraphics[width=0.33\linewidth]{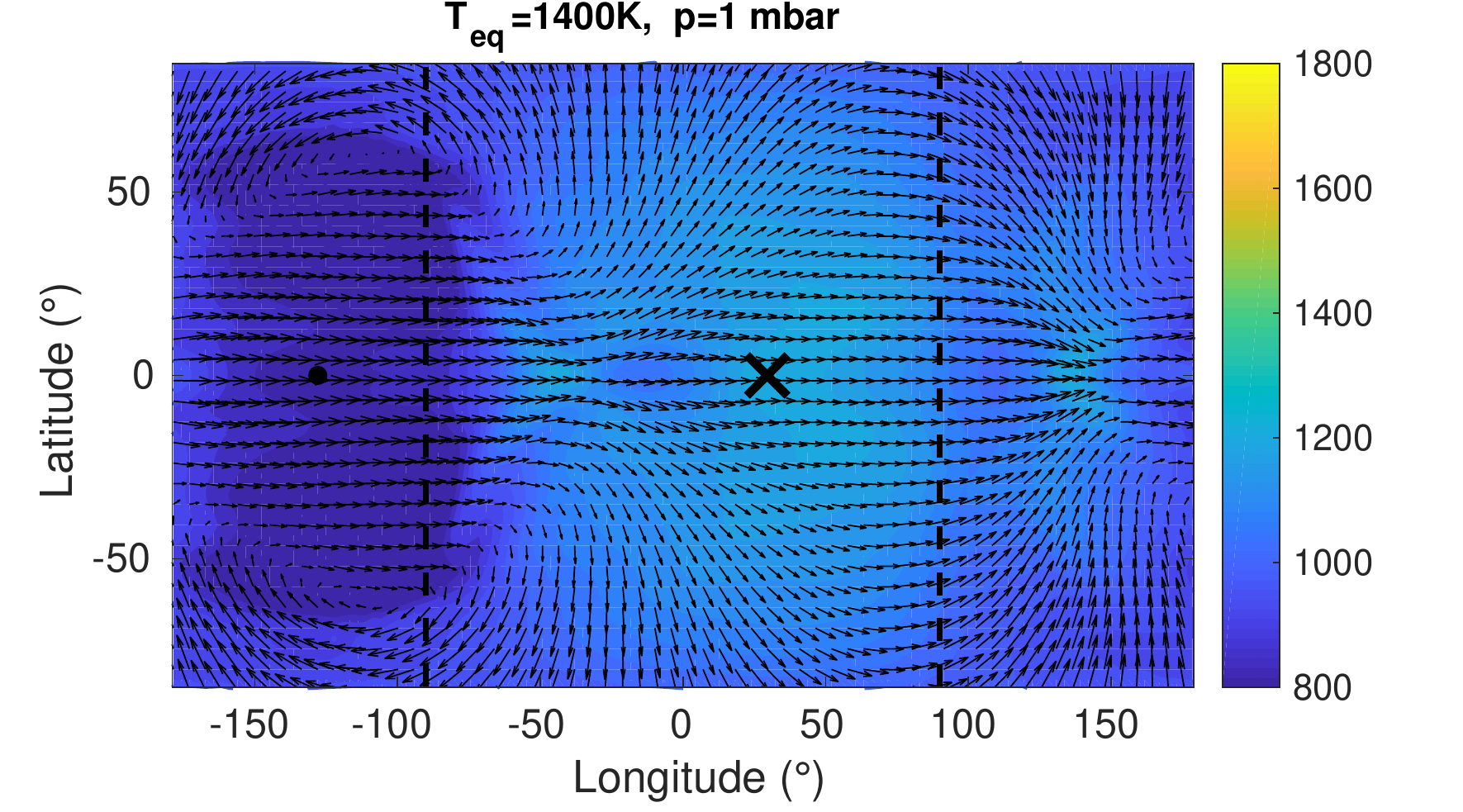}
\includegraphics[width=0.33\linewidth]{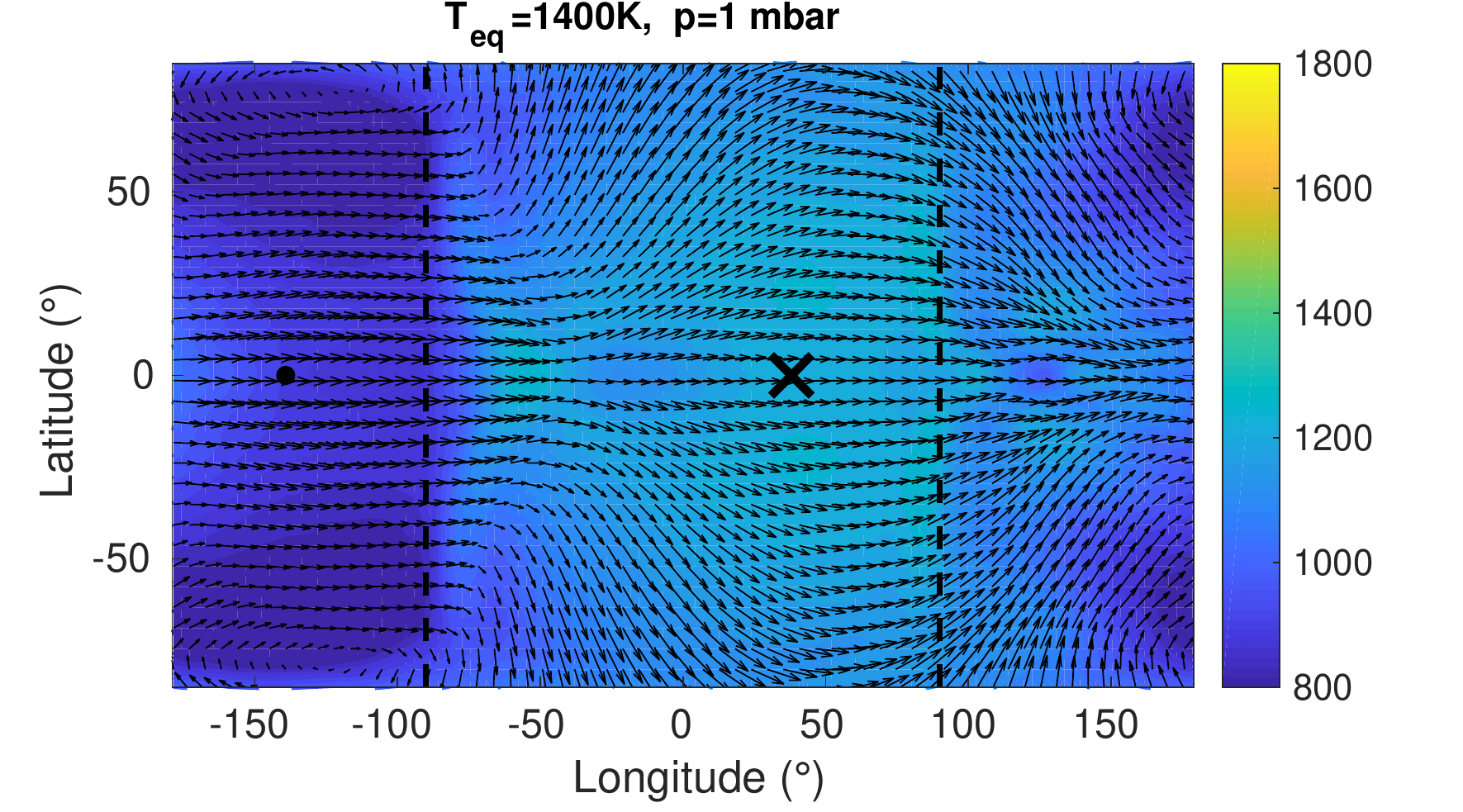}
\includegraphics[width=0.33\linewidth]{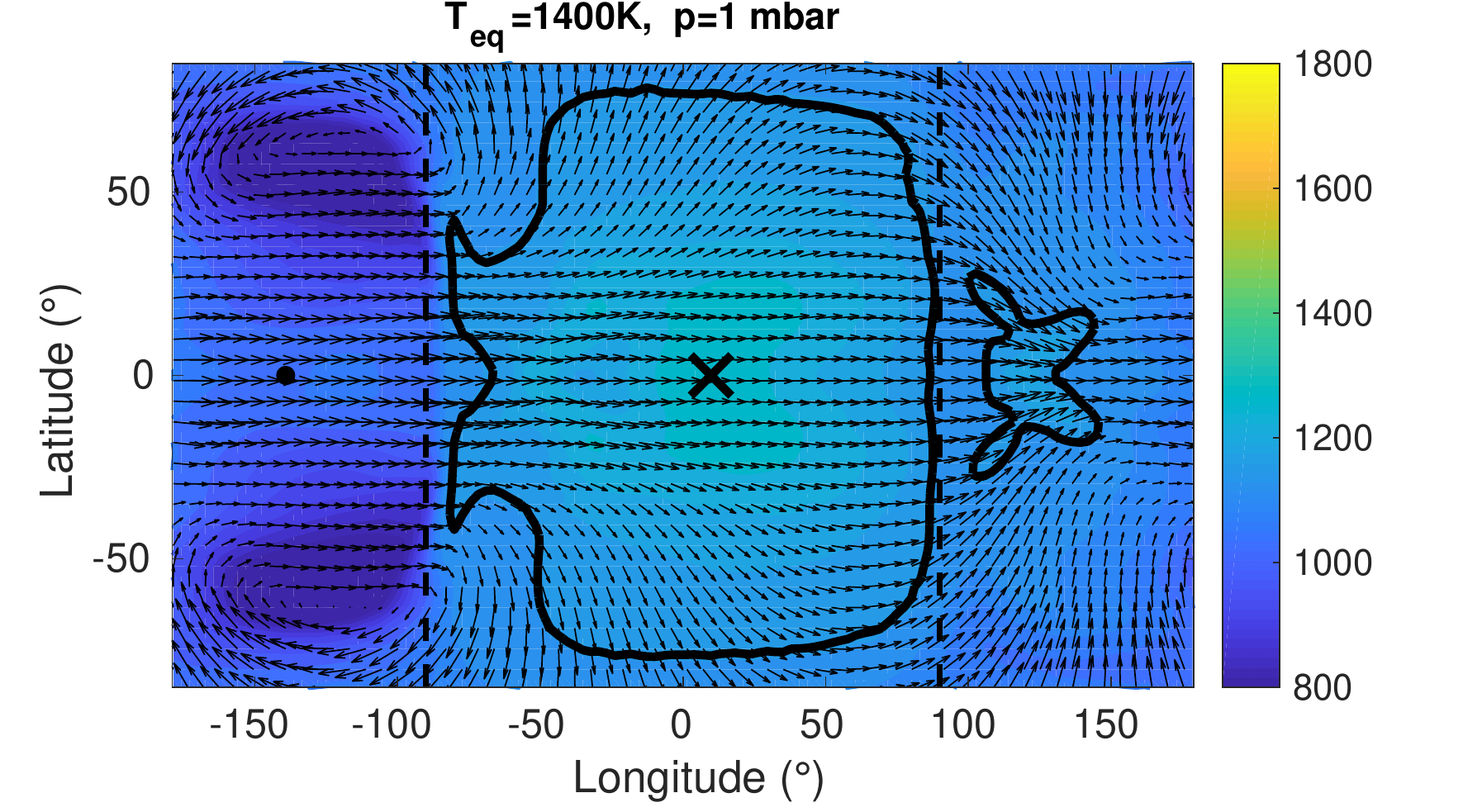}

\includegraphics[width=0.33\linewidth]{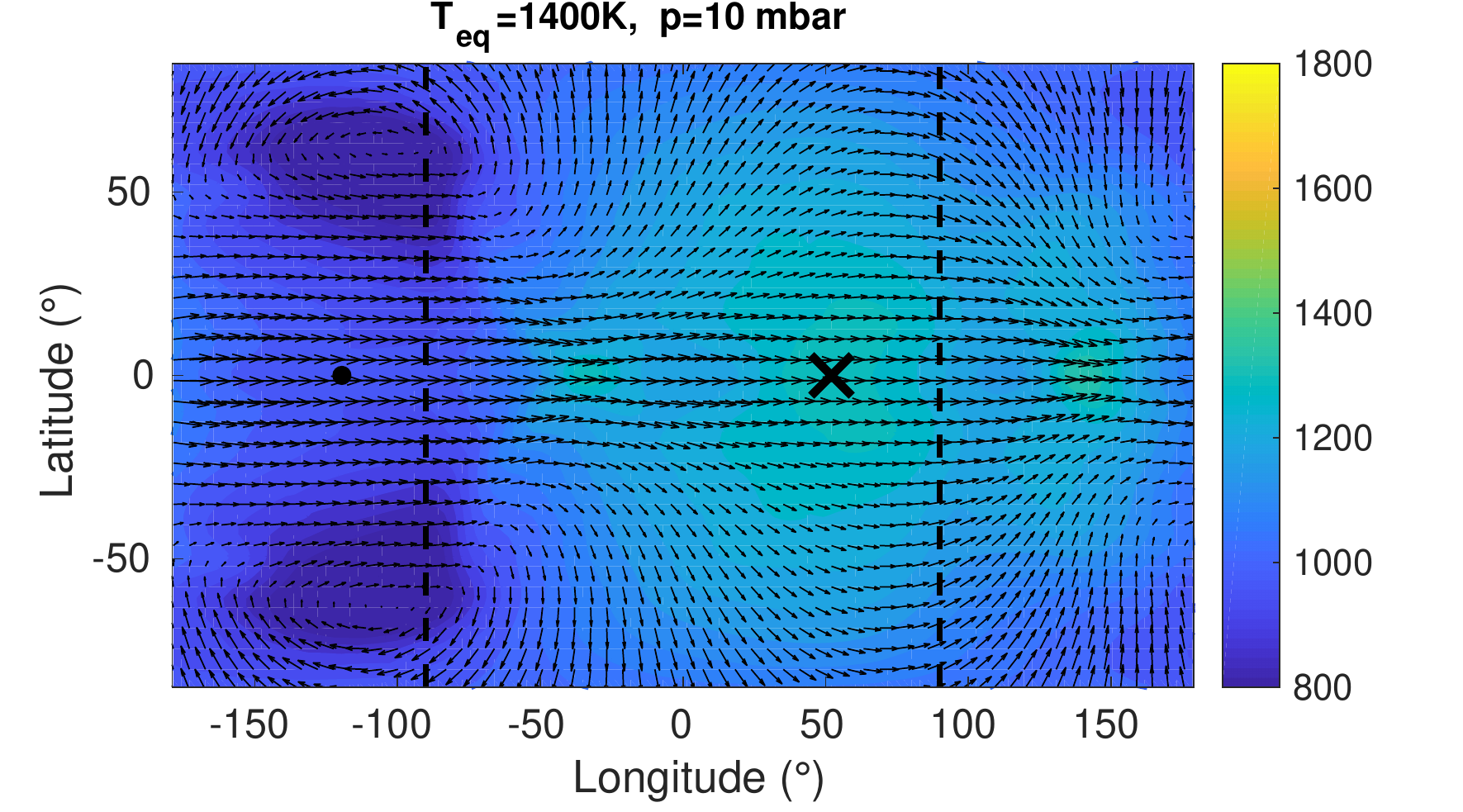}
\includegraphics[width=0.33\linewidth]{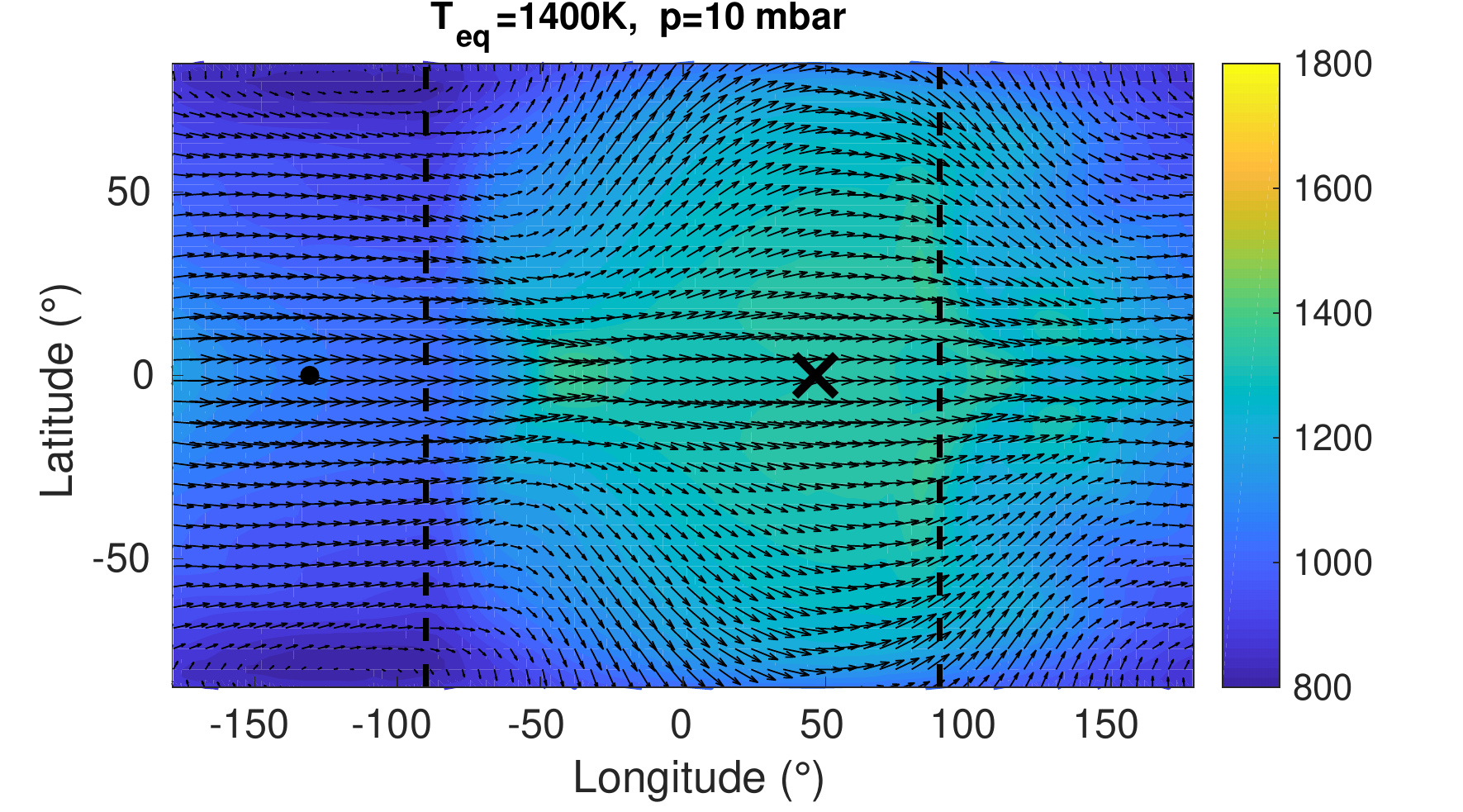}
\includegraphics[width=0.33\linewidth]{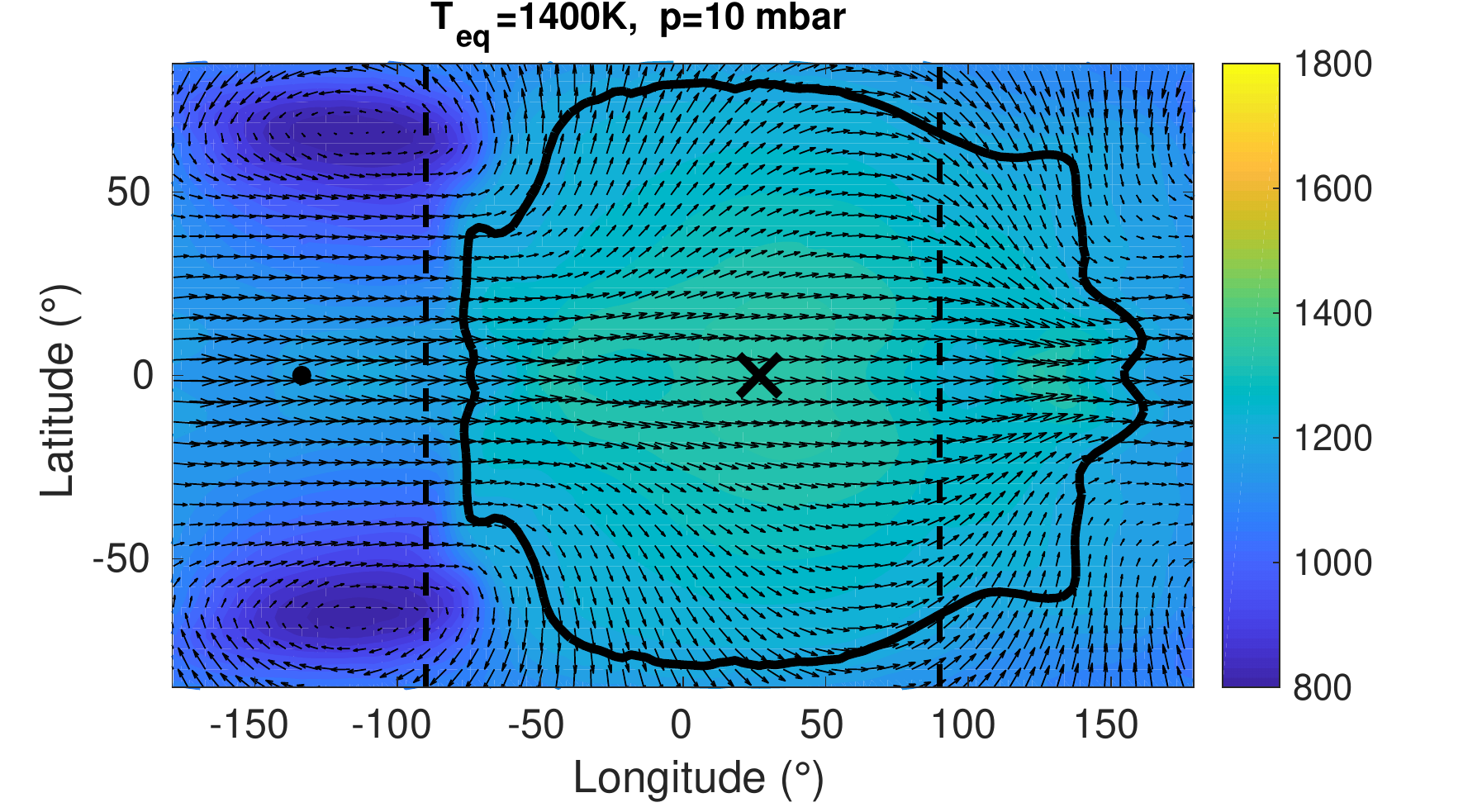}

\includegraphics[width=0.33\linewidth]{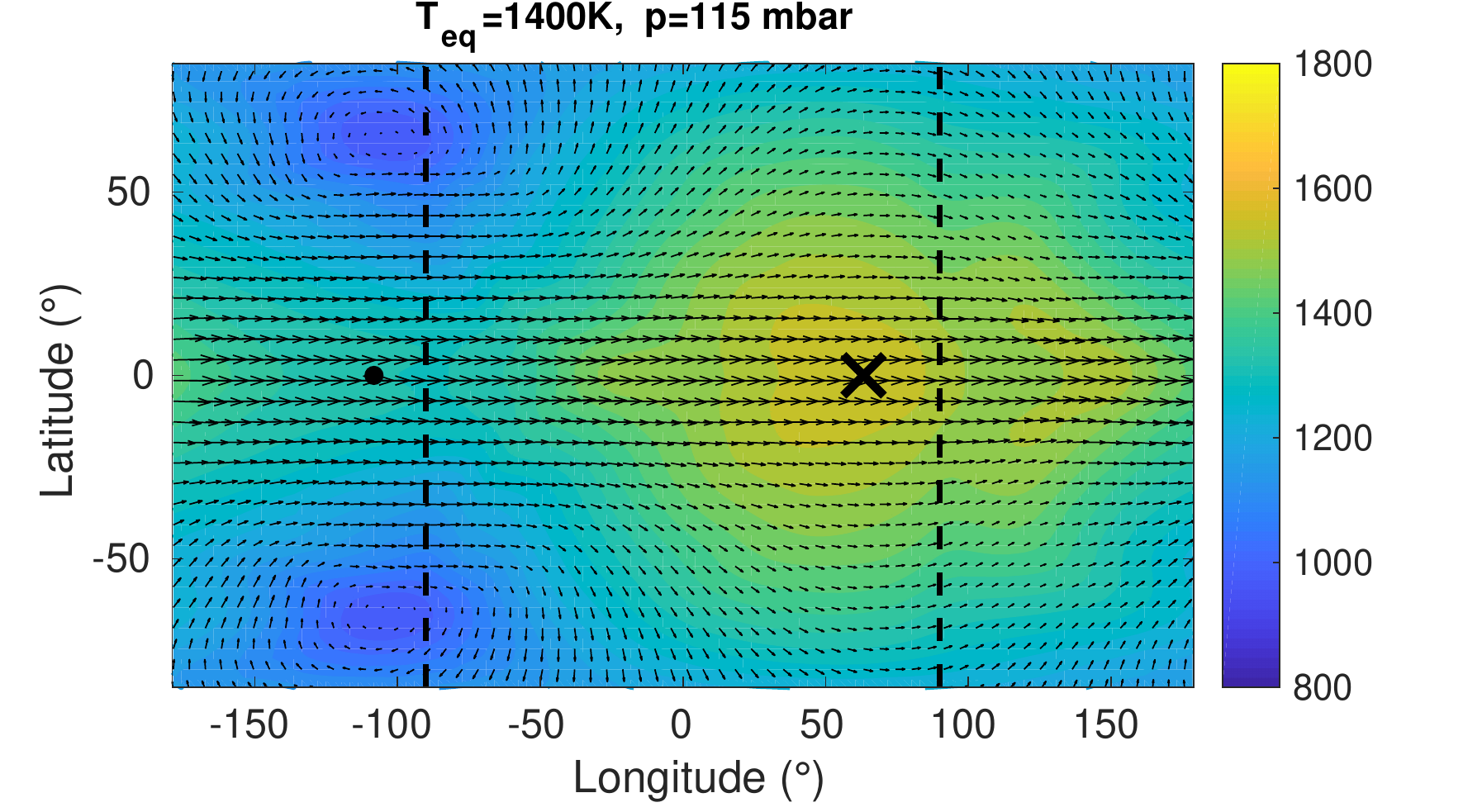}
\includegraphics[width=0.33\linewidth]{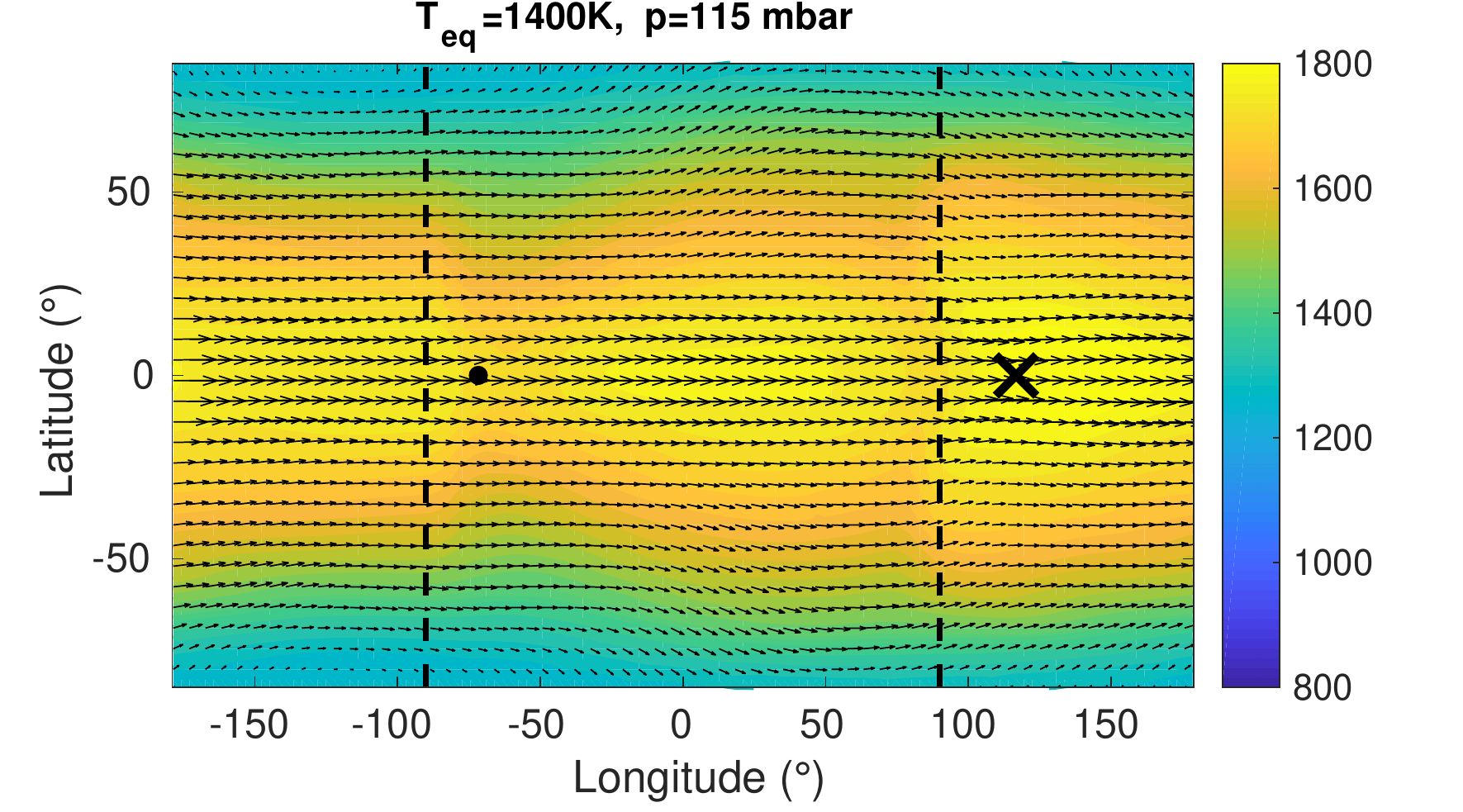}
\includegraphics[width=0.33\linewidth]{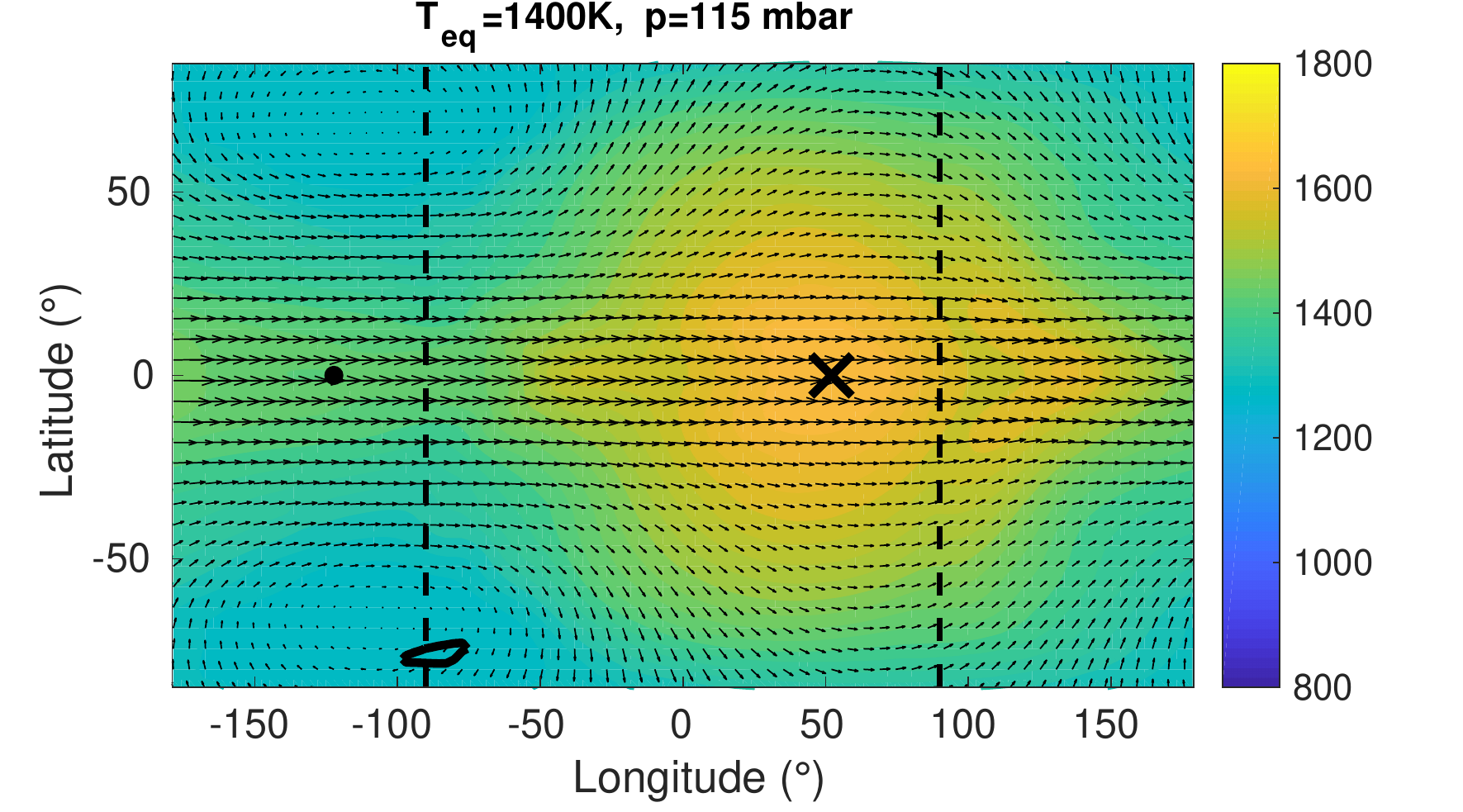}

\includegraphics[width=0.33\linewidth]{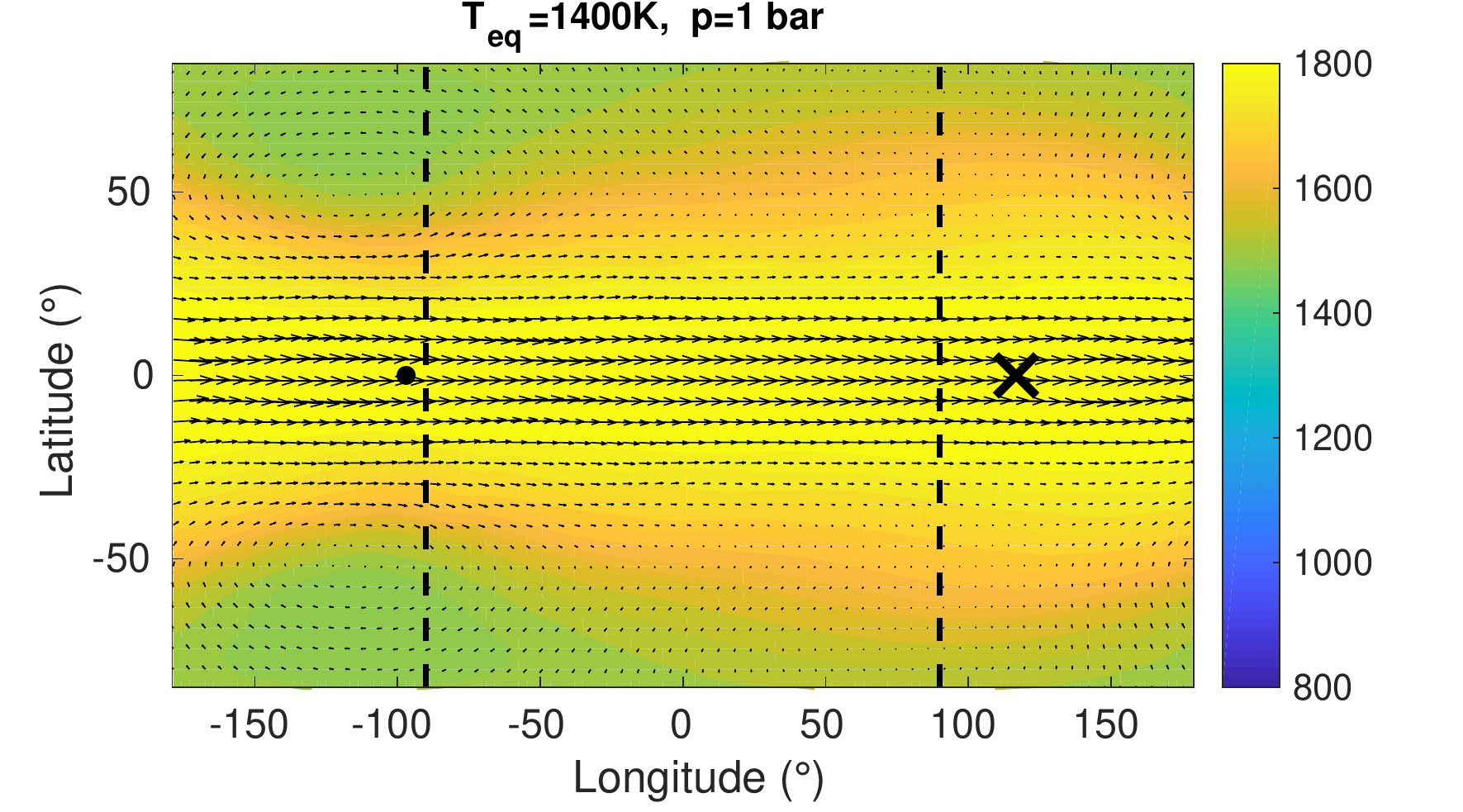}
\includegraphics[width=0.33\linewidth]{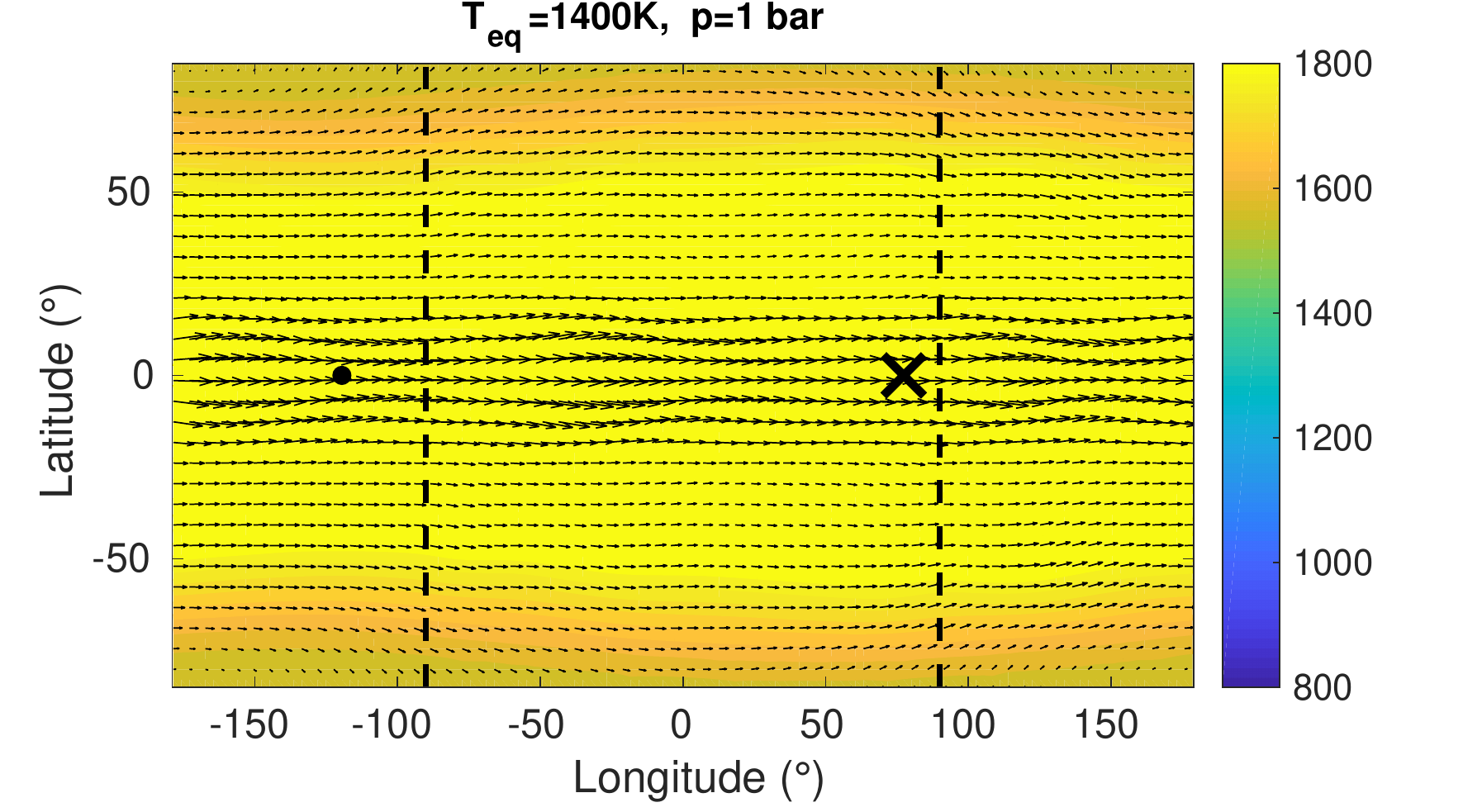}
\includegraphics[width=0.33\linewidth]{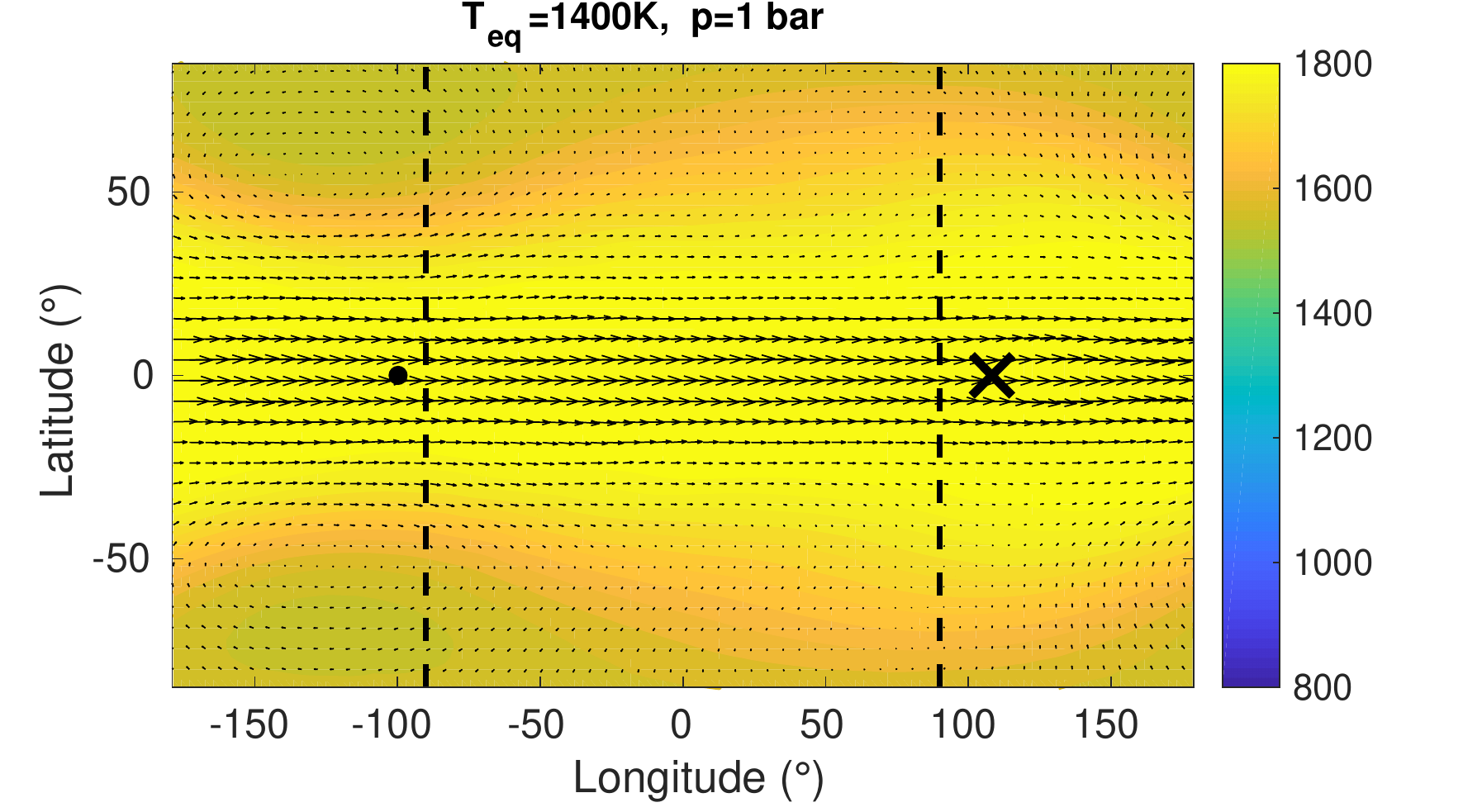}

\caption{Temperature and wind map of the $T_{\rm eq}=1400K$ models at different pressure levels (rows). The left column is the cloudless case, the middle column is the nightside cloud case whereas the right column is the temperature-dependent MnS cloud case. The central longitude of the hottest (resp. coldest) hemisphere is marked as a cross (resp. dot). In the temperature-dependent MnS cloud case the condensation curve of the MnS clouds are 
overlayed.}
\label{fig::Temp1400}
\end{figure*}

\begin{figure*}

\includegraphics[width=0.33\linewidth]{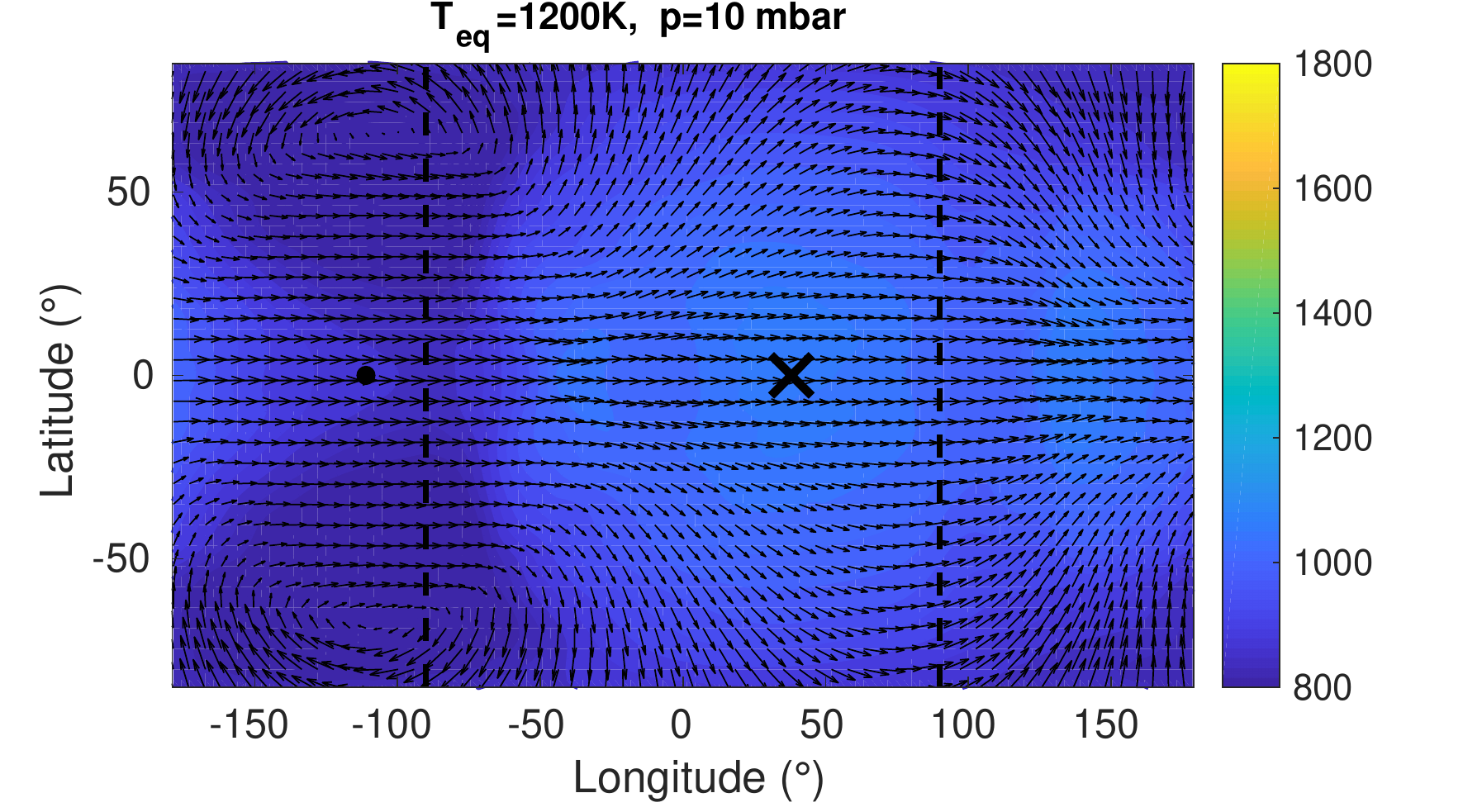}
\includegraphics[width=0.33\linewidth]{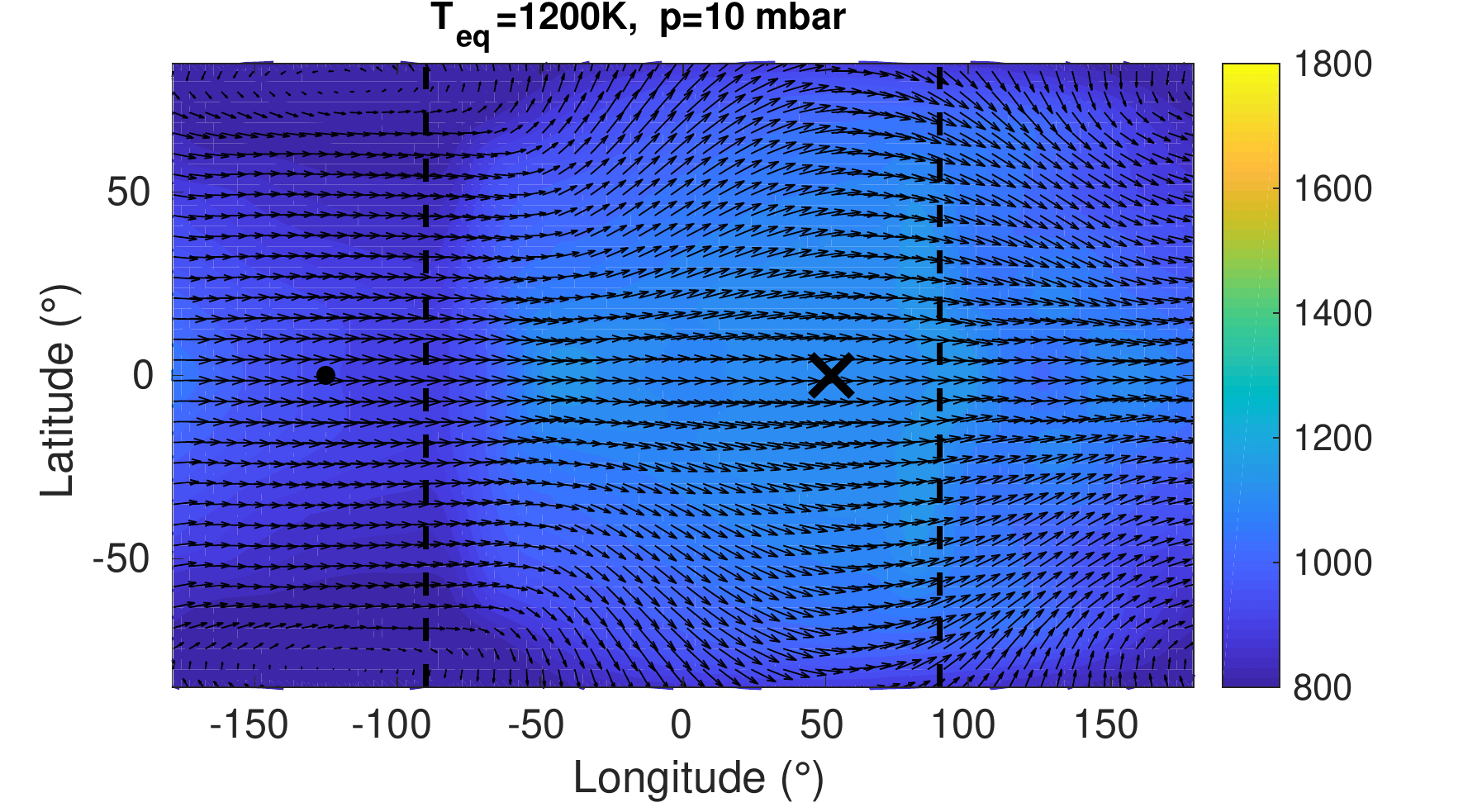}
\includegraphics[width=0.33\linewidth]{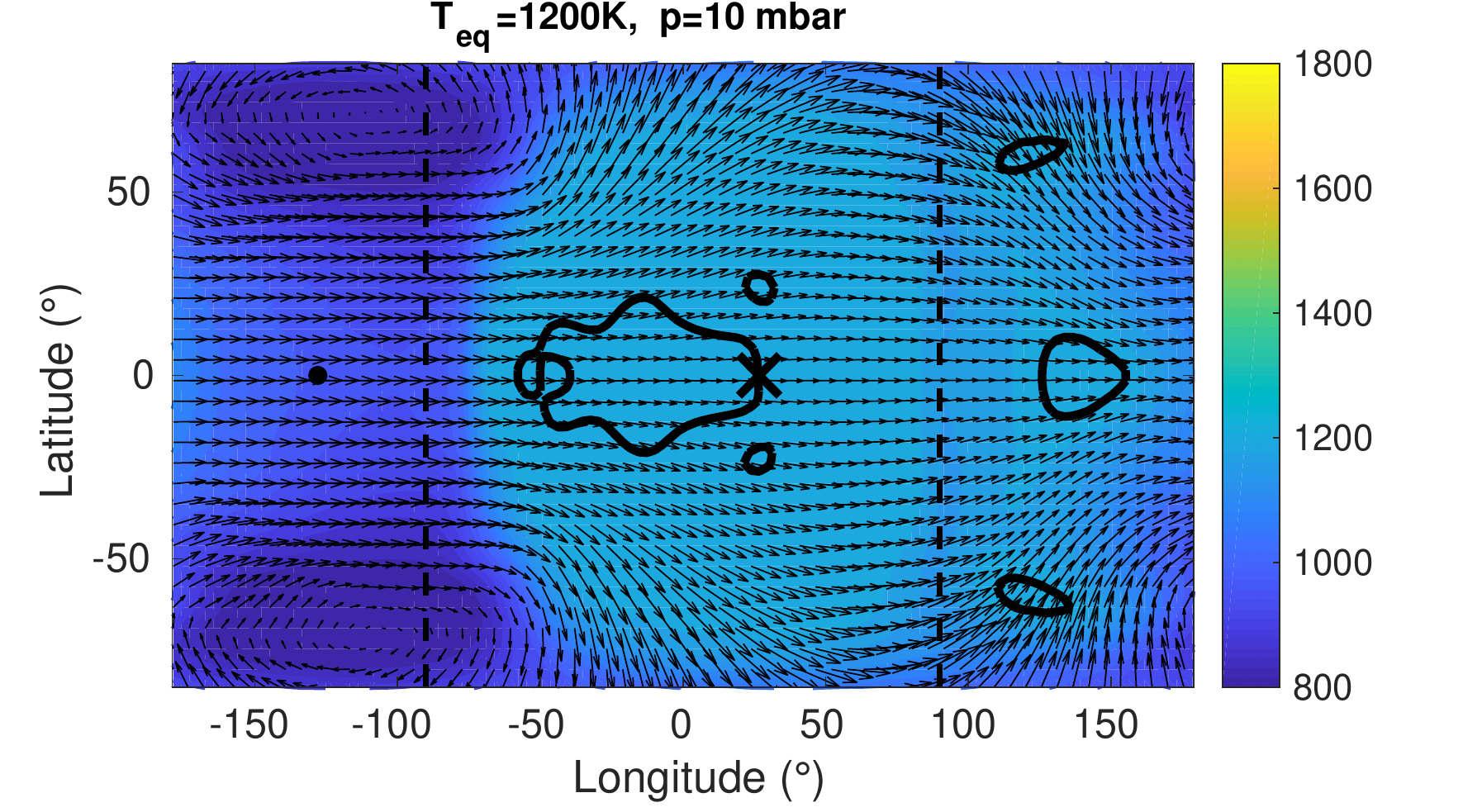}

\includegraphics[width=0.33\linewidth]{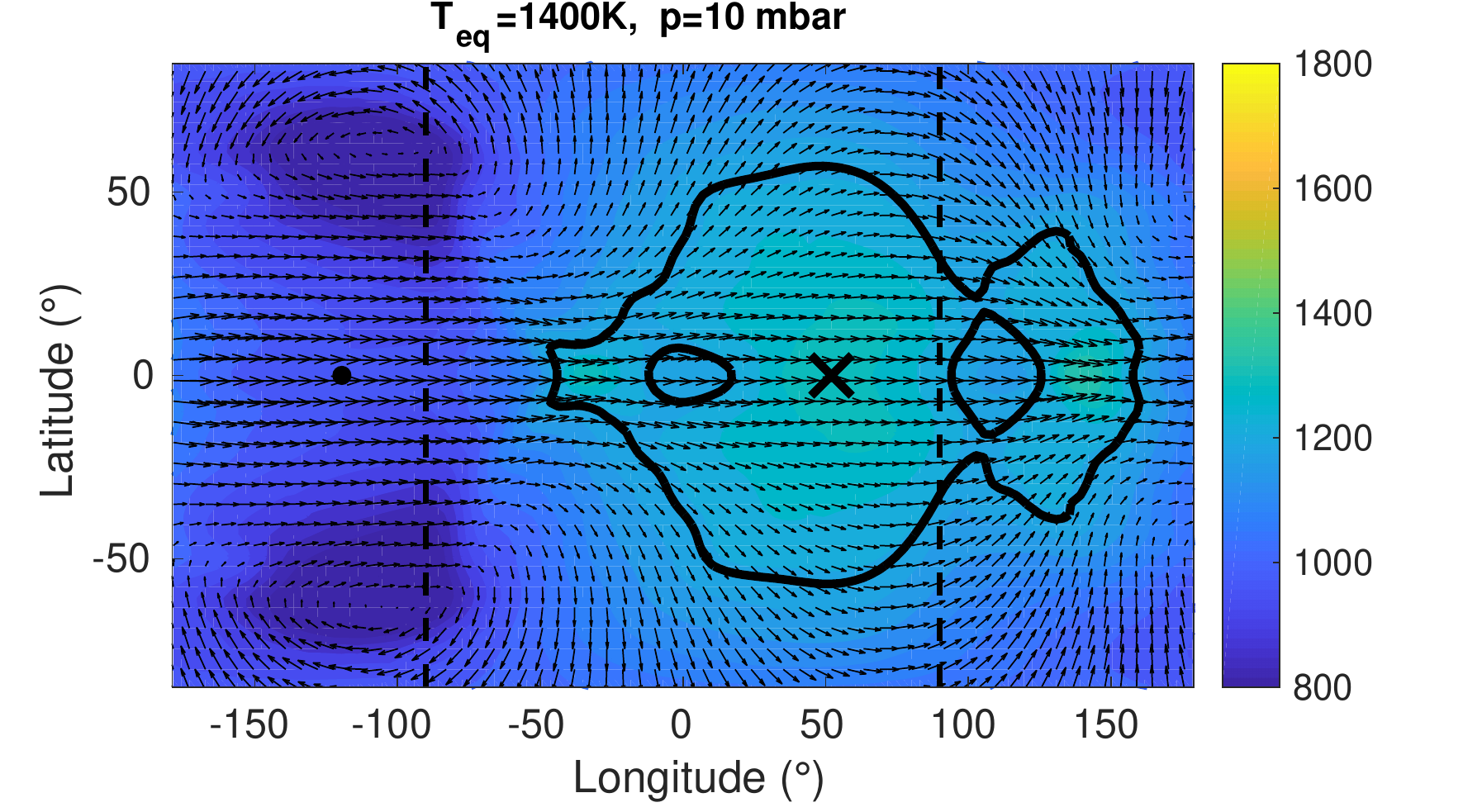}
\includegraphics[width=0.33\linewidth]{Images/Teq1400-g10-TL-NS-Clouds-MnS-prescription-2-timeAv-temp-xy-0001036800-29-Limb-T4-tit.pdf}
\includegraphics[width=0.33\linewidth]{Images/Teq1400-g10-TL-MnS-1microns-timeAv-temp-xy-0001036800-29-Limb-T4-tit-With-MnS-Cond.pdf}

\includegraphics[width=0.33\linewidth]{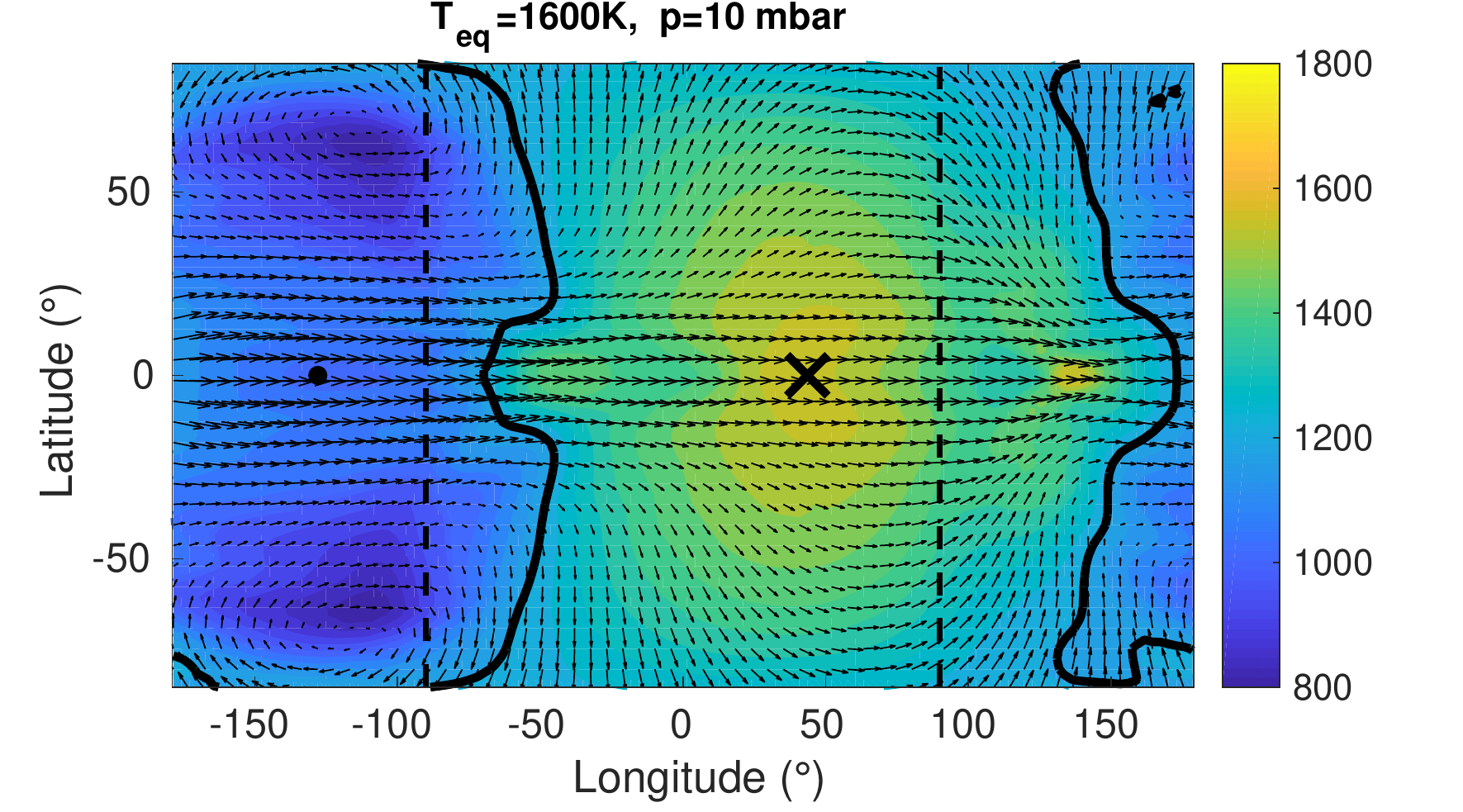}
\includegraphics[width=0.33\linewidth]{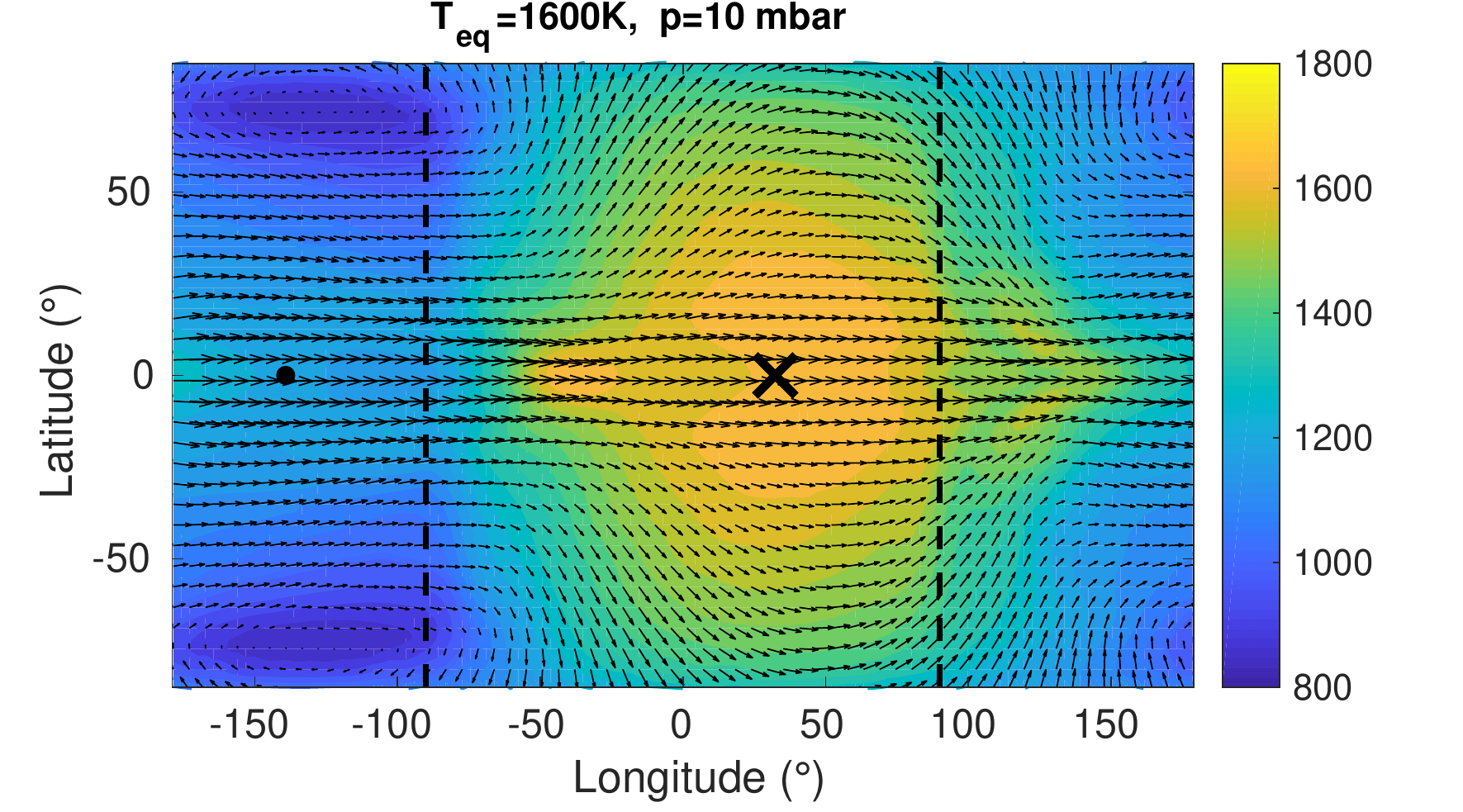}
\includegraphics[width=0.33\linewidth]{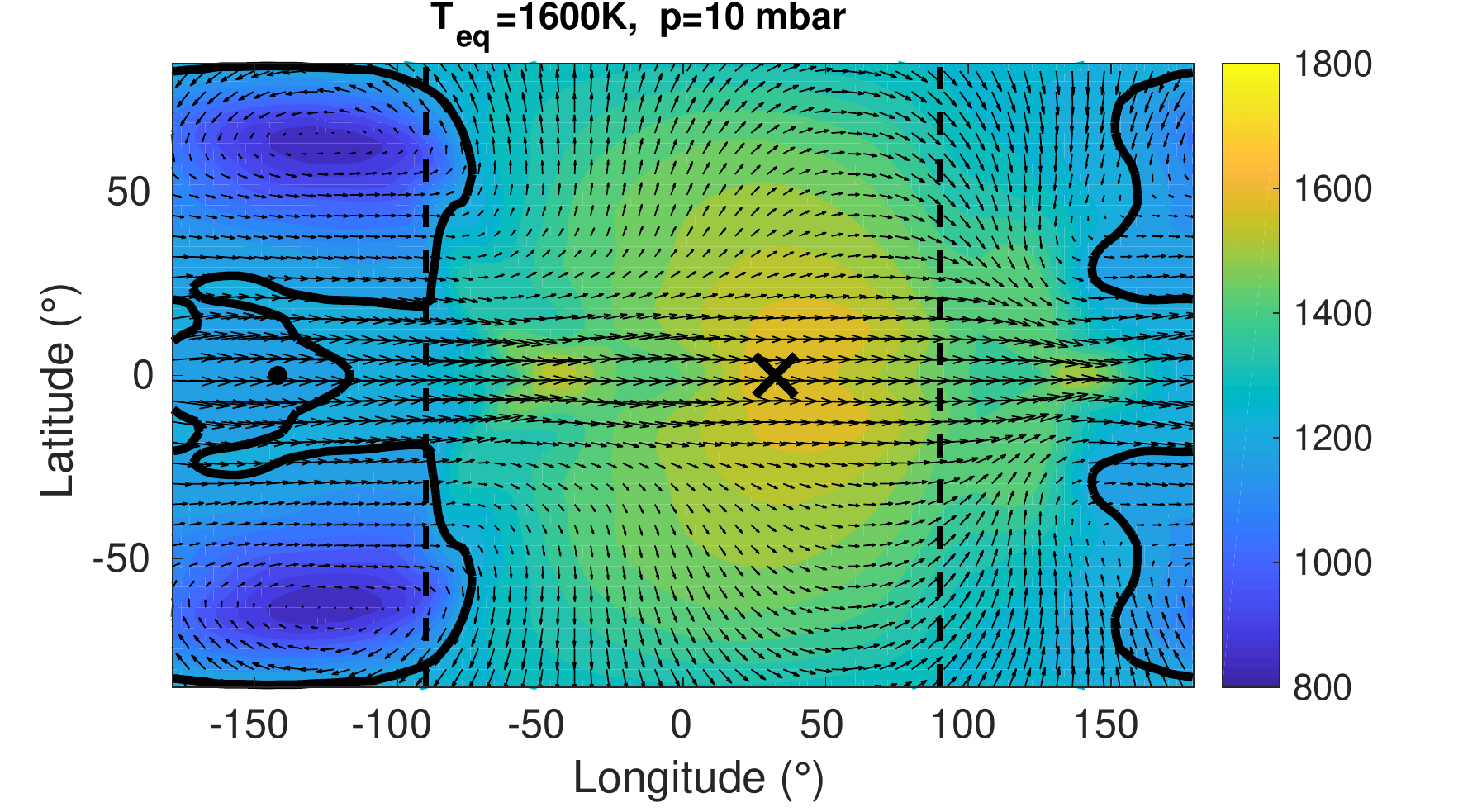}

\includegraphics[width=0.33\linewidth]{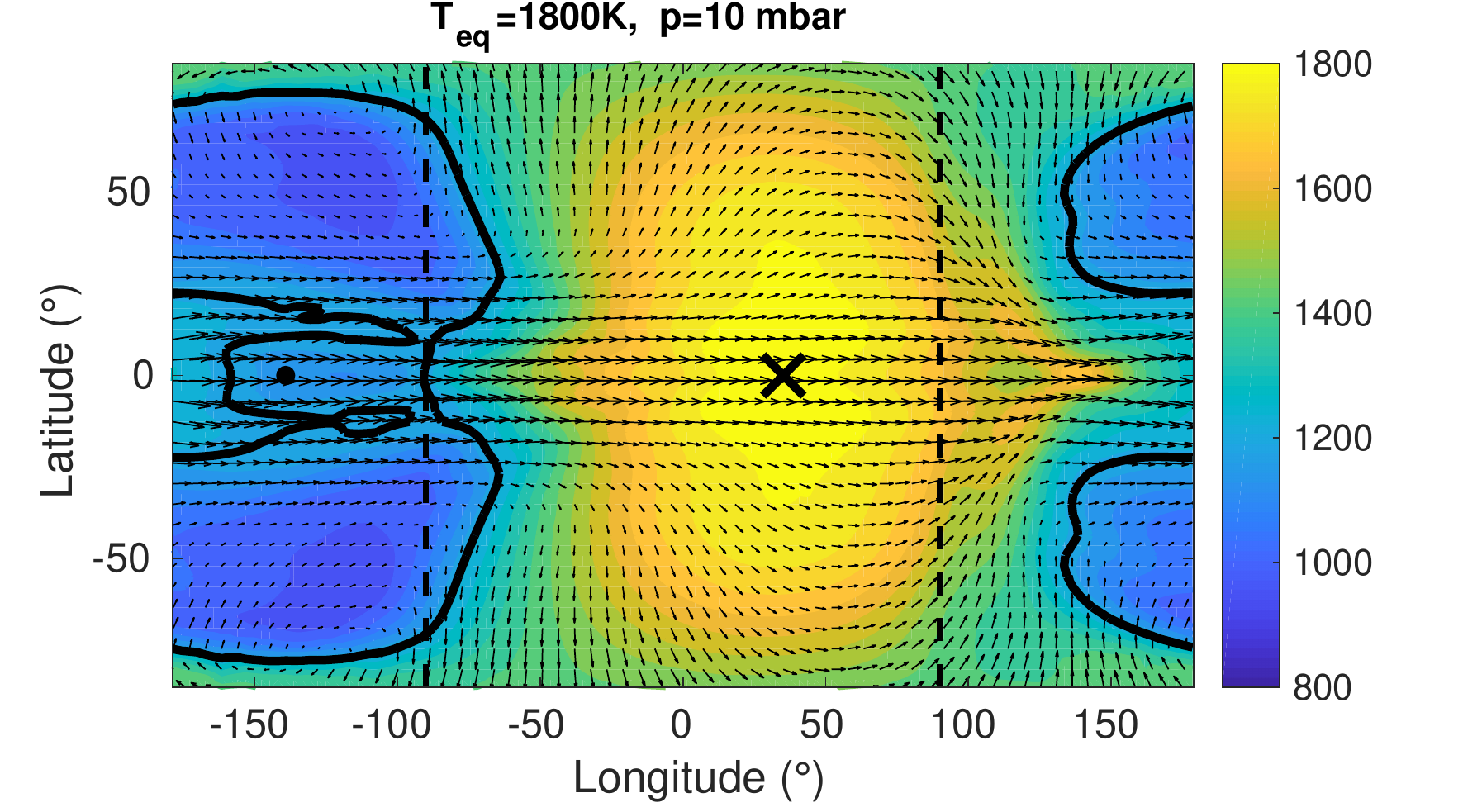}
\includegraphics[width=0.33\linewidth]{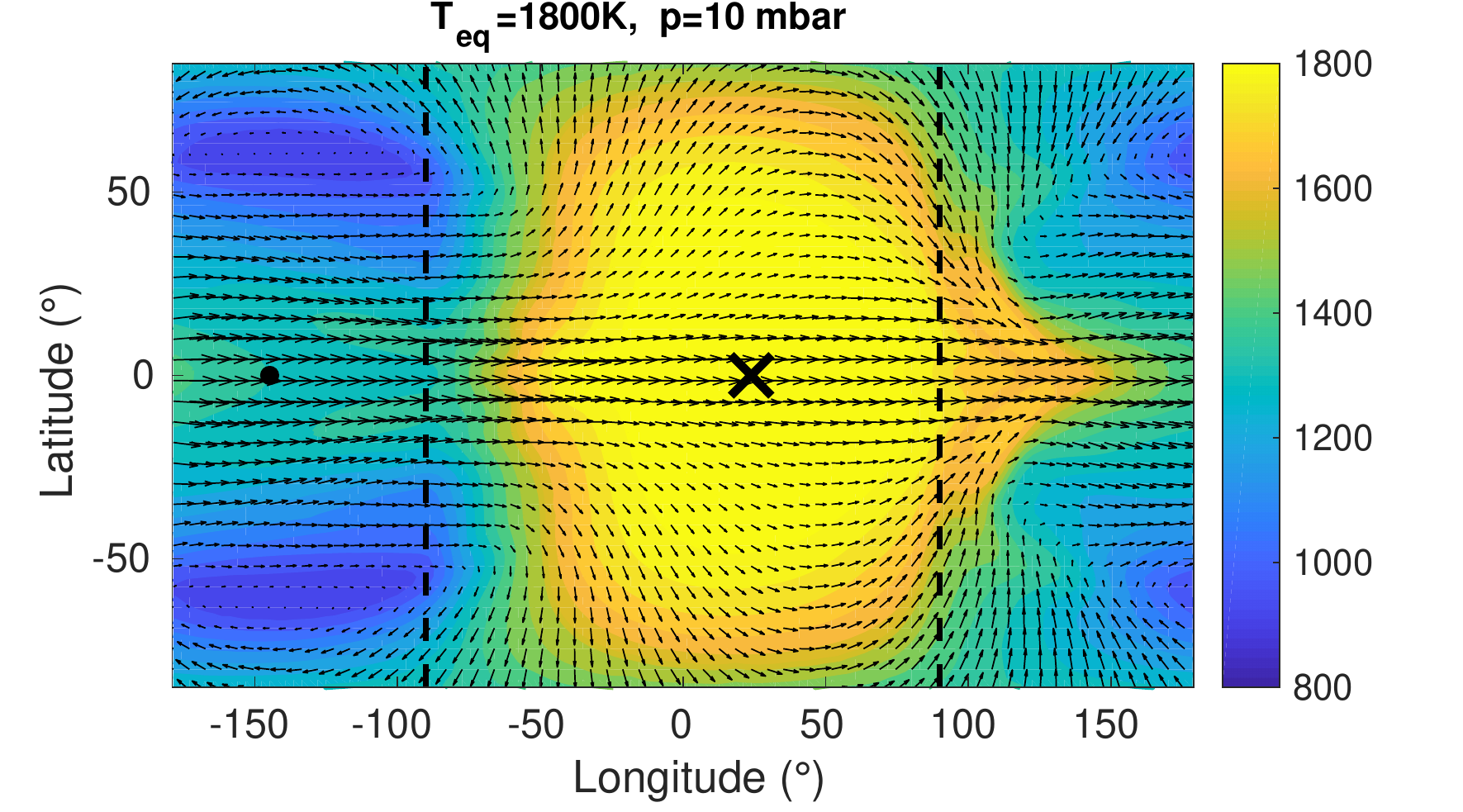}
\includegraphics[width=0.33\linewidth,trim=0 1 1 1]{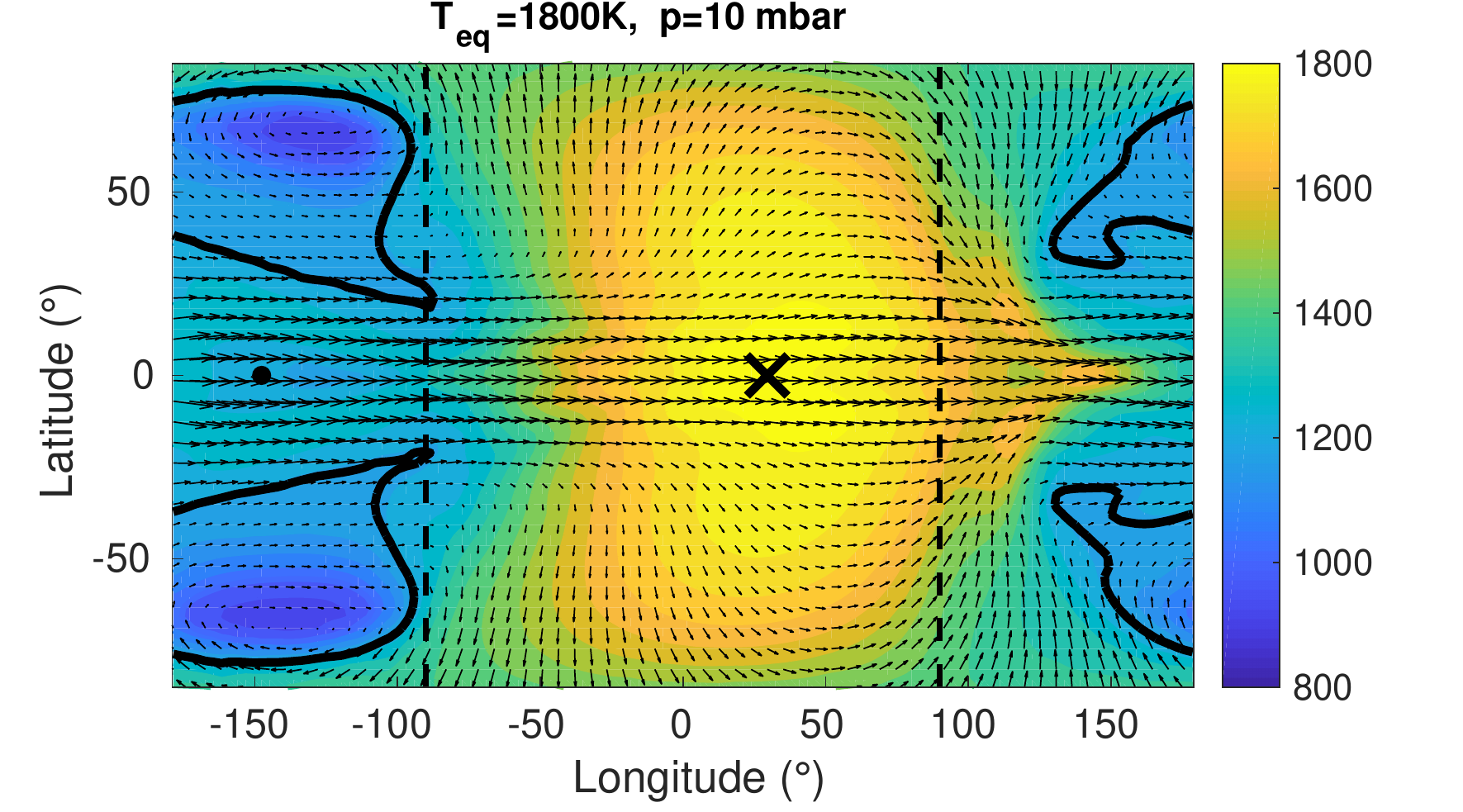}

\includegraphics[width=0.33\linewidth]{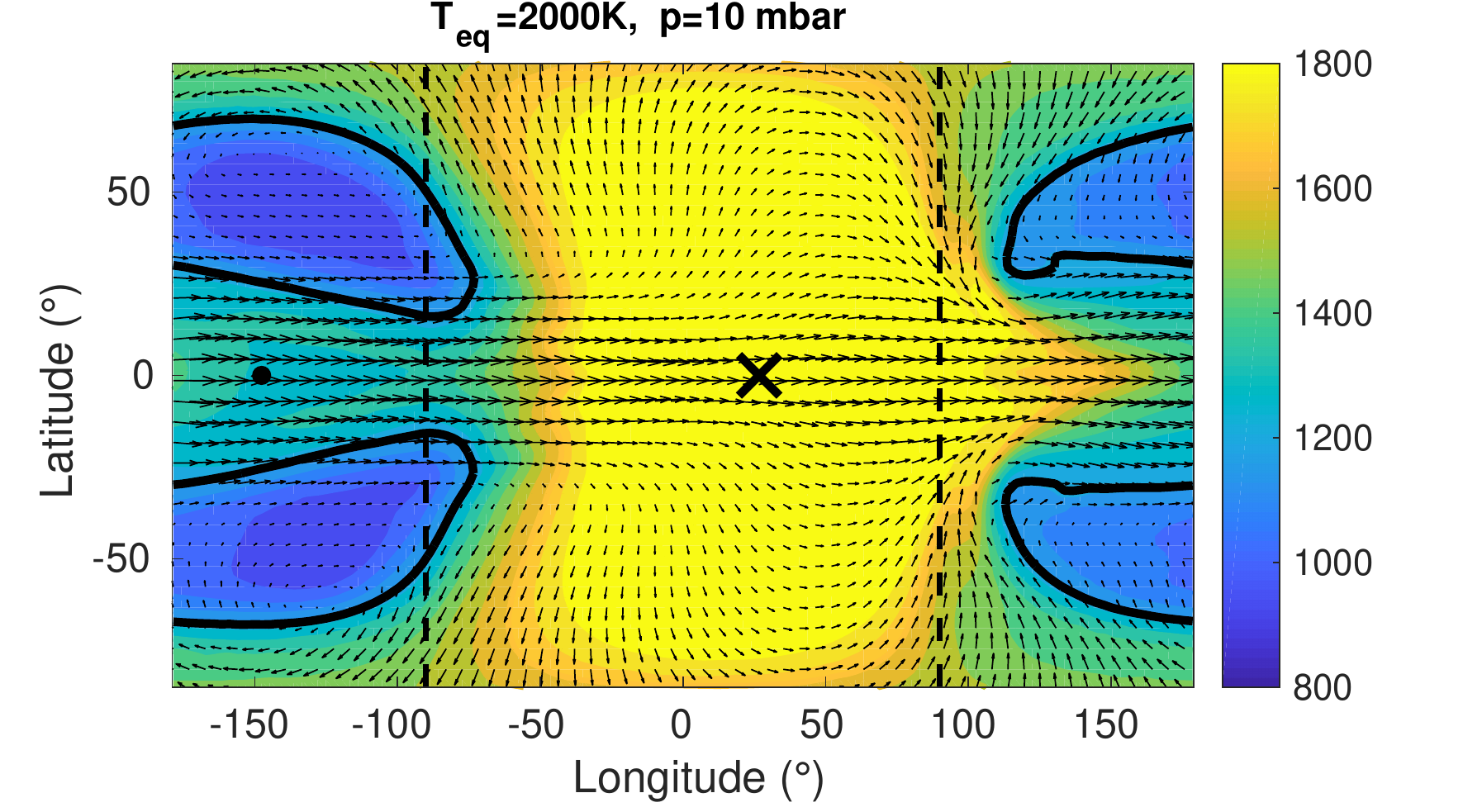}
\includegraphics[width=0.33\linewidth]{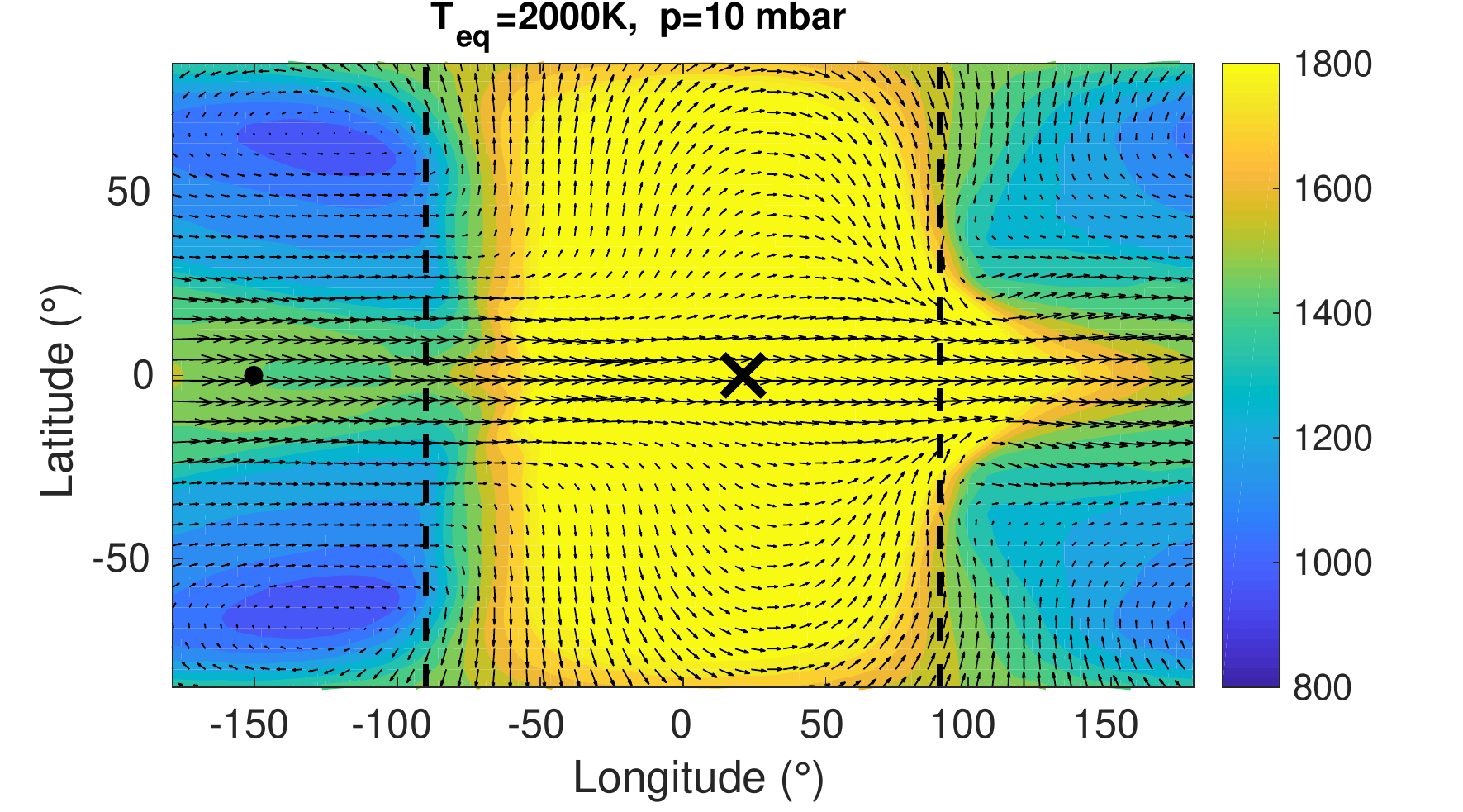}
\includegraphics[width=0.33\linewidth]{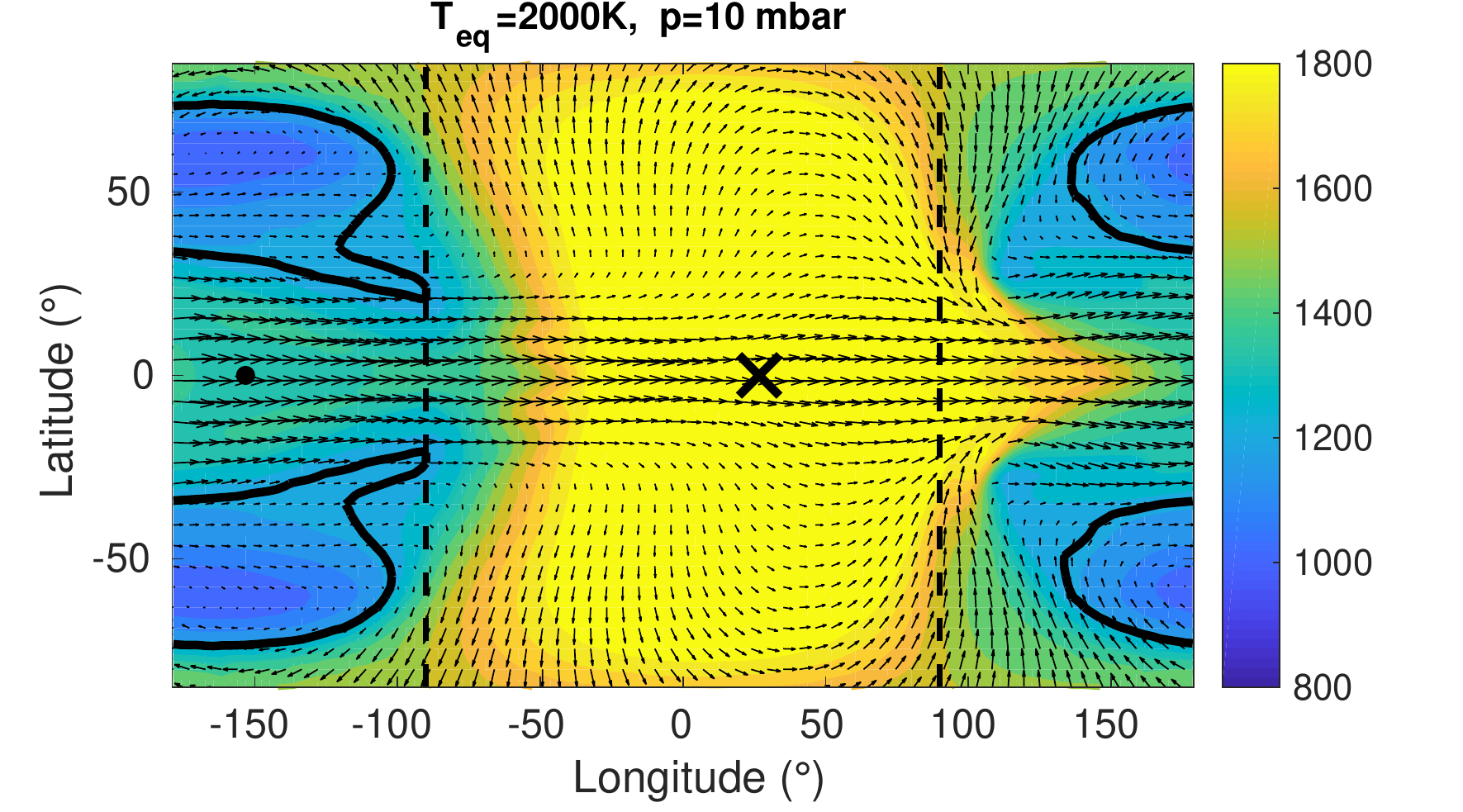}

\hfill
\caption{Temperature and cloud maps at 10mbar for models with different equilibrium temperature without clouds (left), with radiatively active nightside clouds (centre) and with radiatively active temperature-dependent MnS clouds (right). The cross shows the centre of the hottest hemisphere and the dot shows the coolest one calculated using a $T^4$ average. The plain line on the leftmost and rightmost panels show the condensation temperature of MnS. These clouds are in both day and nightside at low equilibrium temperature but only on the nightside at high equilibrium temperatures.}
\label{fig::TempMapMnSTeq}
\end{figure*}

\subsection{Cloud models}
\label{Sec::Methods_clouds}
{ In addition to our cloudless simulations we perform extra series of calculations under two different assumptions: prescribed nightside clouds and temperature-dependent clouds. We also show a third model were clouds are added to the cloudless model only when calculating the final spectra (e.g. post-processed clouds). }

The prescribed nightside clouds model assumes that cloud particles are present at all longitudes and latitudes on the nightside between 200 mbar and the top of the model ($\mu$bar). We further assume that our clouds are made of manganese sulphide (MnS) with the mixing ratio of the cloud constituents given by the solar manganese abundance ($n({ \rm Mn})=10^{-6.4}n({\rm H})$). The cloud particles are assumed to have a density of $4 g/cm^3$ and the number of particles is calculated through mass conservation by assuming that all cloud particles are spheres of radius $a$. 

{ The vertical extent of our cloud is the maximum one for the species considered. Although real cloud should have a more limited extant~\citep{Lee2016,Lines2018}, is was shown by~\citet{Roman2019} that the vertical extant of the clouds does not affect qualitatively the behaviour of the simulations.}

{ We choose manganese sulphide clouds for two reasons. First, they are proposed as a good alternative to silicate clouds in planets cooler than $1600\,\rm K$ by~\citet{Parmentier2016}. Second, they have a smaller total abundance and hence a smaller opacity than silicate clouds, which allows for a better numerical stability of the model. Overall, we do not believe that our results for the nightside clouds are strongly dependent on the choice of cloud species, as long as optically thick clouds are formed.}

The temperature-dependent cloud model assumes that the amount of cloud forming is determined locally so that the partial vapour pressure of MnS is always equal to the saturation vapour pressure. As discussed in \citet{Parmentier2016}, this corresponds to a situation where vertical atmospheric mixing is dominant over any other microphysical timescale or dynamical transport. Such a model provides a reasonable estimate of the horizontal distribution of clouds in hot Jupiters but likely overestimates the total amount of clouds forming. { The results from the temperature-dependent cloud model is going to be very dependent on the choice of cloud species, since this choice would change the condensation curve and hence the exact location of the clouds. Nonetheless, we use them here as an illustration to understand how temperature-dependent clouds of any kind (e.g. silicate or metallic cloud species) can modify the observable trends with equilibrium temperature.}

In both cloud models the clouds take an active part in the radiative balance of the global circulation model from both scattering and absorption of radiation. In our last set of models, the clouds are, however, radiatively inactive in the hydrodynamic simulation and only added when calculating the final spectrum. We describe these as post-processed clouds following~\citet{Parmentier2016}. Although these models are not energetically self-consistent (i.e. they usually underestimate the outgoing flux compared to a radiatively coupled simulation), they are useful to provide a quick estimate of the potential radiative importance of clouds of different chemical compositions in shaping the observations. 

\subsection{Differences with other studies}
 Our modelling framework is significantly different than recent work studying the effect of clouds on the atmospheric circulation of hot Jupiters. { We differ from~\citet{Mendonca2018a}, \citet{Roman2017} and \citet{Roman2019} who used semi-grey radiative transfer whereas we use non-grey radiative transfer. Compared to~\citet{Dobbs-Dixon2013}  who used a grey infrared cloud opacity, we use an opacity from Mie scattering calculation that decreases at long wavelength and show that this non-grey effects of the clouds are key to understand their observational effects. 
 
Our model is most similar to the one of~\citet{Lines2019} who used a non-grey global circulation model of HD 209458b coupled with the 1D cloud model Eddysed, that naturally form non-grey clouds on the nightside of the planet. However, their Eddysed model leads to clouds with a very low single-scattering albedo (due to their high iron content), whereas the MnS cloud optical properties used here have a much larger single-scattering albedo.}

Our cloud model is less detailed than the ones of~\citet{Lee2016} or~\citet{Lines2018} that use first principles to calculate the cloud properties for individual planets. Instead we follow the approaches of ~\citet{Parmentier2016} or ~\citet{Roman2017} where we either prescribe the location of the clouds and their properties or let the species-specific condensation temperature dictate the longitudinal cloud distribution. This allows us to investigate the effect of clouds over a larger range of planetary and cloud parameters, trying to explain trends seen in the population of planets rather than the detailed observations of a given planet.

\section{Atmospheric circulation for a range of equilibrium temperatures}
\label{sec::GCMResults}

When clouds are present in a planetary atmosphere, they strongly modify the atmospheric opacities. As a consequence, the thermal structure of the atmosphere changes to maintain planet-wide radiative equilibrium. Clouds usually have two main, but opposite effects. By increasing the thermal opacities, they increase the greenhouse effect and warm up the atmosphere below them.  By increasing the scattering at optical wavelengths, they increase the albedo of the planet, reducing the atmospheric temperature. On a hot Jupiter, the balance between the greenhouse and the albedo effect of the clouds will strongly depend on the cloud distribution: dayside clouds can act both via the albedo and greenhouse effects whereas nightside clouds can only act via the greenhouse effect. 

\subsection{Atmospheric circulation}
\label{sec::AtmCirc}

We now describe the general behaviour of our three sets of global circulation models: cloudless, nightside clouds, and temperature-dependent cloud. Figure~\ref{fig::Temp1400}  depicts the vertical structure of temperature and winds over the globe for the $T_{\rm eq}=1400\,\rm K$ model. Pressures of 1  to 100 mbar are shown, which straddles  the infrared photospheric levels across  most wavelengths. Figure ~\ref{fig::TempMapMnSTeq} shows how these temperature and wind structures vary with irradiation level across a wide range of models, from $T_{\rm eq}=1200$ to  $2000\, \rm K$.

All our models develop an atmospheric flow qualitatively similar to previously published models of hot Jupiters over this range of equilibrium temperatures~\citep[e.g.][]{Komacek2017}: a super-rotating jet forms at the equator whereas cold vortices form at the mid-latitude in the nightside. For all three sets of models both the dayside temperature and the day/night temperature contrast on isobars increases with equilibrium temperature. This confirms the qualitative findings of ~\citet{Perez-Becker2013a}, ~\citet{Komacek2017} and~\citet{Komacek2018} obtained with grey radiative transfer.

When nightside clouds are present (see middle column of Figures~\ref{fig::Temp1400} and~\ref{fig::TempMapMnSTeq}), the greenhouse effect of the clouds is dominant and both the dayside and the nightside of the planet warm up, { similar to what is seen for the Eddysed model of HD209458b by~\citet{Lines2019} and the semi-grey model of WASP-43b in~\citet{Mendonca2018a}}. Because the nightside warms more than the dayside, the day/night temperature difference on isobars is greatly reduced. The eastward shift of the hottest hemisphere is reduced by $\approx 10^{\circ}$ but stays largely between $25$ and $60^{\circ}$. 

The flow pattern is also affected by the presence of nightside clouds. The most noticeable difference is the strong reduction in the strength of the mid-latitudes nightside cold vortices. These vortices are usually the coldest point of these simulations and warm up when their strength is reduced. Because they can play an important role setting the global chemical abundance through 3D quenching processes~\citep{Drummond2018a,Drummond2018b}, a change in their strength could have important implication on the dayside and limb atmospheric abundances measured during transit and secondary eclipses. 

 We now look at the models including temperature-dependent MnS clouds (right panel of Figure~\ref{fig::TempMapMnSTeq}), where the spatial distribution of the MnS clouds are determined by the actual saturation of MnS. In short, if the temperature is hotter than the condensation temperature, then the cloud must disappear. As a consequence, the cloud distribution on isobars follows temperature contours. Rather than an exact day/night contrast, the clouds are present in part of the western side of the dayside and are not present in the western side of the nightside. The cloud map can be well approximated to a day/night contrast shifted eastward, with the cloudy/cloudless hemispheres shifted eastward compared to the day/night hemisphere. The eastward shift of the cloud map is directly linked to the eastward shift of the temperature map.

The change in the thermal structure is qualitatively similar but smaller than in the prescribed nightside clouds models. Both the dayside and the nightside get warmer, the planetary-scale temperature contrast on isobars gets smaller and the hot spot shift is slightly reduced but never reaches zero. The 1200K model is an exception where the temperature-dependent MnS clouds have a stronger effect on the temperatures than in the the nightside clouds model. That is because the clouds of the nightside cloud models were chosen based on the nightside cloud coverage of the 1400K case temperature-dependent MnS cloud case. In the 1200K case the nightside cloud model has clouds with a base at lower pressures than in the temperature-dependent cloud model, having therefore a smaller effect on the thermal profiles. 

The radiative feedback of the clouds also changes the spatial distribution of the clouds that would be expected from the cloudless model. By warming up the atmosphere, the clouds create a more inhospitable dayside and { their abundance in the dayside is reduced~\citep[see also][]{Charnay2015a,Oreshenko2016,Roman2019,Lines2019}}. The cloud radiative feedback increases the dayside temperature by $100-200\,\rm K$. As a consequence, the distribution of clouds on a given model with temperature-dependent clouds looks similar to the distribution of clouds in the model without radiative feedback with an equilibrium temperature $100-200\,\rm K$ hotter. This confirms the finding from Fig. 9 from \citet{Parmentier2016} that the radiative feedback of the clouds should not affect the behaviour of the trends seen in the phase curve offset vs. equilibrium temperature but merely shift them by $\approx 100-200\, \rm K$ in equilibrium temperature. 

\subsection{Heat redistribution}
\label{sec::quantitative}

The main role of the atmosphere is to transport the heat deposited on the dayside to the nightside of the planet. The heat redistribution can be estimated directly from observations, either from estimating the dayside brightness temperature from a secondary eclipse spectrum and assuming a prior on the Bond albedo, or through measuring both the dayside and nightside brightness temperatures with a phase curve measurement~\citep[see e.g.][]{Schwartz2017}. 

We now estimate the heat redistribution from our models by comparing the dayside temperature of the planet to its equilibrium temperature. Given that the albedo is small for both the cloudless case and the nightside cloud case the dayside temperature is a good proxy for the redistribution. We first calculate the dayside effective temperature, $T_{\rm day}$ by integrating over all wavelengths the flux emitted from the dayside and finding the temperature of the blackbody emitting the same total flux. We then define the heat redistribution parameter, $f$, as the ratio of the dayside brightness temperature to the equilibrium temperature to the power of 4~\citep[see e.g.][]{Arcangeli2018}. This is related to the redistribution efficiency $\epsilon$ defined in~\citet{Cowan2011} by $\epsilon=1+0.6\left(1-f\right)$. $f=1$ or $\epsilon=1$ is a planet wide energy redistribution, $f=2$ or $\epsilon=0.4$ corresponds to a dayside only heat redistribution whereas the limiting value $f=2.666$ or $\epsilon=0$ corresponds to no heat transport.

\begin{figure}
\includegraphics[width=\linewidth]{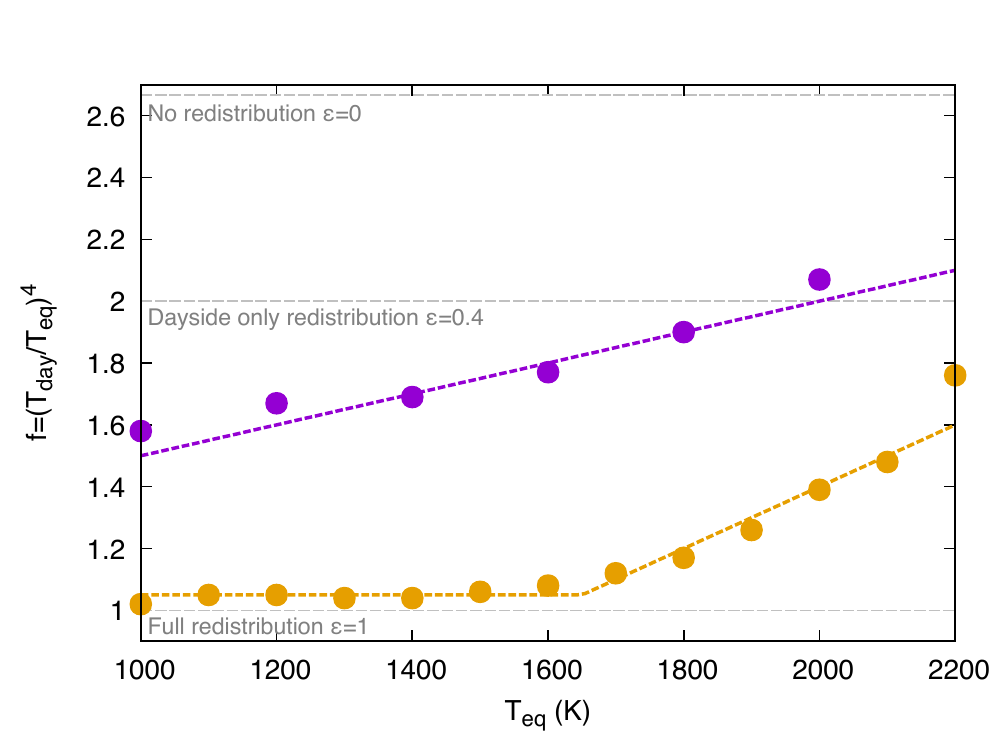}
\caption{Heat redistribution parameter $f=(T_{\rm day}/T_{\rm eq})^4$ for models with (purple) and without (orange) nightside clouds. We additionally show the fits provided in section.~\ref{sec::quantitative}. When nightside clouds are present the heat transport is much less efficient.}
\label{fig::Ujet}
\end{figure}

The value of $f$ calculated from our global circulation model outputs is given in the bottom panel of Figure~\ref{fig::Ujet}. Cloudless hot Jupiters with temperatures cooler than $1650\,\rm K$ are very good at redistributing their energy and the redistribution parameter is constant and close to 1, corresponding to a very efficient day-to-night redistribution and a dayside temperature close to the equilibrium temperature. For larger equilibrium temperatures the heat redistribution becomes poorer and $f$ increases linearly with equilibrium temperature. Our redistribution in the cloudless case is very close to the one calculated from the semi-grey model of~\citet{Perna2012} and ~\citet{Komacek2017}, showing that the grey approach does capture the main mechanisms determining the day/night heat transport for cloudless atmospheres (see Appendix~\ref{AppA} for more details). 

For the models with nightside clouds the heat redistribution is much poorer, with values of $f$ ranging from $1.6$ at low temperatures and reaching a dayside only redistribution for equilibrium temperature of 2000 K. When nightside clouds are present, heat transported from the dayside cannot escape easily on the nightside and is returned to the dayside via the circulation. This change is very strong, as seen in Fig.~\ref{fig::Perna}, it corresponds to the heat transport efficiency obtained by~\citet{Komacek2017} for the unrealistically short drag timescales of $10^3s$.

The heat redistribution parameter can be used as a good prior for the expected dayside temperature of hot Jupiters, which can be particularly useful for 1D retrieval models~\citep[e.g.][]{Arcangeli2018}. We show below analytical fits to the values derived from the GCM. In the cloudless case we have:
\begin{equation}
f=\left\{\begin{array}{ll}
                  1.05 &\textrm{ for } 1000\, \rm K<T_{\rm eq}<1650\, \rm K\\
                  10^{-3}T_{\rm eq}-0.6 &\textrm{ for } 1650\, \rm K<T_{\rm eq}<2200\, \rm K.
                \end{array}
              \right.
\label{eq::fFit}
\end{equation}
And in the case with nightside cloud we obtain: 
\begin{equation}
f=1+0.5\frac{T_{\rm eq}}{10^3}.
\end{equation}

\subsection{Atmospheric timescales}
The efficiency of the heat redistribution is thought to be determined by a competition between advective or wave driven heat transport and the radiative losses~\citep{Showman2002,Perez-Becker2013a,Komacek2016,Zhang2018}. A hotter planet should be more efficient at loosing heat through radiation, lowering the heat transport, but this might be balanced by a stronger atmospheric circulation. We now estimate these timescales from the GCM outputs in order to understand which mechanisms are responsible for the trends in heat redistribution seen in Fig~\ref{fig::Ujet}.

 { We define the advective timescale as the timescale it takes for a parcel of gas to be advected across one hemisphere by the equatorial zonal jet :
\begin{equation}
\tau_{\rm adv}\approx\frac{\pi R_{\rm p}}{U_{\rm jet}}\propto\frac{1}{T_{\rm eq}}.
\end{equation}
We define the radiative timescale as the time needed for a parcel of gas to loose a significant portion of energy  by radiation \citep[see][]{Showman2002}):
\begin{equation}
\tau_{\rm rad}\approx\frac{P_{\rm photo}}{g}\frac{c_{\rm p}}{4\sigma T_{\rm photo}^3}\propto \frac{P_{\rm photo}}{T_{\rm photo}^3},
\label{eq::RadTime}
\end{equation}
Finally we define the wave timescale as the timescale for gravity waves to travel across a hemisphere of the planet in the isothermal case~\citep{Perez-Becker2013a}:
\begin{equation}
\tau_{\rm wave}\approx \frac{\pi R_p}{\sqrt{k_{\rm B}T_{\rm photo}/\mu}}\propto \frac{1}{\sqrt{T_{\rm photo}}}.
\label{eq::RadTime}
\end{equation}
In the above equations} $R_{\rm p}$ is the planetary radius, $U_{\rm jet}$ the mean wind speed at the equator, $c_{\rm p}$ the heat capacity at constant pressure, $\sigma$ the Stefan-Boltzman constant and $\mu$ the mean molecular weight. $P_{\rm photo}$ and $T_{\rm photo}$ are the photospheric pressure and temperature. 

\begin{figure}
\includegraphics[width=\linewidth]{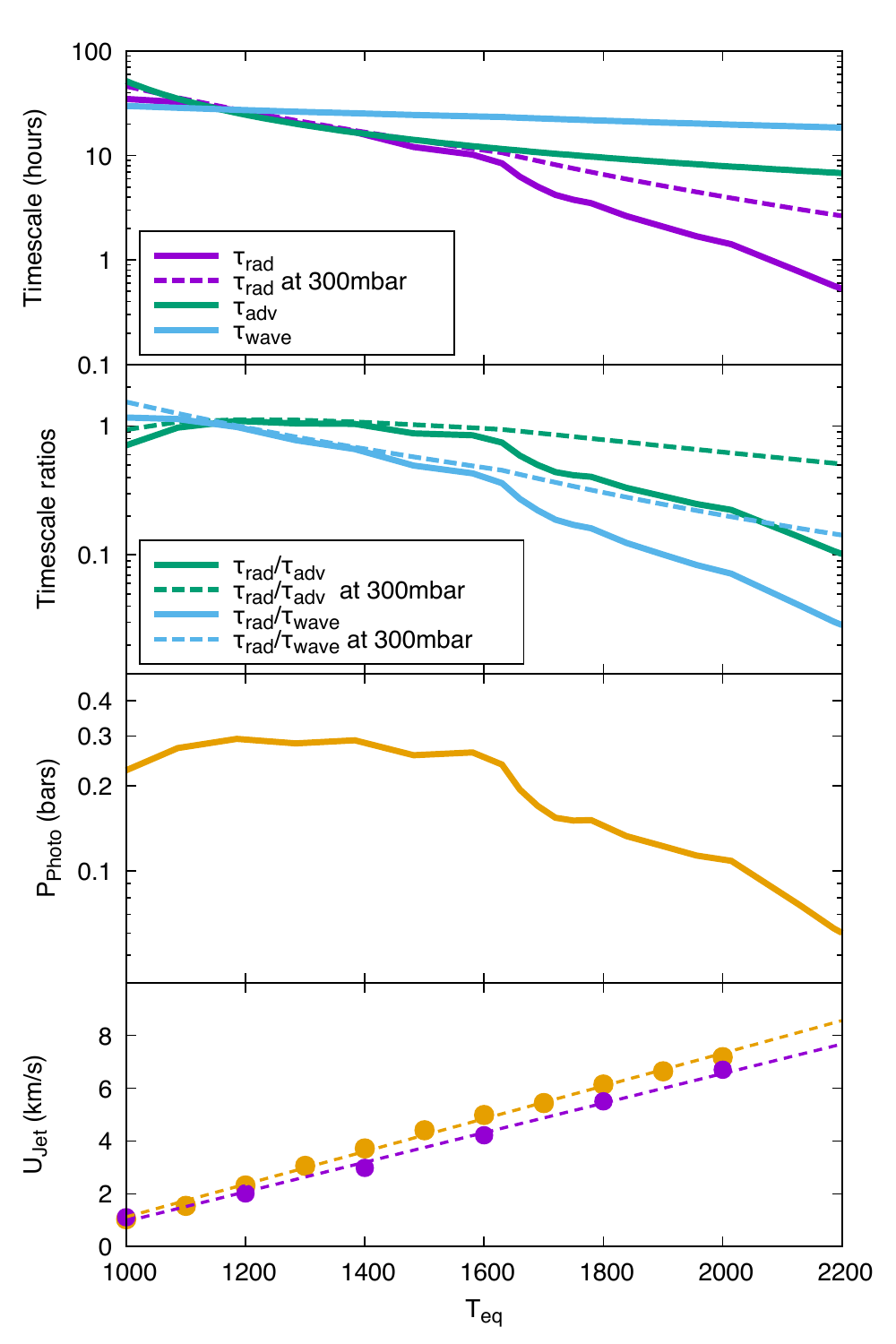}
\caption{Relevant timescales (top panel), timescale ratios (second panel), photospheric pressure (third panel) and averaged equatorial jet speed (bottom panel). The plain lines show quantities calculated with the photospheric pressure shown in the bottom panel whereas the dashed lines assume a constant, 300mbar, photospheric pressure.}
\label{fig::Timescales}
\end{figure}

In order to estimate the advective timescale we first estimate the jet speed directly from the GCM models by taking the average zonal velocity between latitudes $+/-20$ degrees around the equator. Because the jet is approximately constant between 1bar and 1mbar, we evaluate the jet speed at 0.1 bar. As shown in the bottom panel of Figure~\ref{fig::Timescales}, for the cloudless case the mean jet speed increases linearly with temperature, which is consistent with the semi-grey models from~\citet{Komacek2017} and the models of~\citet{Showman2011}. We can fit the linear relationship with: 
\begin{equation}
U_{\rm jet}=6.2T_{\rm eq}-5075\,{\rm m/s}.
\label{eq::UjetFit}
\end{equation}

When nightside clouds are present, the jet speed is slightly reduced. This is likely due to the hotter nightside temperatures: the nightside clouds reduce the temperature gradient on isobars, which reduces the pressure gradient on horizontal surfaces which decreases the wind speed~\citep{Komacek2016}. The jet speed in the nightside clouds case can be approximated by:
\begin{equation}
U_{\rm jet}=5.62T_{\rm eq}-4644\,{\rm m/s}.
\end{equation}

We now estimate the dayside photospheric pressure. { For this we use the 1D radiative/convective models ScCHIMERA of~\cite{Gharib-Nezhad2019}  with a dayside averaged irradiation scaled by the redistribution factor obtained from the GCM, so that both the 1D dayside average and the 3D dayside bolometric brightness temperatures are the same.} { The model uses similar opacities than those used in the radiative transfer calculations of our GCM simulations (see Fig. 8 of ~\citet{Piskorz2018}  for a comparison of the codes) and the bolometric contribution functions are easier to calculate}. We obtain the photospheric pressure by looking at the pressure of the maximum of the bolometric contribution function. At least two competing effects determine the photospheric pressure.  First, as the temperature increases, the Planck function shifts towards smaller wavelengths where the opacity is smaller~\citep[see e.g. the no TiO/VO case of figure A1 of][]{Parmentier2015}, hence increasing the photospheric pressure. On the other hand, increasing the temperature increases the thermal broadening of the lines, lowering the photospheric pressure. As seen in the bottom panel of Fig.~\ref{fig::Timescales} the photospheric pressure remains approximately constant at 200 mbar between 1000K and 1600K, but for larger equilibrium temperatures it decreases with increasing temperature, reaching 60 mbar at 2200K. 

We now show the radiative, advective and wave transport timescale for the cloudless model in the top panel of Fig.~\ref{fig::Timescales}. As expected, the radiative timescale has a stronger dependence with equilibrium temperature than the advective or the wave transport timescale. We used  equation~\ref{eq::UjetFit} for the jet speed, we assumed that the photospheric temperature was the dayside brightness temperature, related to the equilibrium temperature through the heat redistribution factor given by eq.~\ref{eq::fFit}. Finally we either used a constant photospheric pressure (dashed lines) { or the calculated photospheric pressure (plain lines).}

All timescales decrease with temperature, although the radiative timescale has a stronger variation with temperature than the advective timescale and the wave timescale. As shown by comparing the dotted and plain lines, correctly estimating the photospheric pressure is fundamental to accurately compute the radiative timescale. Indeed, the decrease of the radiative timescale is as much due to the increase in temperature as it is due to the decrease in photospheric pressure. 

In most of our parameter space we have $\tau_{\rm rad}<\tau_{\rm adv}<\tau_{\rm wave}$. As a consequence, in the equatorial regions we expect that the competition between advection of heat by the super-rotating jet and the radiative losses will determine the efficiency of the heat transport. In the mid-latitudes, however, where the Coriolis forces become large enough so that the circulation is no more dominated by the super-rotating jets and the wave timescale might become shorter than the advective timescale. 

The ratio of these timescales can be seen in the second panel of Fig.~\ref{fig::Timescales}. The radiative to advective timescale ratio is close to unity for temperatures lower than 1600K and decreases by a factor 10 between 1600K and 2200K. This is remarkably similar to the behaviour of the heat redistribution parameter shown in Fig.~\ref{fig::Ujet}, pointing out that the transport of heat by the jet transport is the dominant mechanism determining the day/night heat transport in our models. We can see that the variation of photospheric pressure with temperature accounts for a factor 5 in reduction of the advective to radiative timescale and is therefore an important factor in determining the day/night heat transport in hot Jupiters. 

\section{Comparison with observations: dayside and nightside temperatures.}
\label{sec::ComparingNS}
\begin{figure*}
\includegraphics[width=\linewidth]{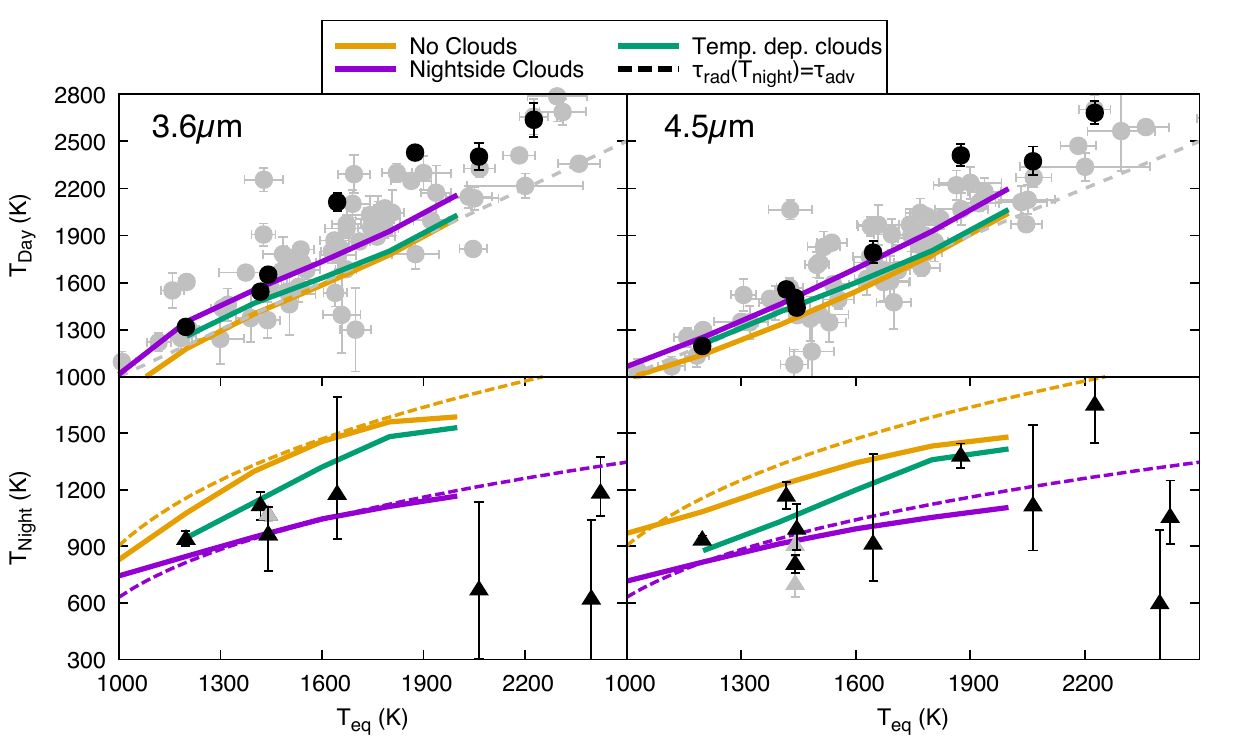}
\caption{Day (top)  and nightside (bottom) brightness temperatures at $3.6\mu m$ (left) and $4.5\mu m$ (right). Black points are from phase curve measurements~\citep{Beatty2018} whereas grey points are from secondary eclipse measurements taken from~\citep{Baxter2020}. We compare the data to our three sets of simulations: cloudless (orange), nightside cloud (purple) and temperature-dependent cloud (green). We also show as dashed line the nightside temperature estimate based on equation~\ref{eq::Tnight} with a photospheric pressure of 250mbar for the clear sky case and 100mbar for the case with nightside clouds. The grey points are the~\citet{Morello2019} and the ~\citet{Mendonca2018a} data reductions of WASP-43b. The grey line is the 1:1 line for reference.}
\label{fig::DayNightTemp}
\end{figure*}
We now compare the predictions from our three models to the dayside and nightside brightness temperatures derived from Spitzer phase curves observations. We show in Figure~\ref{fig::DayNightTemp} the dayside and nightside brightness temperatures at $3.6$ and $4.5\mu m$ estimated by~\citet{Beatty2018}. We calculate the brightness temperature from the model by adjusting a blackbody spectrum to the band-integrated flux emitted in the direction of the observer.

Overall, we see that our cloudless model underestimates the dayside brightness temperatures and overestimates the nightside ones. When nightside clouds are present they increase the brightness temperature in the dayside and decrease it in the nightside leading to a better match of the observations. {A similar behaviour was seen in~\citet{Lines2019} for HD209458b, in ~\citet{Mendonca2018a} for WASP-43b and in~\citet{Roman2017} when comparing their western terminator and western terminator+nightside cloud models of Kepler-7b.}

When looking at the models in more detail we see that the dayside brightness temperatures of the cloudless cases increase linearly with equilibrium temperature. This is surprising because the heat redistribution becomes less efficient with increasing temperature and thus the effective temperature should increase with increasing temperature. However, the brightness temperature in a given bandpass can increase faster or slower than the effective temperature. { That is because the effective temperature is determined by the mean brightness temperature over the spectral range span by the Planck function.} As the equilibrium temperature increases, the emission of the planet is shifted towards shorter wavelengths where the opacity is smaller. The difference between the effective temperature and the brightness temperature is therefore expected to increase with equilibrium temperature, which leads to a brightness temperature that increases linearly, and not faster, with equilibrium temperature (see section~\ref{sec::BrightnessTemp} for a more detailed discussion).

The nightside temperature varies much less with equilibrium temperature both in the cloudless case and in the nightside cloud case. The flatness of the trend in the observations has been previously described as a smoking gun for the presence of clouds on the nightside of the planet. To explain it~\citet{Keating2019} and ~\citet{Beatty2018}  hypothesize that the emission spectrum is probing the cloud top temperature set by the condensation curve of a given cloud species for all planets. {However, it is unclear why the top of the clouds should stay at the same temperature for two reasons. First as the temperature increases the cloud deck should move to lower pressures where the condensation temperature is going to be cooler, not similar, so the brightness temperature would decrease with increasing equilibrium temperature rather than stay constant. Second, whereas the condensation curve determines the temperature of the cloud base, the cloud top, seen by the observation, can be situated several scale heights higher than the cloud base. The exact vertical extent of the cloud depends on the complex interaction between mixing and particle settling, which could vary significantly with equilibrium temperature. }

We therefore propose another explanation for the weak dependence of the nightside temperature on equilibrium temperature. It is not is not set by the cloud condensation curve, but by the strong dependency of the radiative timescale with temperature in equation~\ref{eq::RadTime}. We postulate that the nightside temperature is the temperature for which a parcel of gas does not have enough time to cool more before being brought back to the dayside by the atmospheric circulation. To quantify this, we equate the radiative timescale on the nightside with the advective timescale: 
\begin{equation}
\tau_{\rm rad}\left(T_{\rm night}\right)\approx\tau_{\rm adv}
\end{equation}
Which can be expressed as: 
\begin{equation}
T_{\rm night}\approx\left(\frac{P_{\rm photo}}{g}\frac{c_{\rm p}}{4\sigma}\frac{U_{\rm jet}}{\pi R_{\rm p}}\right)^{1\left/3\right.}.
\label{eq::Tnight}
\end{equation}
By replacing the jet speed by equation~\ref{eq::UjetFit} we find that :
\begin{equation}
T_{\rm night}\propto \left(T_{\rm eq}\right)^{1/3}.
\label{eq::propto}
\end{equation}

As seen by the dashed lines of Figure~\ref{fig::DayNightTemp}, this overly simplified model predicts a very shallow variation of the nightside temperature with equilibrium temperature, matching well the dependence predicted by both the nightside clouds and the cloudless simulations. Therefore we conclude that the nearly constant nightside temperature observed in hot Jupiters can be explained by the strong dependence of the radiative timescale with temperature.

{ The exact dependence of the nightside temperature with equilibrium temperature depends on the scaling between the wind speed and the equilibrium temperature. The linear scaling found depends on our choice of not adding any additional Rayleigh drag in the model. Most mechanism that have been proposed to slow down the wind act more strongly on the fast winds (e.g. instability driven dissipation) and on hot planets (e.g. ohmic drag)~\citep{Koll2018}. Therefore, all these mechanisms should flatten even more the dependence of the nightside temperature on equilibrium temperature. In the limit of constant wind, our toy model predicts a constant nightside temperature. Additionally, if the wave timescale were used instead of the advective timescale \ref{eq::propto} would become:
\begin{equation}
T_{\rm night}\propto \left(T_{\rm eq}\right)^{1\left/\right.6},
\end{equation}
where the nightside temperature is even less dependent on the equilibrium temperature than before.

The scaling described here also applies to the energy balance model of~\citet{Bell2018} when heat transport by thermal dissociation is neglected. Their model solves for the advection of a parcel of gas that radiates energy as a blackbody at the local temperature (e.g. proportionally to $T^4$). Similarly to our model hotter gas radiates energy faster. As shown in their Figure 5 the energy balance model also predicts a very flat relationship between nightside and equilibrium temperature ~\citep[see also][]{Keating2019}.} 

In the clear case, however, the nightside temperatures predicted by the global circulation models are too high compared to the observations. Conversely the nightside temperatures from the models with nightside clouds are more in line with the observations. This is likely due to the nightside clouds moving the photospheric to lower pressures where the radiative timescale is shorter.

{ As is discussed in~\citet{Taylor2020}, the measured brightness temperature is not always representative of the real local temperature. Particularly, when clouds with a significant single scattering albedo are present, they change the emissivity of the photosphere to a value $\varepsilon<1$. This would affect our reasoning two ways. First, the atmosphere would be less efficient to cool down radiatively, which would lead to a modified expression for the radiative timescale:
\begin{equation}
\tau_{\rm rad}\approx\frac{P_{\rm photo}}{g}\frac{c_{\rm p}}{4\varepsilon\sigma T_{\rm photo}^3}
\label{eq::RadTimeEps}
\end{equation} 
Second, the relationship between the observed brightness temperature and the real temperature would be modified as follow:
\begin{equation}
T_{\rm observed}=\frac{\sqrt{\varepsilon}}{1+\sqrt{\varepsilon}}T_{\rm real}.
\label{eq::Emissivity}
\end{equation}
Overall that would lead to an additional factor in equation~\ref{eq::Tnight}:
\begin{equation}
\frac{2\varepsilon^{1/6}}{1+\sqrt{\varepsilon}}
\end{equation}
The two effects of the emissivity partially cancel each other, leading to variations of less than $10\%$ for the measured nightside temperature for atmospheric emissivities larger than $0.15$. Thus we do not expect emissivity to play a significant role in determining the observed brightness temperature of hot Jupiter nightsides.  
}

Finally, we note that if only one cloud were to fit all the hot Jupiter observations~\citep[e.g.][]{Gao2020a}, the cloud cover on the nightside would change with equilibrium temperature. As seen in the temperature dependent MnS cloud model of Figure~\ref{fig::DayNightTemp}, this would lead to a larger variation of the nightside temperature with equilibrium temperature, with the model behaving like the nightside cloud model at low temperature and like the cloudless model at high temperature. 

 As a conclusion, while the fact that the nightside temperature is constant with equilibrium temperature is due to the strong dependence of the radiative timescale with temperature, nightside clouds are necessary to lower the photospheric pressure and decrease the nightside temperature.
 
\section{How cloudy nights shape the phase curves of hot Jupiters}
\label{sec::Effect}

\begin{figure}
\includegraphics[width=\linewidth]{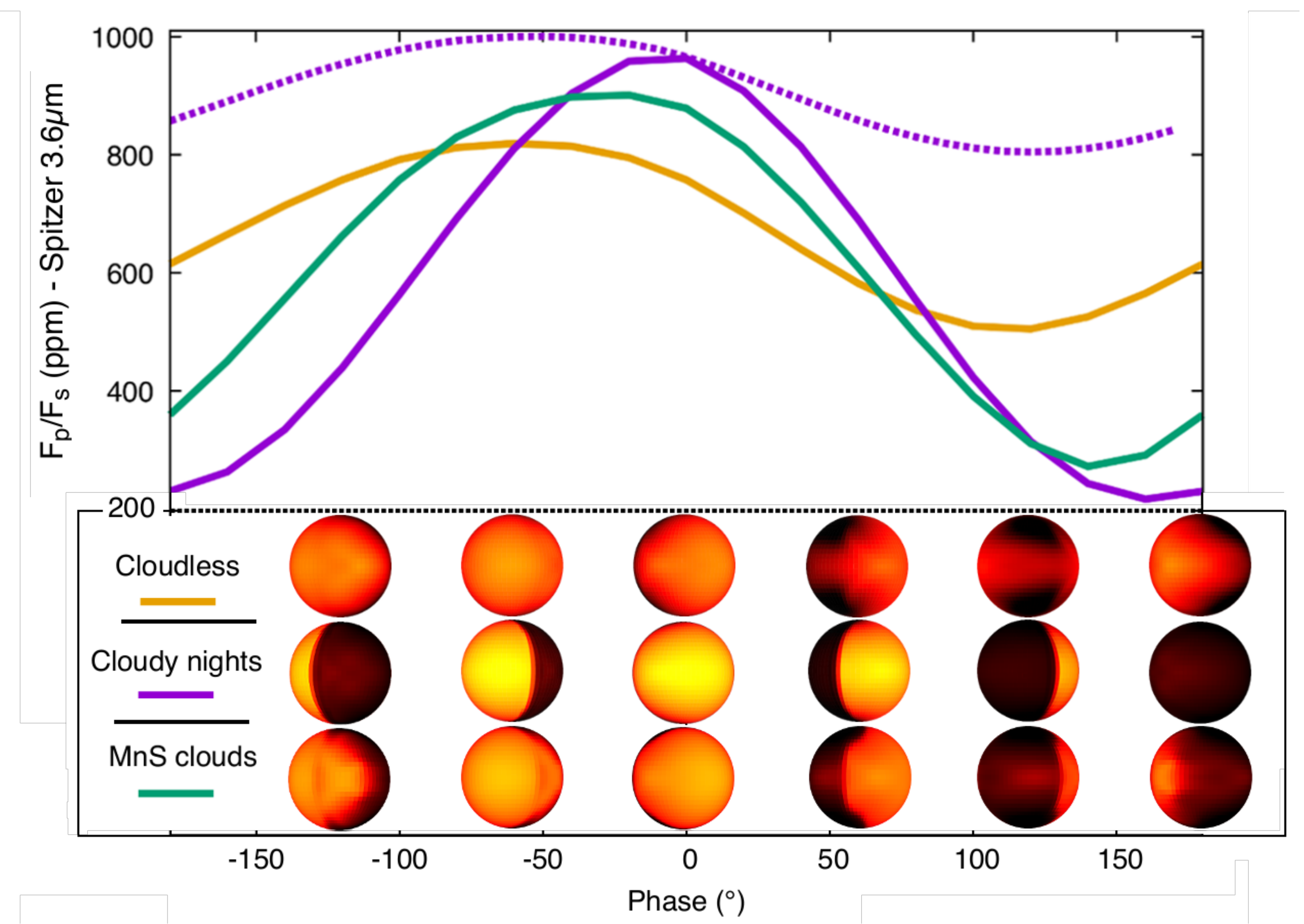}
\caption{Spitzer $3.6\mu m$ phase curve of $T_{\rm eq}=1400\,\rm K$ global circulation models without clouds (golden line), with nightside MnS clouds (purple line) or with temperature-dependent MnS clouds (green line). The dotted line is calculated based on the temperature map from the model with nightside MnS clouds but with the cloud opacity neglected in the radiative transfer calculation. The bottom panels show the flux distribution at $3.6\mu m$ on the planet hemisphere facing the observer for the three main models. At each phase the phase curves are the integral of the flux maps of the bottom panel.}
\label{fig::PhaseCurve}
\end{figure}

\subsection{Broadbands phase curve}
We have seen in the previous sections that the presence of nightside clouds decreases the day/night temperature contrast on isobars and decreases the hot spot offset. However, the link between atmospheric properties and phase curve offset is not always immediate, particularly when clouds are present. Indeed the phase curve at a given wavelength is given by both the thermal structure and the opacity structure of the planet. In order to disentangle both effects we will first look in more details at the $T_{\rm eq}=1400\,\rm K$ case. 

We show in Fig.~\ref{fig::PhaseCurve} the phase curve in the Spitzer bandpass for our three different models: cloudless, nightside clouds, temperature-dependent MnS clouds. The nightside clouds have two effects on the phase curve. First, they dramatically increase the phase curve amplitude. Second, they significantly reduce the phase curve offset from $\approx 60^\circ$ to almost zero. This seems in apparent contradiction with the temperature map of Figure~\ref{fig::Temp1400}. Indeed, on isobars the day/night temperature variations and the eastward shift of the hottest hemisphere is \emph{larger} when nightside clouds are present than in the cloudless model. 

To understand this apparent paradox, we show in Fig.~\ref{fig::PhaseCurve} the flux map at different phases in the prescribed nightside clouds and the cloudless cases. Each point in the phase curve is the average of the total intensity of the flux maps. In the cloudless case, there is a strong flux emerging from the western part of the nightside. As a consequence the brightest hemisphere is not the dayside hemisphere but is the hemisphere visible at phase $-60^\circ$, in agreement with the temperature map. When nightside clouds are present, almost no flux is coming from the nightside. At phase $-60^\circ$, when the brightest point of the planet is in plain sight, the lack of nightside flux compensates entirely for the increase in flux due to the Earth-facing geometry of the hottest spot, and the hemispherically averaged flux is smaller than the flux at phase zero.

To understand better why the clouds produce a larger phase curve amplitude and reduce the offset whereas the temperature maps shown in Fig.~\ref{fig::TempMapMnSTeq} would have predicted the opposite behaviour, we calculate a phase curve using the thermal structure from the global circulation model including nightside clouds but omit the cloud opacities when calculating the phase curve. As expected, the resulting phase curve (dotted line of Fig.~\ref{fig::PhaseCurve}) has the same flux value at phase zero, when only the dayside is visible. Additionally the resulting phase curve shows a much stronger offset and a much reduced amplitude compared to the case where the opacity of clouds is taken into account in the phase curve calculation. This confirms that the reason for the small offset and the large amplitude is to be found in the radiative effects of the clouds rather than their dynamical effect.

\begin{figure*}
\includegraphics[width=0.33\linewidth]{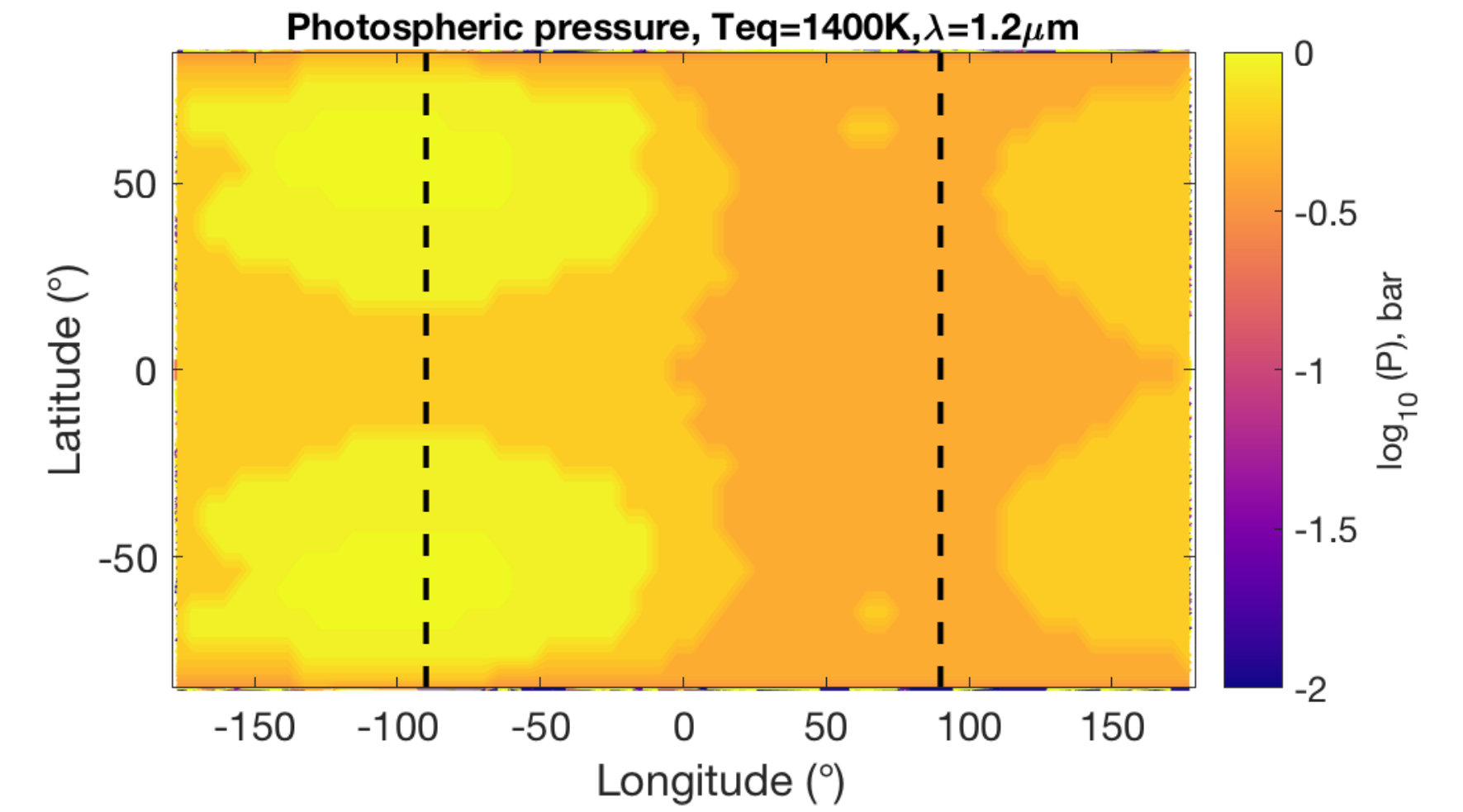}
\includegraphics[width=0.33\linewidth]{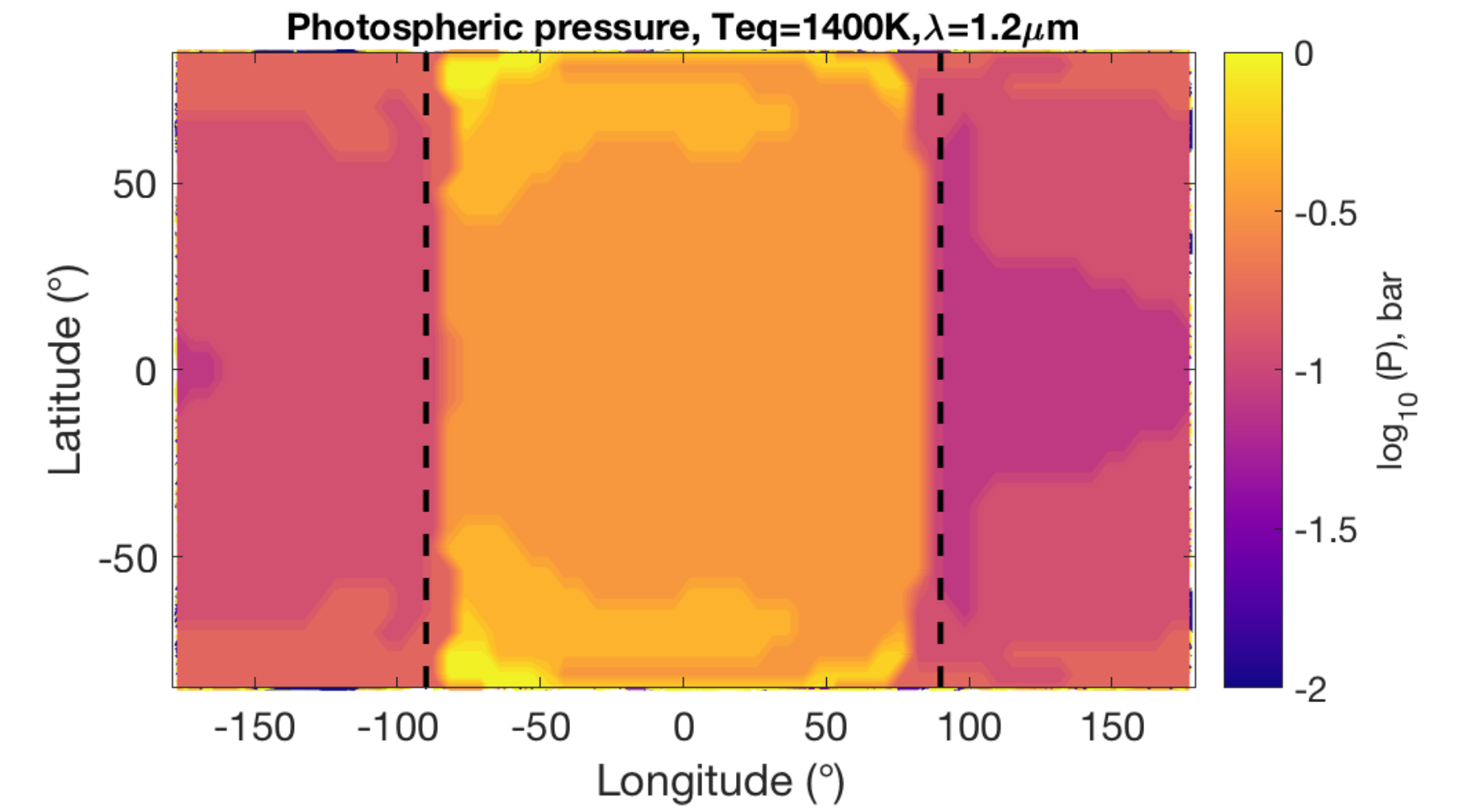}
\includegraphics[width=0.33\linewidth]{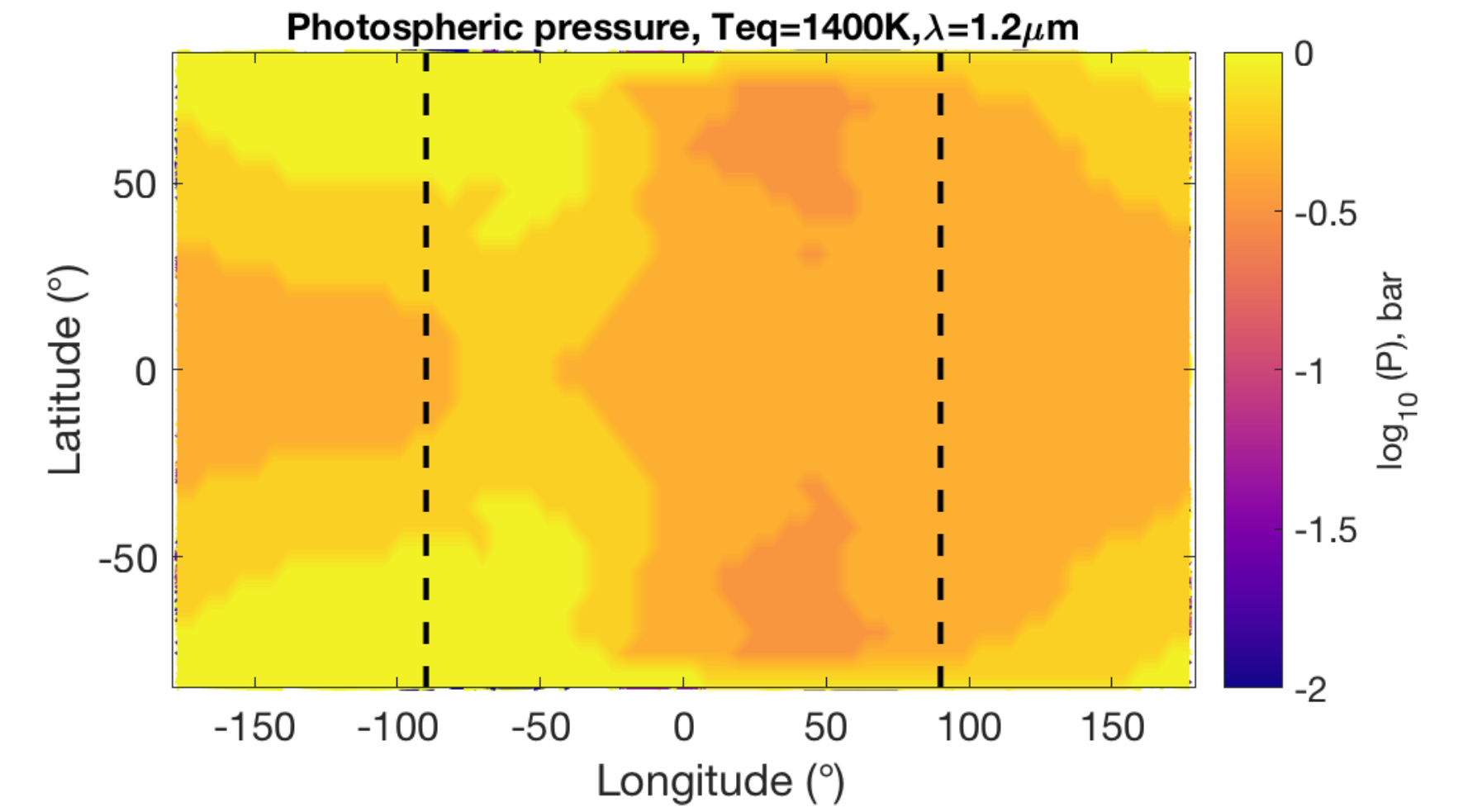}

\includegraphics[width=0.33\linewidth]{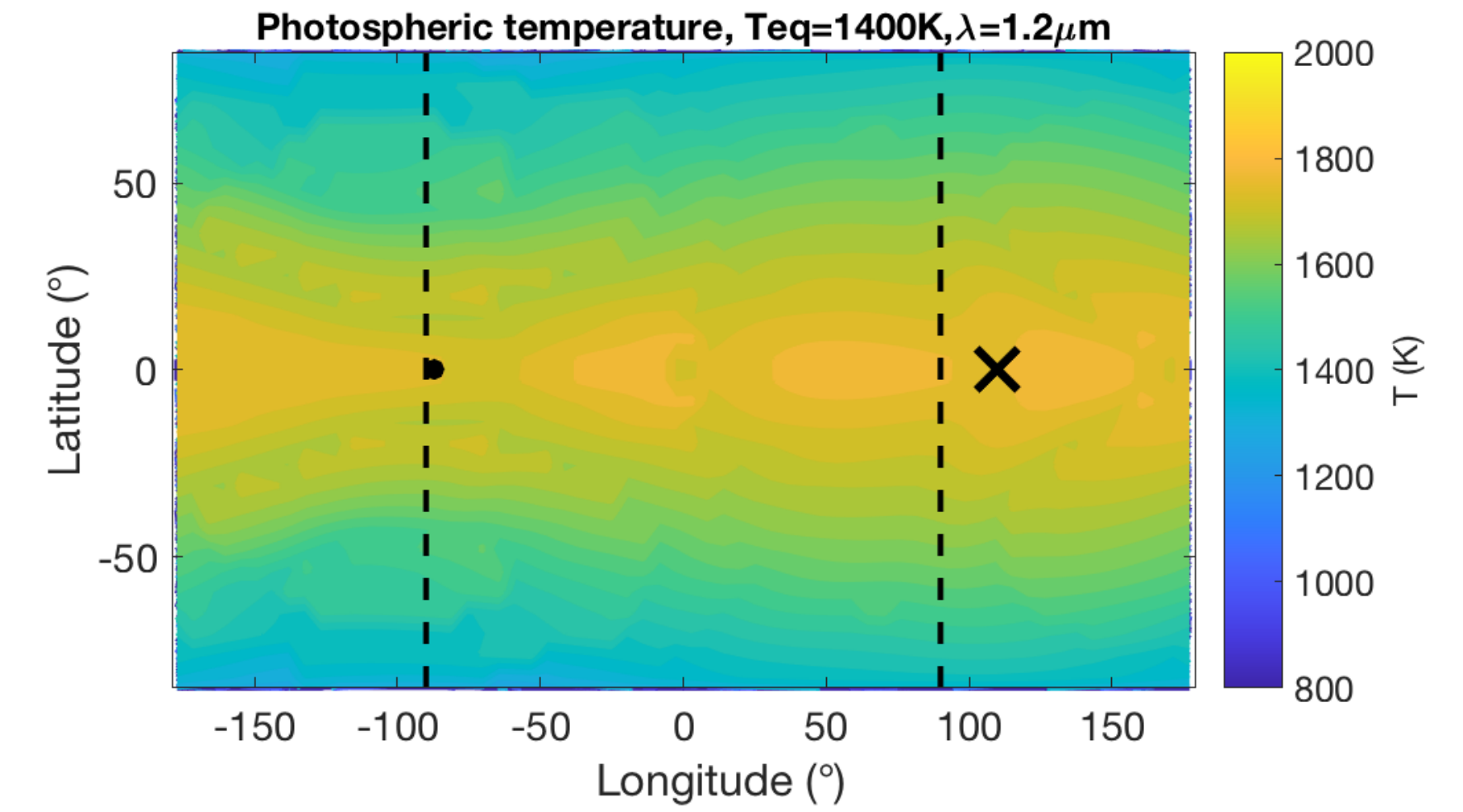}
\includegraphics[width=0.33\linewidth]{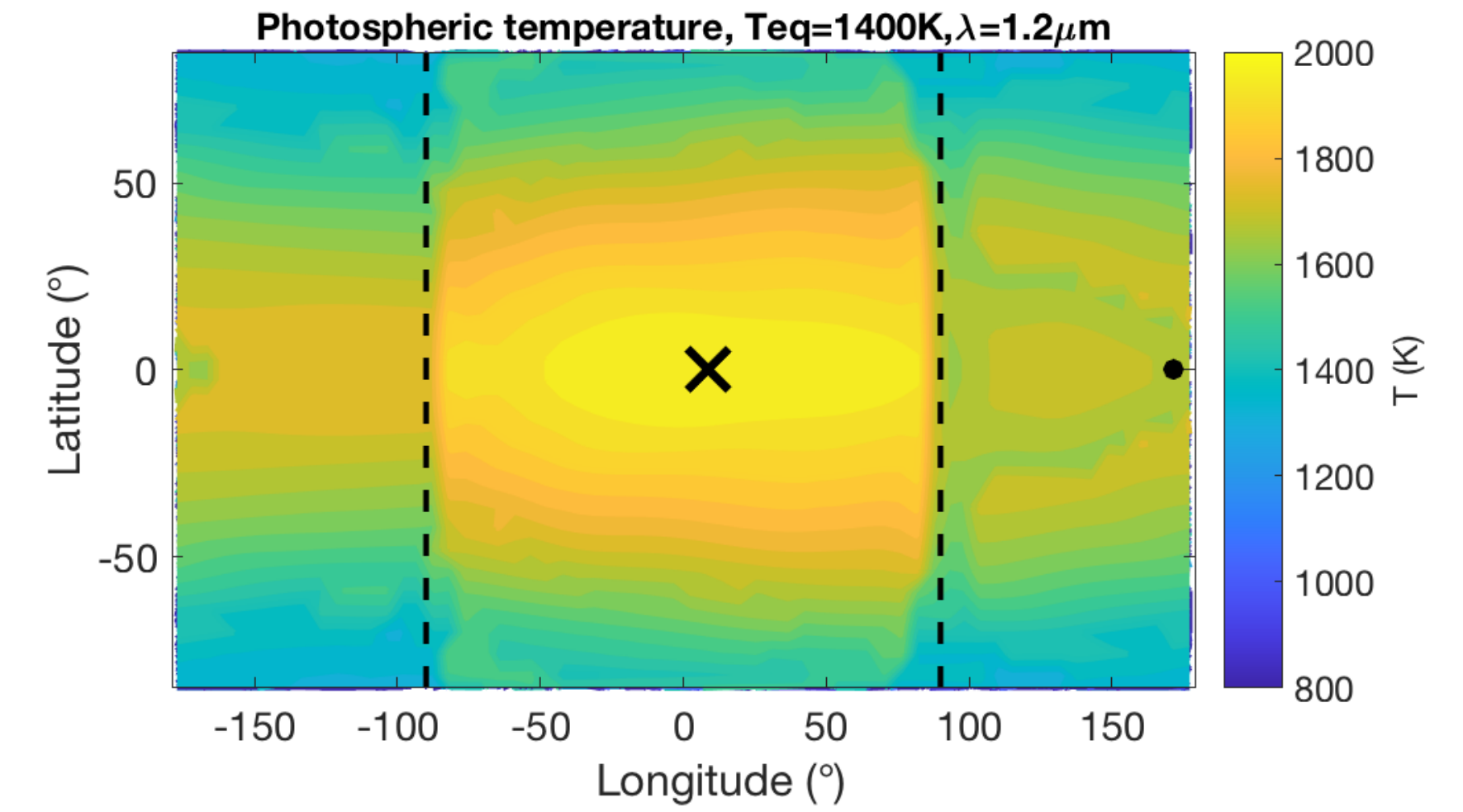}
\includegraphics[width=0.33\linewidth]{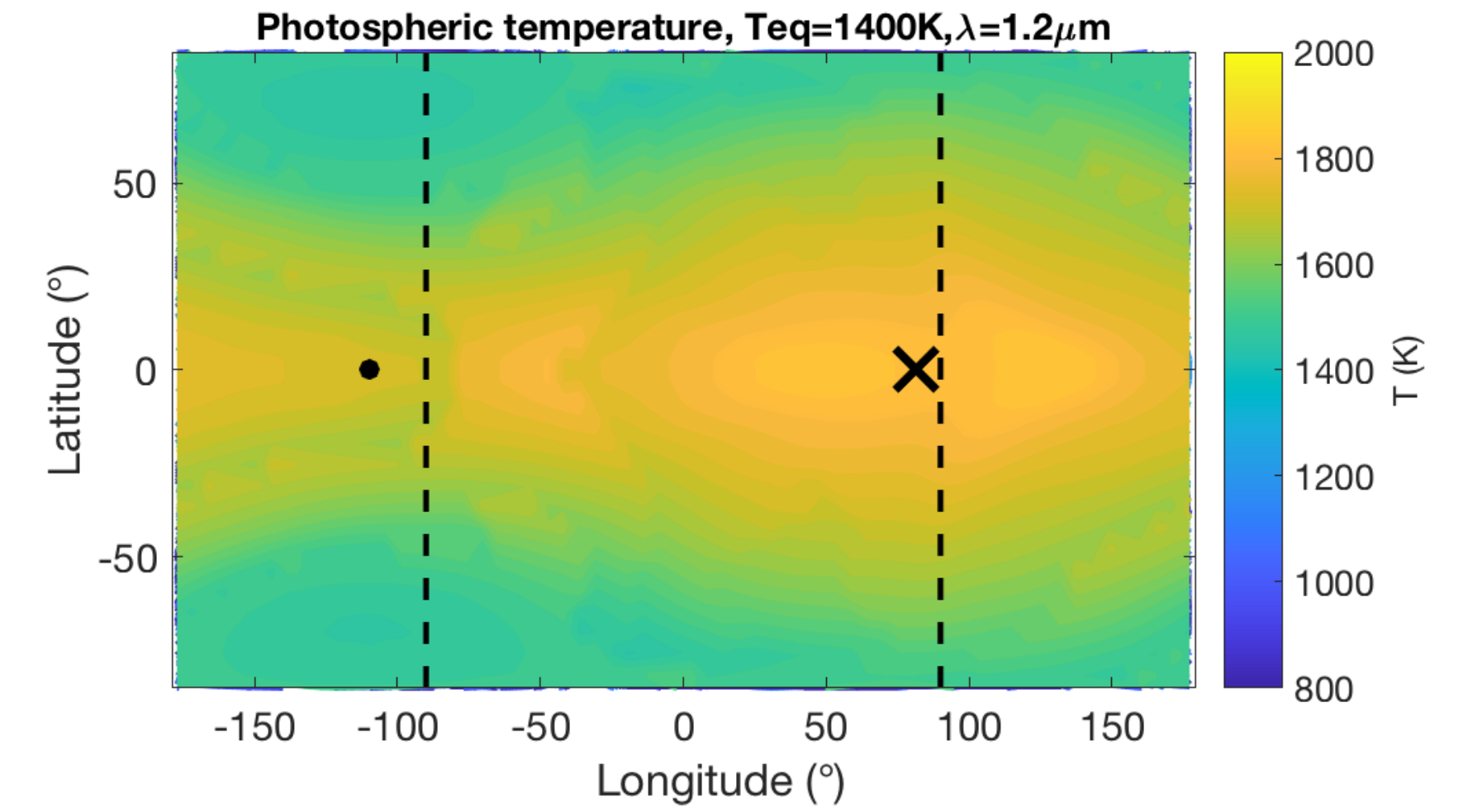}

\includegraphics[width=0.33\linewidth]{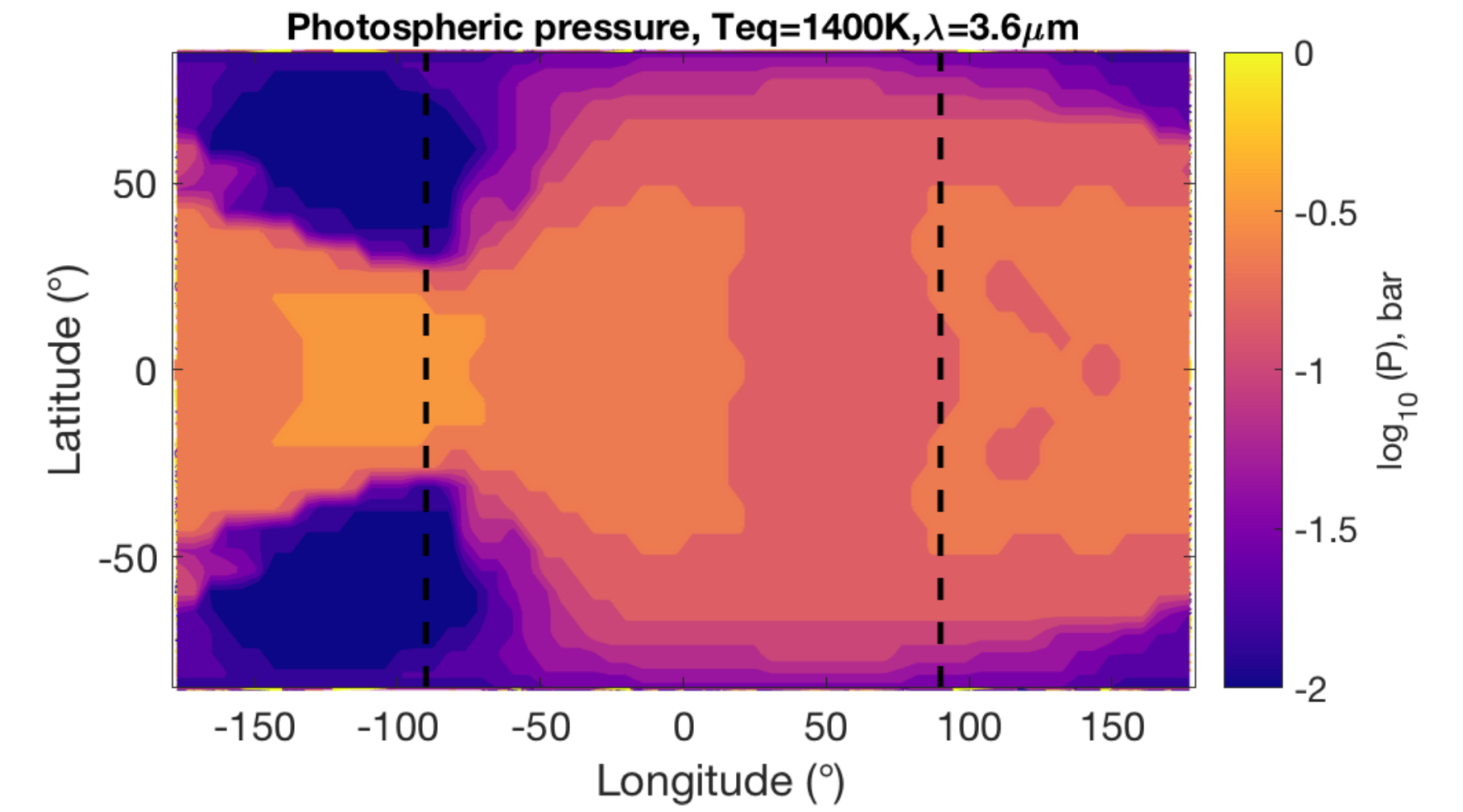}
\includegraphics[width=0.33\linewidth]{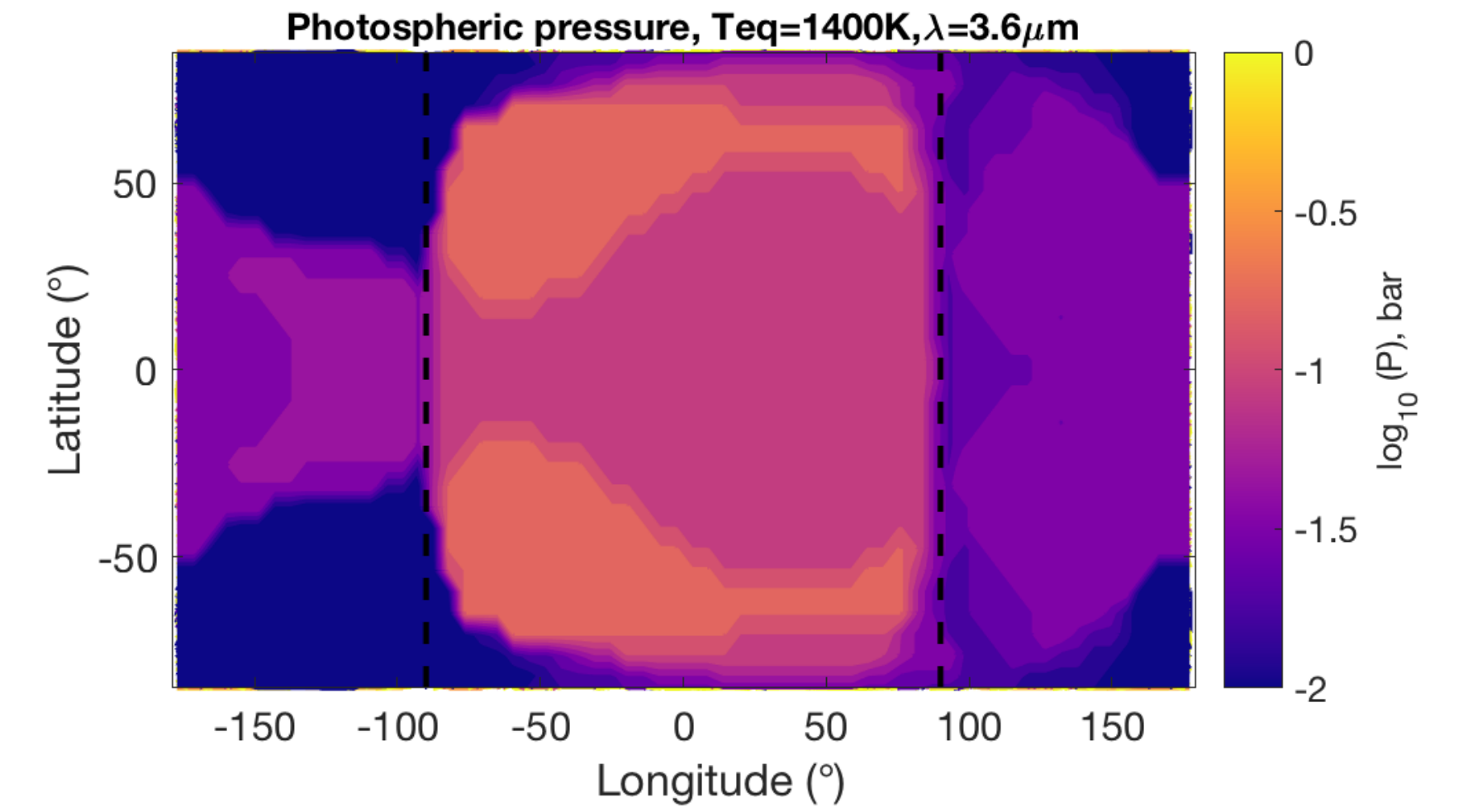}
\includegraphics[width=0.33\linewidth]{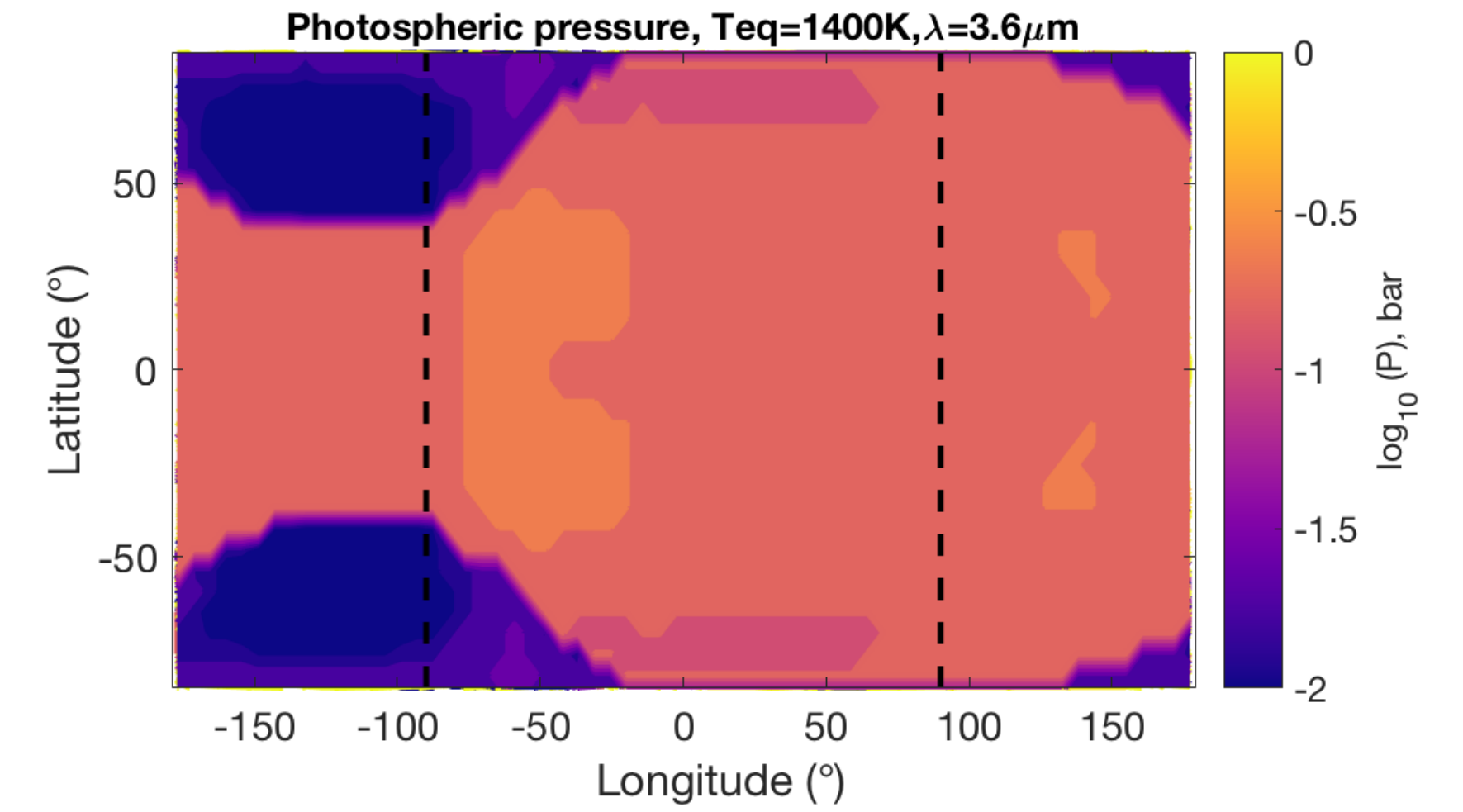}

\includegraphics[width=0.33\linewidth]{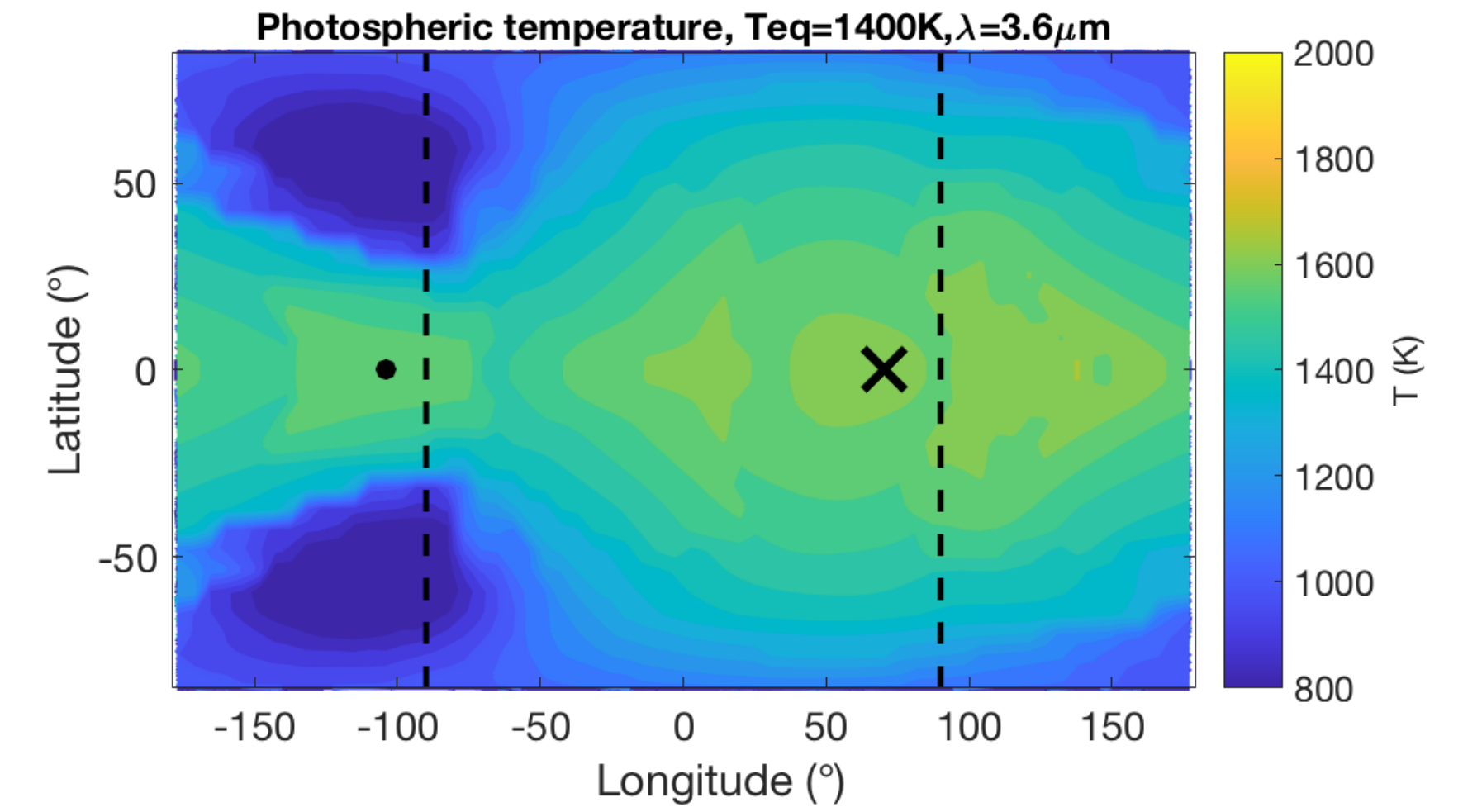}
\includegraphics[width=0.33\linewidth]{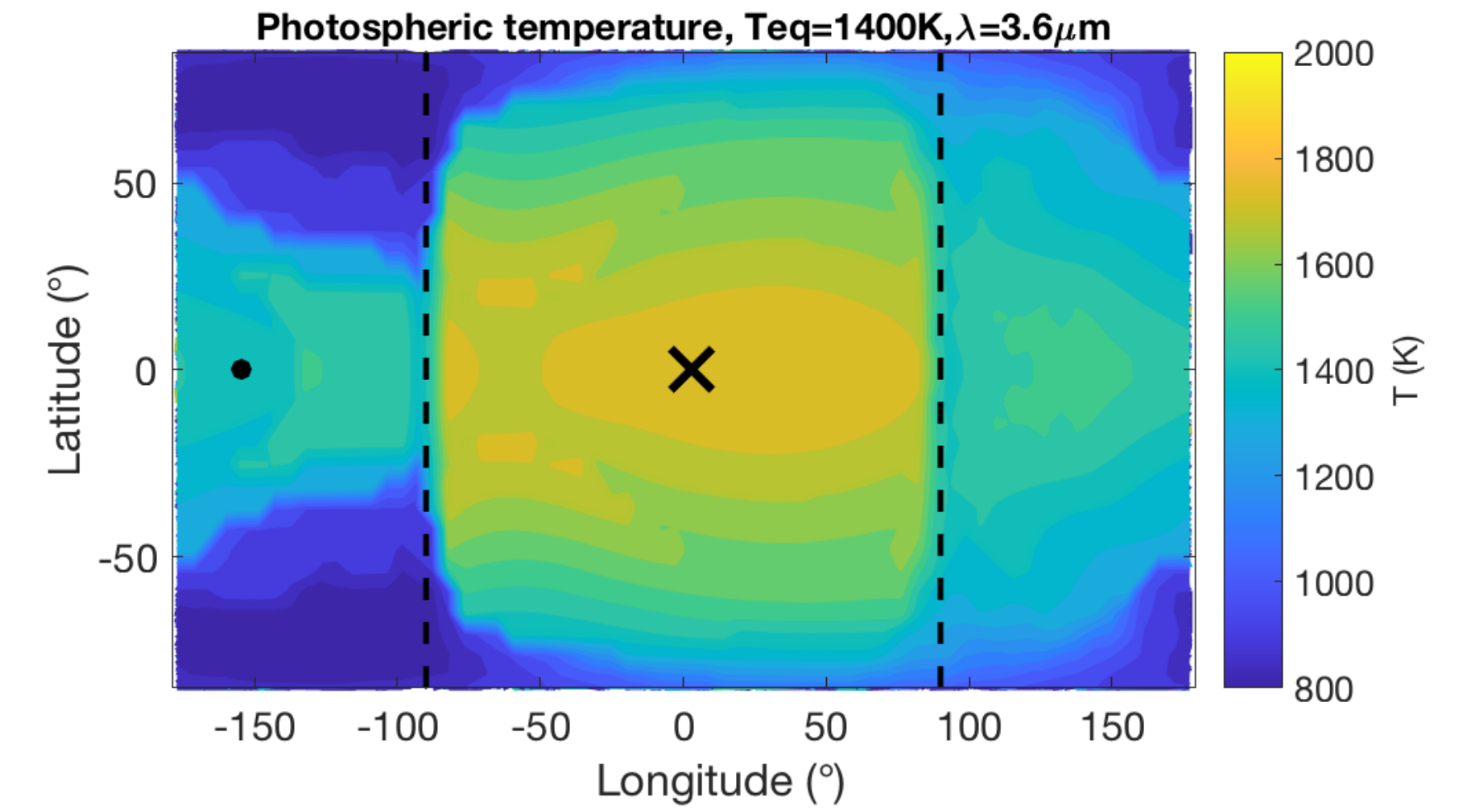}
\includegraphics[width=0.33\linewidth]{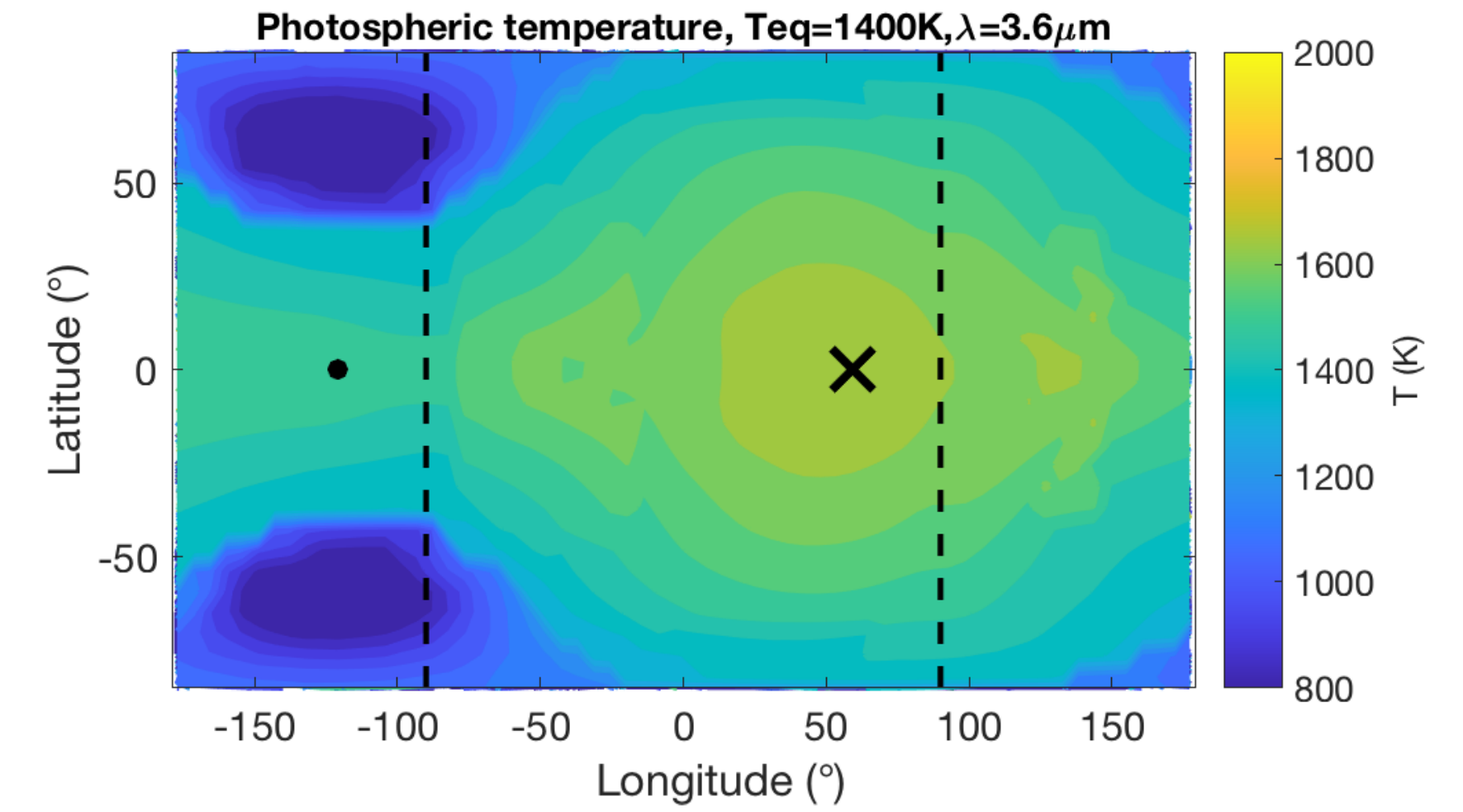}

\includegraphics[width=0.33\linewidth]{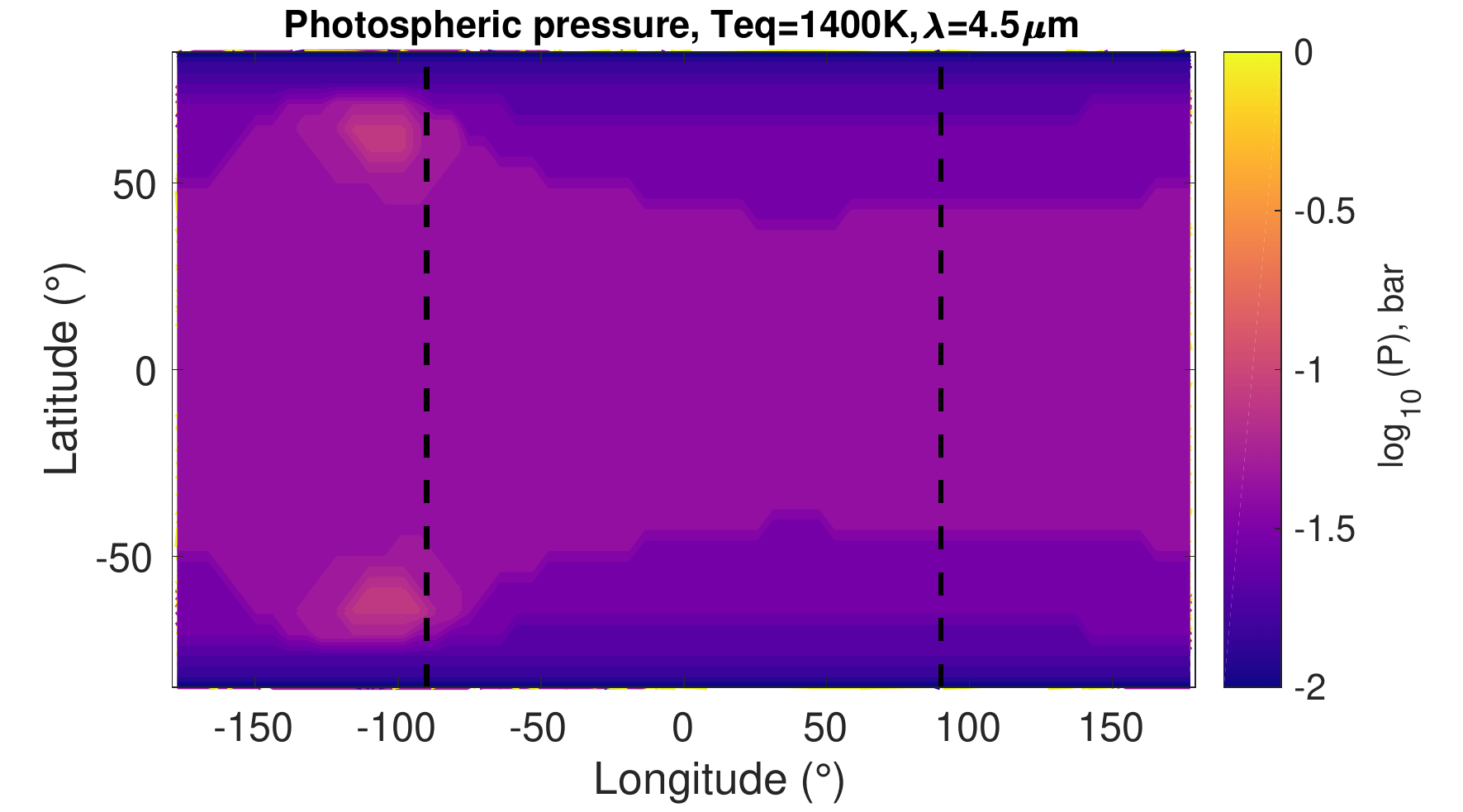}
\includegraphics[width=0.33\linewidth]{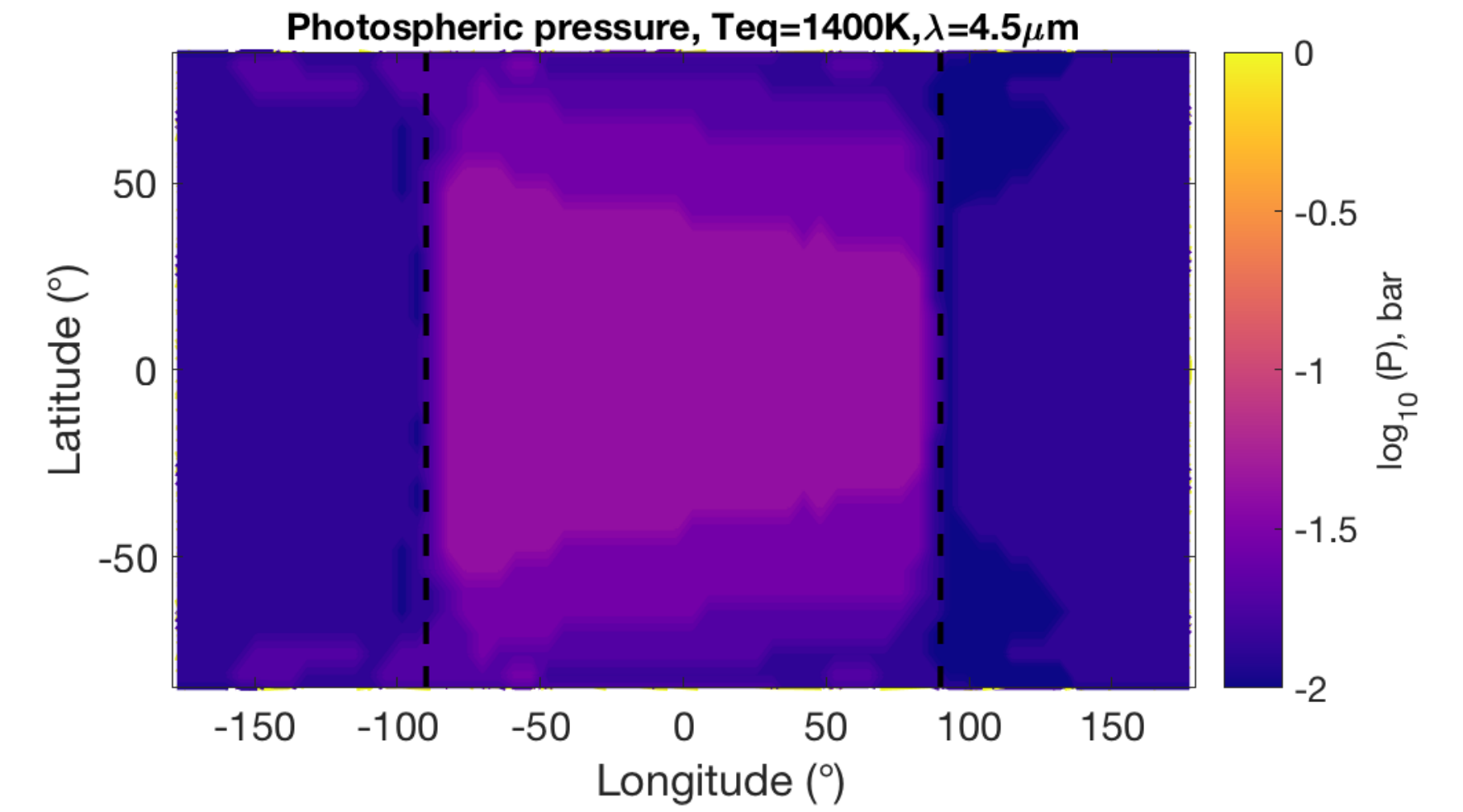}
\includegraphics[width=0.33\linewidth]{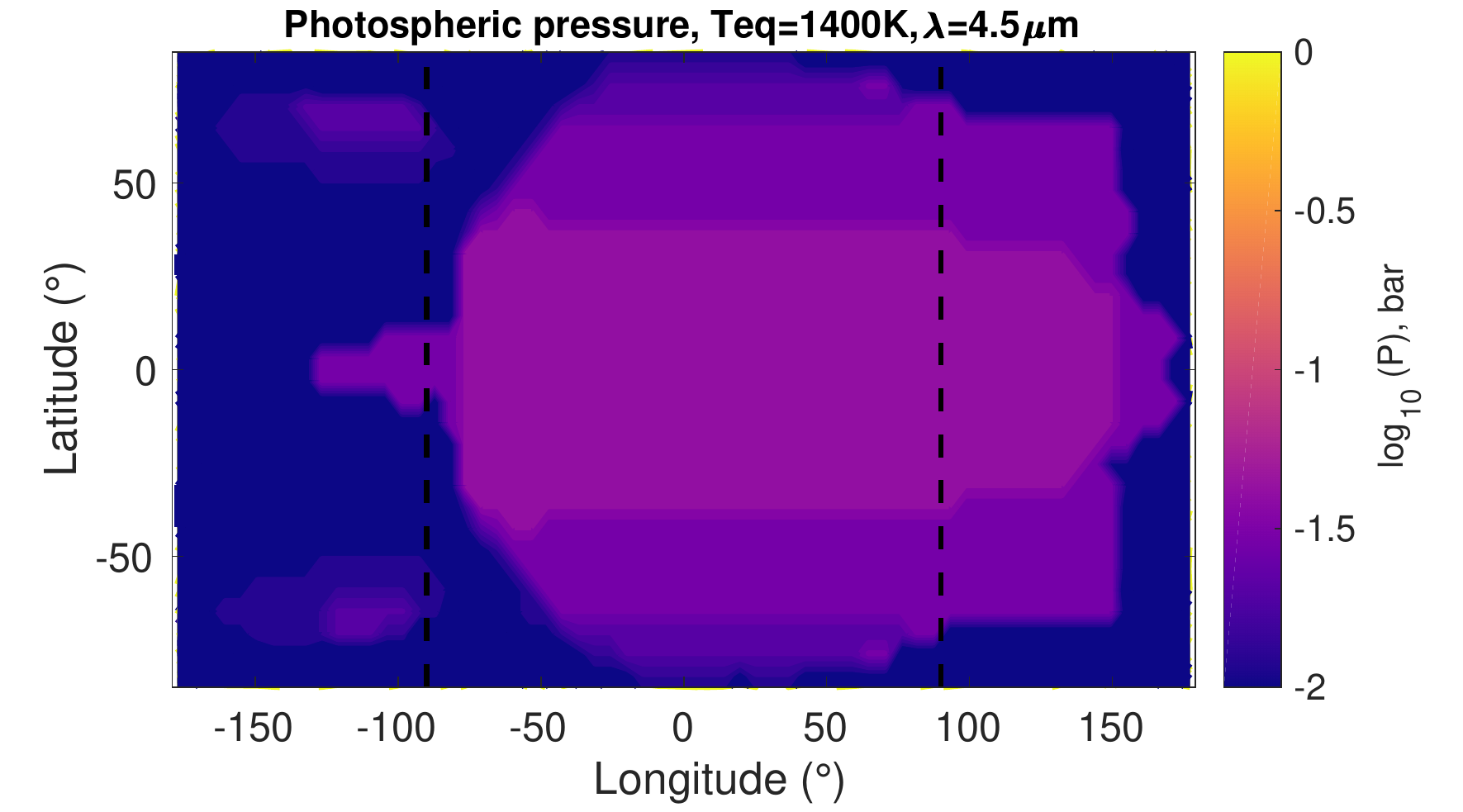}

\includegraphics[width=0.33\linewidth]{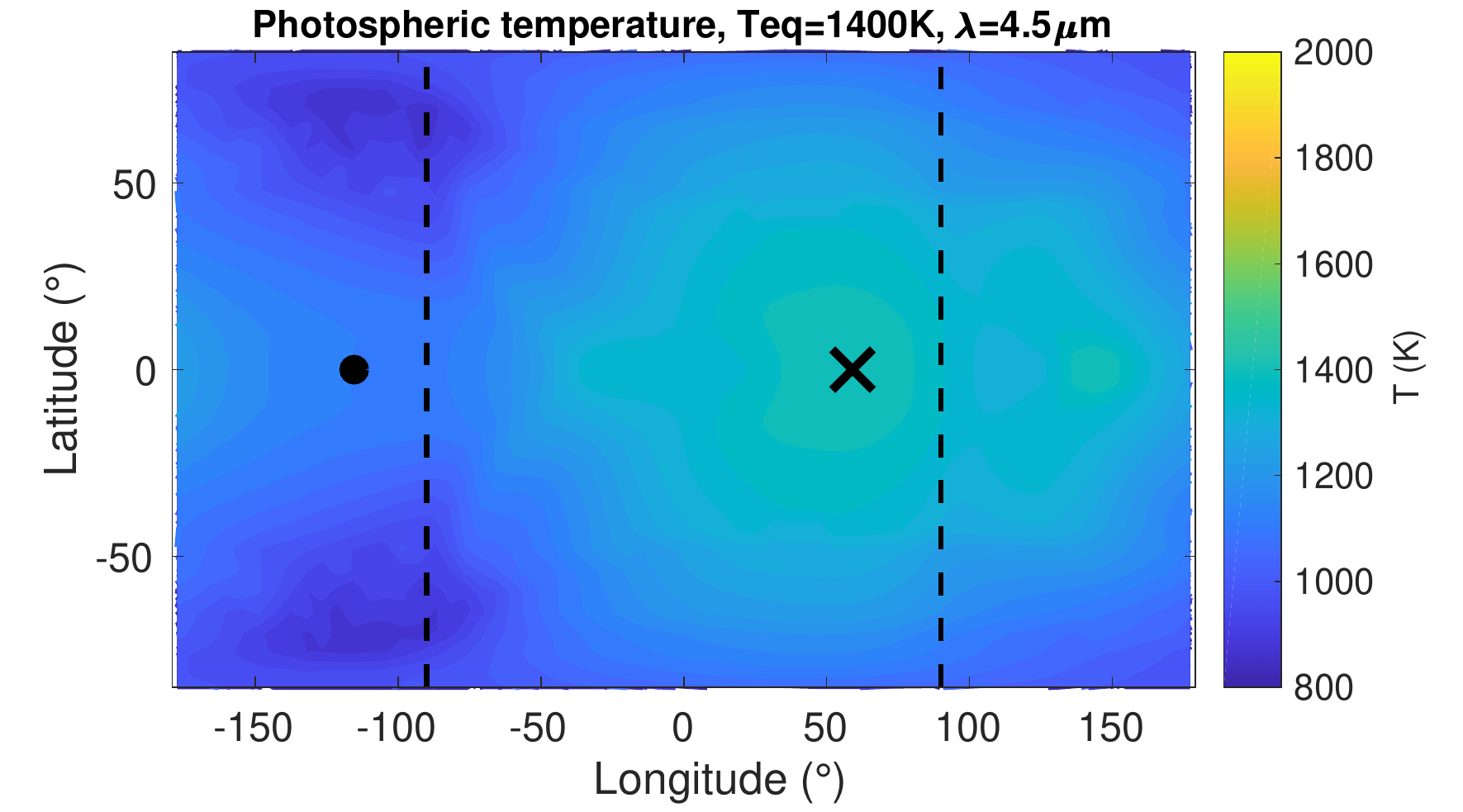}
\includegraphics[width=0.33\linewidth]{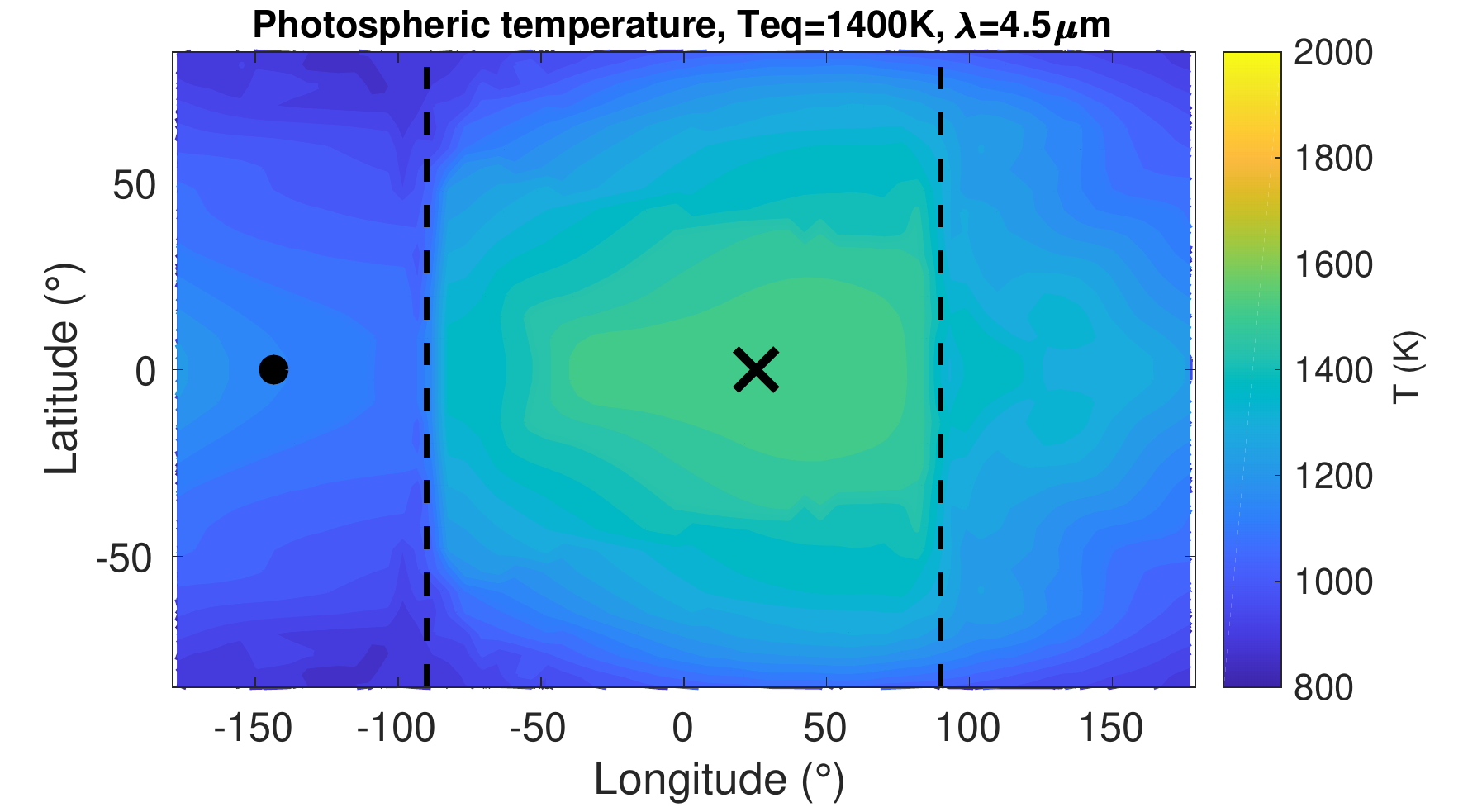}
\includegraphics[width=0.33\linewidth]{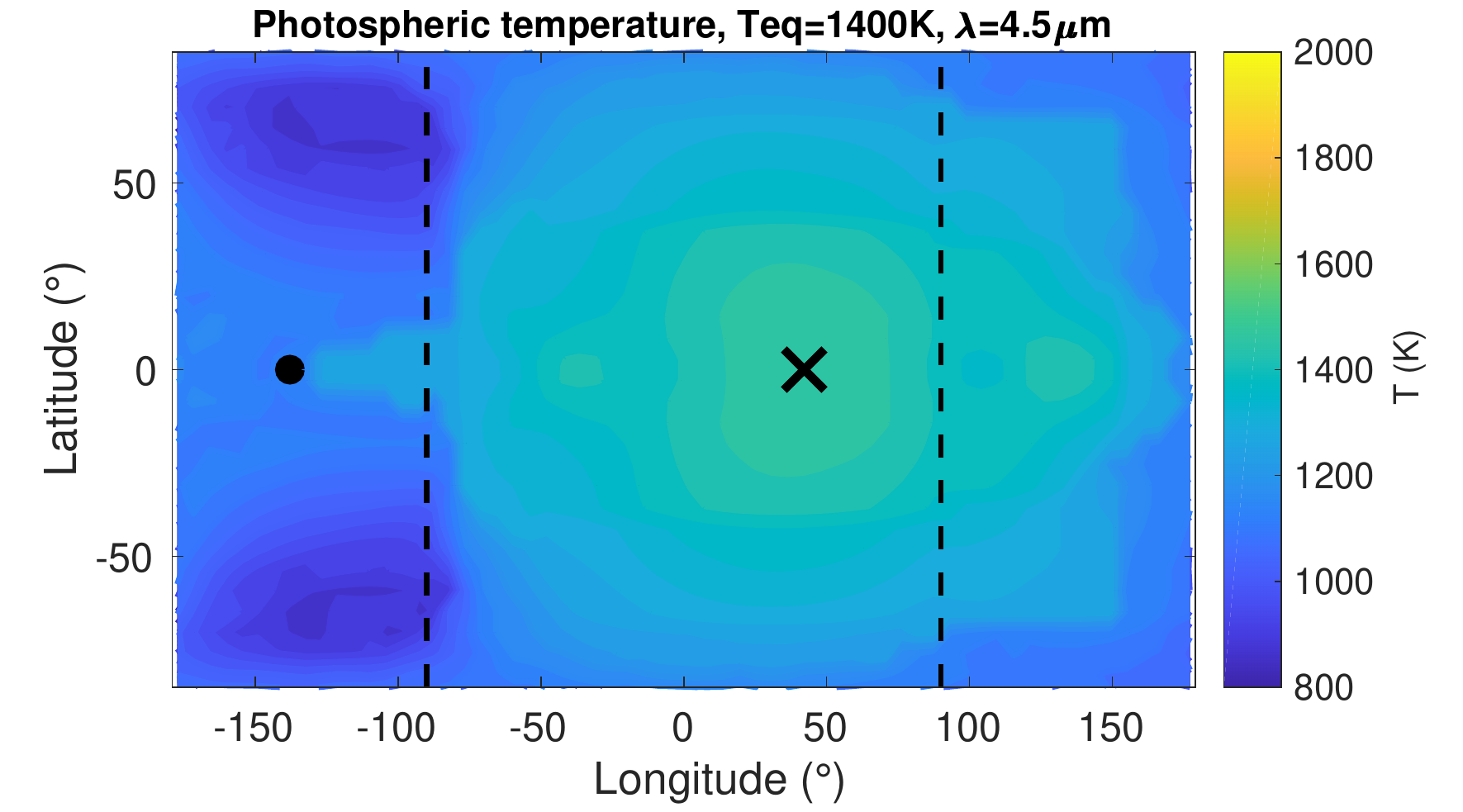}
\caption{Photospheric pressure and photospheric temperatures at three different wavelengths for the cloudless case (left), the nightside cloud case (middle) and the temperature-dependent cloud (right). The cross marks the central longitude of the hottest projected hemisphere at the photosphere. The dot marks the central longitude of the coolest projected hemisphere at the photosphere.}
\label{fig::Photosphere}
\end{figure*}

We now turn to Fig.~\ref{fig::Photosphere} which shows the map of the photospheric pressures and temperatures at different wavelengths for the cloudless, nightside clouds, and temperature-dependent MnS clouds. The main effect of the clouds is to raise the nightside photosphere by several scale heights, from $\approx300\rm mbar$ to $\approx30\rm mbar$ at $3.6\mu m$. As a consequence, even though the nightside is hotter at a given pressure when clouds are present, because the pressures probed are smaller, the photospheric temperature is cooler. Importantly the photospheric temperature maps show a sharp gradient between the day and the nightside due to the sharp change in photospheric pressure. The hottest photospheric hemisphere is then centred on the substellar point, even though the hottest hemisphere on isobars is centred $\approx 60^\circ$ east of the substellar point. Because the phase curve tracks the photospheric temperature and not the isobaric temperature, the phase curve offset is greatly reduced by the presence of clouds. It is worth noting that the photospheric temperature maps looks extremely different to any of the isobaric temperature maps of Fig.~\ref{fig::TempMapMnSTeq}.

The temperature-dependent MnS clouds have a cloud map that is shifted eastward compared to the day/night symmetry. As a consequence, the western part of the nightside is less cloudy than in the prescribed nightside cloud case whereas the western part of the dayside is more cloudy. Additionally the clouds are thinner than the ones considered in the nightside cloud for the $T_{\rm eq}=1400\,\rm K$ case and the nightside is therefore not as dark as in the prescribed nightside cloud models. Overall, the effect of clouds is qualitatively similar but quantitatively smaller than in the nightside cloud case : a reduction of the phase curve offset and an increase in the phase curve amplitude.

As a conclusion, we expect nightside clouds to always affect both the amplitude and the offset of the phase curve at the same time because the two are a consequence of the same physical effect: a dark nightside.

\subsection{Spectral phase curves.}

Clouds often have a non-grey opacity structure. In most cases, due to Mie scattering, the cloud extinction opacity decreases with increasing wavelengths~\citep{Mie1908}. When resonance features are present, the extinction opacity can increase at specific long wavelength. However, for MnS, where no strong resonance features are present, the cloud opacity decreases monotically with wavelength~\citep{Wakeford2015}. 

\begin{figure}
\includegraphics[width=\linewidth]{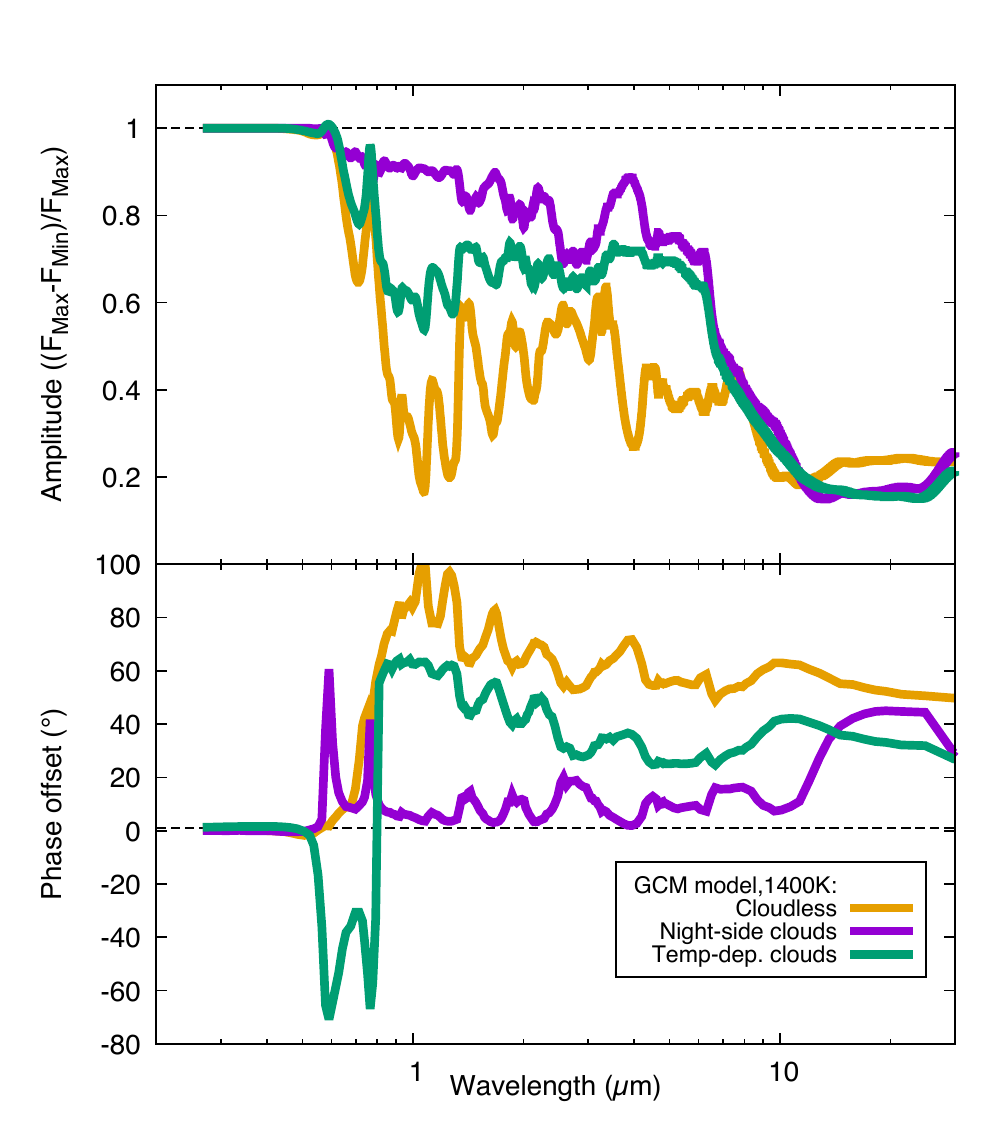}
\caption{Phase curve amplitude (top) and phase curve offset (bottom) for our $T_{\rm eq}=1400\,\rm K$ cloudless (orange line), nightside cloud (purple) and temperature-dependent MnS cloud (green) models.}
\label{fig::SpecVariation}
\end{figure}

As shown in Fig.~\ref{fig::SpecVariation} the phase curve amplitude for the cloudless case shows large spectral variations. At short wavelengths (e.g. $<0.7\mu m)$, the phase curve is dominated by reflected light, leading to a very large day/night brightness contrast. At longer wavelengths (e.g. $>1.5\mu m$) the amplitude is modulated by the molecular bands. Inside molecular bands the phase curve probes lower pressures  where the day/night temperature contrast is larger due to a reduced radiative timescale. Outside molecular bands the atmospheric layers probed are deeper where both the dayside and the nightside brightness temperatures are larger. However the nightside brightness temperature increases more with pressure than the dayside one due to the increased efficiency of heat transport at higher pressures. As a consequence the amplitude of the phase curve is larger inside molecular bands than outside of it. 

When thick nightside clouds are added to the simulations, the phase curve amplitude is increased at almost all wavelengths where the clouds are optically thick. Additionally, the spectral variations of the phase curve amplitude are smaller than in the cloudless case. Interestingly, the phase curve amplitude variations are opposite to those in the cloudless case : the amplitude is smaller inside molecular bands and larger outside of them. This can be explained as follow: on the dayside the photospheric pressure and temperatures are smaller inside a molecular band than outside of it, as in the cloudless case. On the nightside, however, the photospheric pressure is determined by the cloud opacity, which is much more grey, meaning that the photospheric pressure and temperature are similar inside and outside molecular bands. As a consequence, the amplitude of the phase curve is larger outside molecular bands than inside molecular bands, which is the contrary than in the cloudless case.

At wavelengths larger than $\approx 10\mu m$, the $1\mu m$ cloud particles used here become transparent. The phasecurve is therefore sensitive to levels below the clouds, where the thermal structure is homogenised horizontally due to the greenhouse effect of the clouds (see sect.~\ref{sec::AtmCirc}). This explains why in Fig.~\ref{fig::SpecVariation} the phase curve amplitude is reduced when nightside clouds are present at wavelengths larger than $10\mu m$. When the temperature-dependent MnS clouds are used, the amplitude is  intermediate between the nightside cloud and the cloudless model. The spectral variations are smoothed and they often correlate with the cloudless case. 

A similar behaviour is seen from the offset of the phase curve maximum. In the cloudless case the phase curve offset decreases inside a molecular band as the phase curve probes lower pressures where the hot spot is less shifted. In the nightside cloud case, the phase curve offset is strongly reduced at all wavelengths and its spectral variations are anticorrelated with the cloudless case, similarly to the amplitude of the phase curve. The temperature-dependent cloud model is the only one producing a negative offset at short wavelengths. This is due to the small amount of clouds gathering on the cooler western limb of the planet and creating a large albedo variation on the planet dayside. This behaviour, corresponding to the westward offsets observed by the Kepler space telescope was more fully explored in~\citet{Parmentier2016}. At longer wavelengths the offsets are reduced compared to the cloudless case, but not as reduced as in the thick nightside cloud case. 

\section{Comparison with observations: phase curves}
\label{sec::PhaseCurvesobs}
We now compare the outputs from our simulations with the available phase curve observations at different wavelengths. The lightcurve behaviour is systematically illustrated in Figures~\ref{fig::ModelVsObsNSclouds} and~\ref{fig::MnS-Self-consistent} for our various cloud-free and cloudy models. We first describe the behaviour of the cloudless phase curves and then show that the presence of nightside clouds allow for a much better agreement between models and observations.  

We have focused on the lightcurve behaviour at a variety of bandpasses that are inherently interesting and also observationally relevant. All bandpasses have been labelled with their middle wavelength. 0.7$\mu m$ corresponds to the bandpass of the Kepler spacecraft (400-900$\mu m$). It illustrates the optical wavelength behaviour where both reflected and thermally emitted light can be important. 1.5$\mu m$ is the averaged Hubble Space Telescope WFC3/G141 bandpass (1.1-1.7$\mu m$) and represents the near-IR behaviour.  3.6, 4.5, and 8$\mu m$ are the three most important Spitzer/IRAC bandpasses, for which phase curves  have been measured for many planets. 25$\mu m$ microns corresponds to the Spitzer/MIPS instrument and also illustrates the long-wavelength tail of the typical hot-Jupiter Planck function. These longer IR wavelengths are also less likely to be strongly influenced by cloud scattering and thus may exhibit qualitatively different behaviour than shorter IR wavelengths where cloud scattering is critically important.

\subsection{Cloudless phase curves}

We now compare the predictions from our three models to the observed phase curve amplitudes and phase curve offsets, wavelength by wavelength. As can be seen by the solid curves in Fig.~\ref{fig::ModelVsObsNSclouds} the cloudless model fails to match the observations at most wavelengths. It systematically underestimates the amplitude of the phase curve and overestimates the phase offset of the maximum of the phase curve. This is a long standing problem~\citep{Showman2009,Amundsen2016,Roman2017,Parmentier2018,Zhang2018,Lines2019} and different mechanisms have been proposed to reconcile observations and theory: non-synchronous rotation rate~\citep{Showman2009}, increased atmospheric metallicities~\citep{Showman2009,Kataria2015}, disequilibrium chemistry~\citep{Cooper2006,Knutson2012,Mendonca2018b,Drummond2018a,Drummond2018b,Steinrueck2019}, the presence of a drag force reducing the speed of the winds such as magnetic drag~\citep{Perna2010,Perez-Becker2013a,Komacek2016,Arcangeli2019}, shocks~\citep{Heng2012a}), or the presence of clouds~\citep{Kataria2015,Parmentier2016,Komacek2017,Lines2019}. While the following subsections will focus on the role of clouds, we first describe the physics shaping the phase curves of our cloudless models.

\subsubsection{Phase curve amplitude}
\label{sec::PCA}
\begin{figure*}
\includegraphics[width=0.9\linewidth]{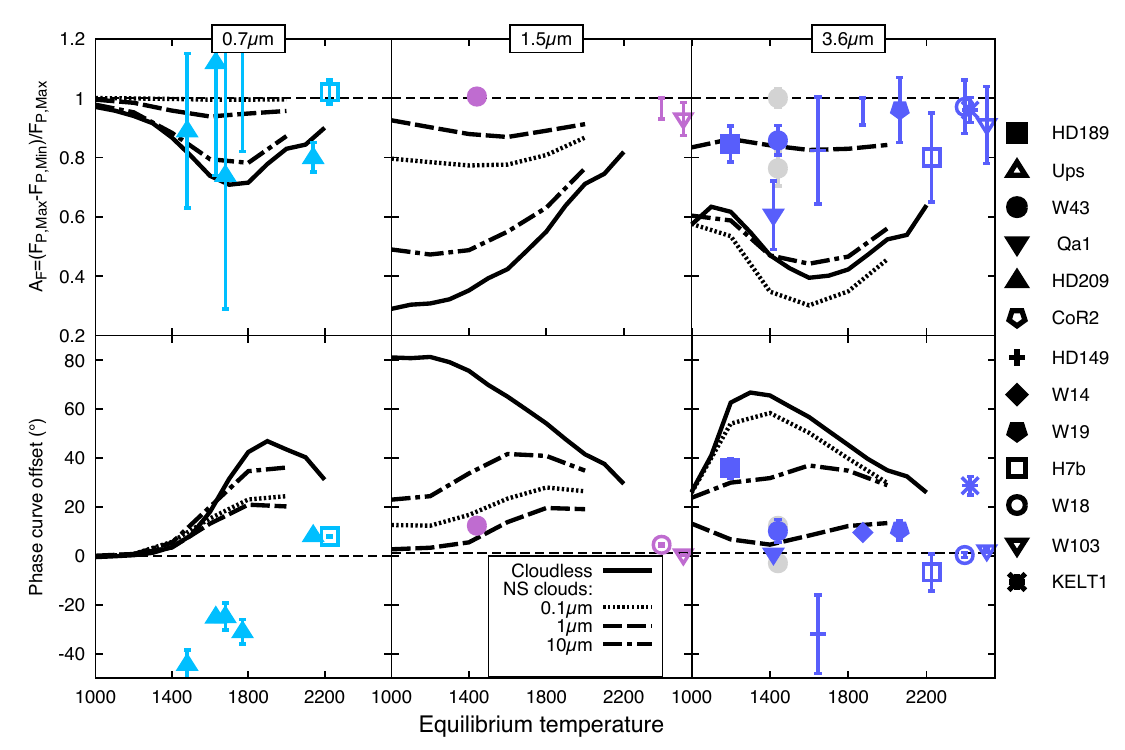}
\includegraphics[width=0.9\linewidth]{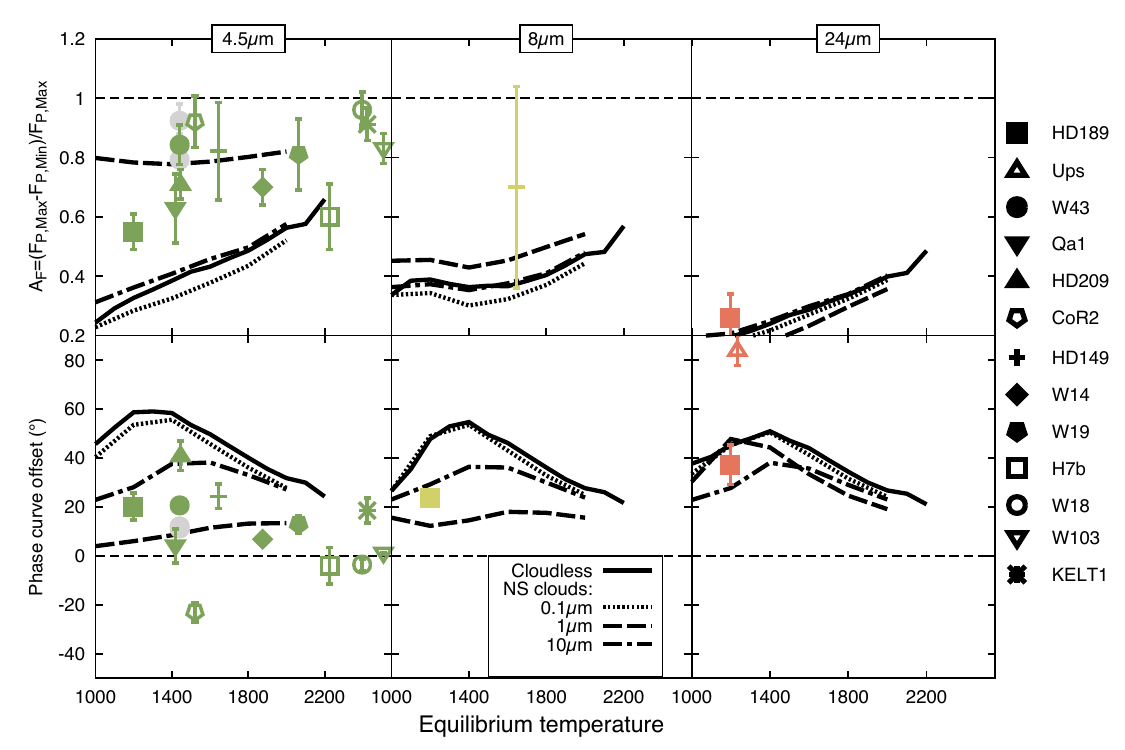}
\caption{Phase curve amplitude (top) and offset (bottom) over 6 different bandpasses. The observation are compared to predictions from cloudless model (plain line) and models with prescribed nightside clouds for three different assumed particle sizes.}
\label{fig::ModelVsObsNSclouds}
\end{figure*}

We find that the phase-curve amplitudes for cloud-free hot Jupiters split into three distinct classes of behavior depending on the wavelength.  In the first class, the phase curve is dominated by thermal emission and exhibits an amplitude that rises monotonically with equilibrium temperature, in agreement with the theories of ~\citet{Perez-Becker2013a,Komacek2016}. This class of behaviour holds at most wavelengths longward of 1 micron where changes in the chemical composition of the atmosphere with temperature do not significantly change the opacities. Fig.~\ref{fig::ModelVsObsNSclouds} illustrates this behaviour at 1.5,  4.5, and 24 microns. When the equilibrium temperature increases, the temperature of the atmosphere increases, leading to a decrease in the radiative timescale of the gas (i.e. the time it takes for a parcel of gas to radiate away its energy). As a consequence, the transfer of heat from the dayside to the nightside becomes less efficient, increasing the temperature contrast at the photosphere.

The second class of behaviour occurs at short wavelengths ($\lambda< 1\mu m$) that are affected by reflected starlight in addition to  thermal emission. The Kepler bandpass ($0.7\mu m$) in Fig~\ref{fig::ModelVsObsNSclouds} illustrates this phenomenon. In the Kepler bandpass the amplitude is large for cool and hot planets but can be as small as 0.7 for planets at intermediate temperatures. The phase curves of cold planets are dominated by reflected light originating, in the cloudless case, from Rayleigh scattering. Therefore, the amplitudes should be close to one, since the nightside does not reflect any light. As the equilibrium temperature increases, thermal emission becomes as important as reflected light in that bandpass, leading to a smaller phase curve amplitude since both the dayside and the nightside are emitting light. Then as the equilibrium temperature increases again, the amplitude increases due to the shortening of the radiative timescale, as in the other phase curves dominated by thermal emission.

The third class of behaviour occurs at specific wavelengths that are influenced by exotic chemical effects, where the phase curve amplitude can exhibit non-monotonic behaviour with equilibrium temperature.  This occurs in particular in the Spitzer 3.6 and 8 micron bands of Fig~\ref{fig::ModelVsObsNSclouds} due to methane and CO opacity but will also happen at all wavelengths where a change in chemical composition with temperature leads to a significant opacity variation. For $T_{\rm eq}<1100\,\rm K$ and for $T_{\rm eq}>1500\,\rm K$, the amplitude of the phase curve rises with equilibrium temperature whereas for $1100\,\rm K<T_{\rm eq}<1500\,\rm K$ the opposite happens. This peculiar behaviour is due to our assumption of local chemical equilibrium that leads to a spatially varying ratio between carbon monoxide and methane molecules on the planet~\citep[see also][]{Dobbs-dixon2017}. Hot Jupiters, however, are not expected to be in local chemical equilibrium since the vertical and horizontal transport of material can be much faster than chemical reactions. In such a case the system is said to be quenched, and homogeneous chemical abundances are expected at the photosphere~\citep{Cooper2006,Visscher2011,Agundez2014,Drummond2018,Steinrueck2019,Drummond2020}. If quenching happens, and as long as CO is the dominant species (rather than CH$_{\rm 4}$), then the amplitude of the phase curve should be monotonically increasing with equilibrium temperature at all wavelengths dominated by thermal emission.
 
 \subsubsection{Phase curve offset}
  
 The offset of the phase curve maximum is shown in the bottom panels of Fig~\ref{fig::ModelVsObsNSclouds}. It always follows a bell curve, with an offset that increases with equilibrium temperature at low equilibrium temperatures and an offset that decreases with equilibrium temperature at larger equilibrium temperatures. Whereas the decrease of the offset with increasing temperature is expected based on the radiative and advective timescales shown in Fig.~\ref{fig::Timescales}, the low temperature behaviour is more puzzling, but is consistent with some of the cases explored in Fig. 9 of~\citet{Komacek2017} 

 Again, the Kepler bandpass stands apart. At low temperatures, when the lightcurve is dominated by reflected light, the amplitude is large and the offset is close to zero because Rayleigh scattering is approximately homogeneous over the dayside of the planet. As the equilibrium temperature increases thermal emission becomes more and more important compared to Rayleigh scattering, leading to an offset that increases with increasing equilibrium temperature. Above $T_{\rm eq}\approx1900\,\rm K$, the decrease of the radiative timescale with equilibrium temperature dominates and the offset decreases with increasing equilibrium temperature.

In these cloudless global circulation models, the offset of the phase curve does not always decreases when the amplitude of the phase curve increases, even in wavelength ranges that are not affected by the day/night variation in chemistry. This is a result that is already apparent in the semi-grey global circulation models of~\citet{Komacek2017} and is in contradiction with the 1D and 2D advective framework proposed by~\citet{Cowan2011},~\citet{Zhang2017} and~\citet{Schwartz2017} that predict such a positive correlation. This points towards different mechanisms setting the eastward shift of the temperature map and the planet-wide temperature contrast. Whereas it is plausible that the former is set by the horizontal advection of heat, the latter is also set by a vertical wave adjustment process that cannot be taken into account easily in the 1D and 2D models of ~\citet{Cowan2011} or ~\citet{Zhang2017}. 

\subsection{The effect of nightside clouds }

We now discuss how the trends in the phase curve offset and the phase curve amplitude change when clouds are present on the nightside of the planet. As shown by the dashed, dot, and dashed-dotted curves in Fig.~\ref{fig::ModelVsObsNSclouds} nightside clouds generally increase the phase curve amplitude and decrease the phase curve offset. Additionally, the presence of nightside clouds flattens the amplitude vs. equilibrium temperature curve, erasing the trends seen in the cloudless case and leading to a flat relationship. As a consequence, if all hot Jupiters have nightside clouds, we do not necessarily expect the phase curve amplitude to increase with the equilibrium temperature.

 The magnitude of the effect depends strongly on the assumed particle size of the clouds. Clouds formed of micron-size particles have the strongest effect on the thermal phase curves of hot Jupiters. This is due to two reasons. First the opacity of micron-sized cloud particles are large around one micron, where the thermal emission of hot Jupiters is peaking, whereas the opacity of smaller particles are usually negligible in the near infrared. As a consequence, clouds with micron-size particles will have the largest possible radiative forcing. Second, for a given cloud mass, the particle number is tied to the particle size. For particles larger than one micron, the number of particles becomes smaller and the total optical depth of the cloud decreases significantly. As an example, particle sizes of ten microns or more would be needed to affect wavelengths as long as the 8 and 24 microns channels, but because the total number of particles drops significantly with particle size, the MnS clouds we choose to model here become relatively transparent if it is made up of these particles. Other clouds, such as silicate based clouds, can have a much larger column mass~\citep{Marley2000} and be more optically thick at large wavelengths. 

When the wavelength of the observation is much larger than the cloud particle size (e.g. 0.1$\mu$m particles at 4.5$\mu$m and 8$\mu$m, and 1$\mu$m particles at 24$\mu$m) the presence of nightside clouds can have the unexpected effect of decreasing the phase curve amplitude. This happens because such clouds are relatively transparent at the wavelength considered and do not strongly alter the photospheric pressure. However, they do affect the thermal structure of the planet and reduce the temperature gradients, leading to a phase curve with a smaller amplitude.

{ Real exoplanet clouds will have a range of particle sizes. Although for the often used log-normal distribution of particle sizes the total opacity is dominated by the largest particles~\citep{Wakeford2015} and can therefore be well represented by a single particle size model, that is not necessarily the case for other particle size distribution~\citep{LeeDphil,Powell2018}. Whereas focusing on single particle size clouds allows for a simpler understanding of the interactions between observable quantities and atmospheric structure, future work is required to examine the robustness of our conclusions by considering complex and varied particle size distributions.}

The models with nightside clouds {brackets almost all of the observations when varying the cloud particle sizes, meaning that the diversity of phase curve parameters can be a consequence of a variety of nightside cloud conditions. Only the observations of Wasp-43b and CoRoT-2b are inconsistent with all our simulations, particularly the large phase curve amplitude of WASP-43b at 1.5 and 4.5$\mu$m} and the negative offset of CoRoT-2b. Models including post-processed silicate clouds, that can have larger optical depth can, however, reach these high phase curve amplitudes (see Sec.~\ref{sec::PostProcess}), but none of our models predict a negative phase curve amplitude for CoRoT-2b. { Additionally, the small offsets and large amplitudes of the Ultra-hot Jupiters (with $T_{\rm eq}>2200\,K$) are shown for completeness but cannot be compared directly to our models as we lack several important opacity sources and the heat transport by H2 dissociation (see methods).} 

Although our model allows for an increase in the phase curve amplitude and a decrease in the phase curve offset compatible with most datapoints, it does not fit simultaneously the phase amplitude and phase offset of specific planets. Particularly, the particle size needed to explain the large amplitude of some points predicts much smaller phase shift than observed. Furthermore, models with nightside clouds only are not able to fit the Kepler observations, since the presence of inhomogeneous clouds on the dayside are needed for this.

\subsection{Models with temperature-dependent MnS clouds}

\begin{figure*}
\includegraphics[width=0.9\linewidth]{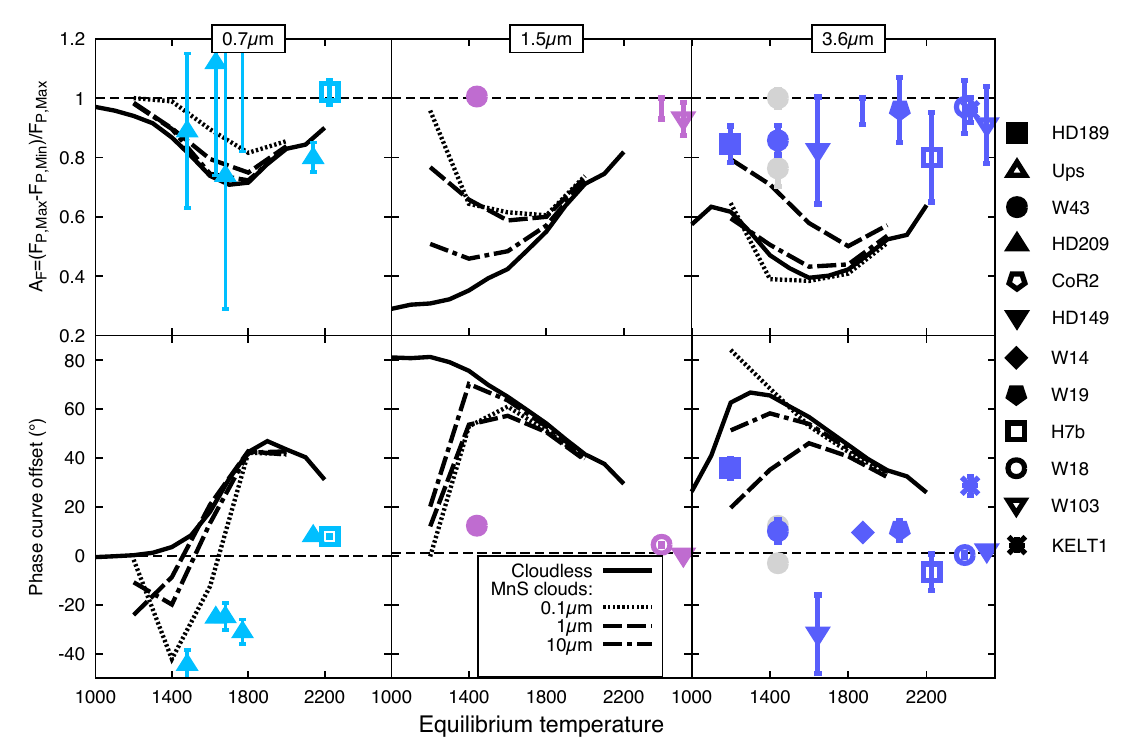}
\includegraphics[width=0.9\linewidth]{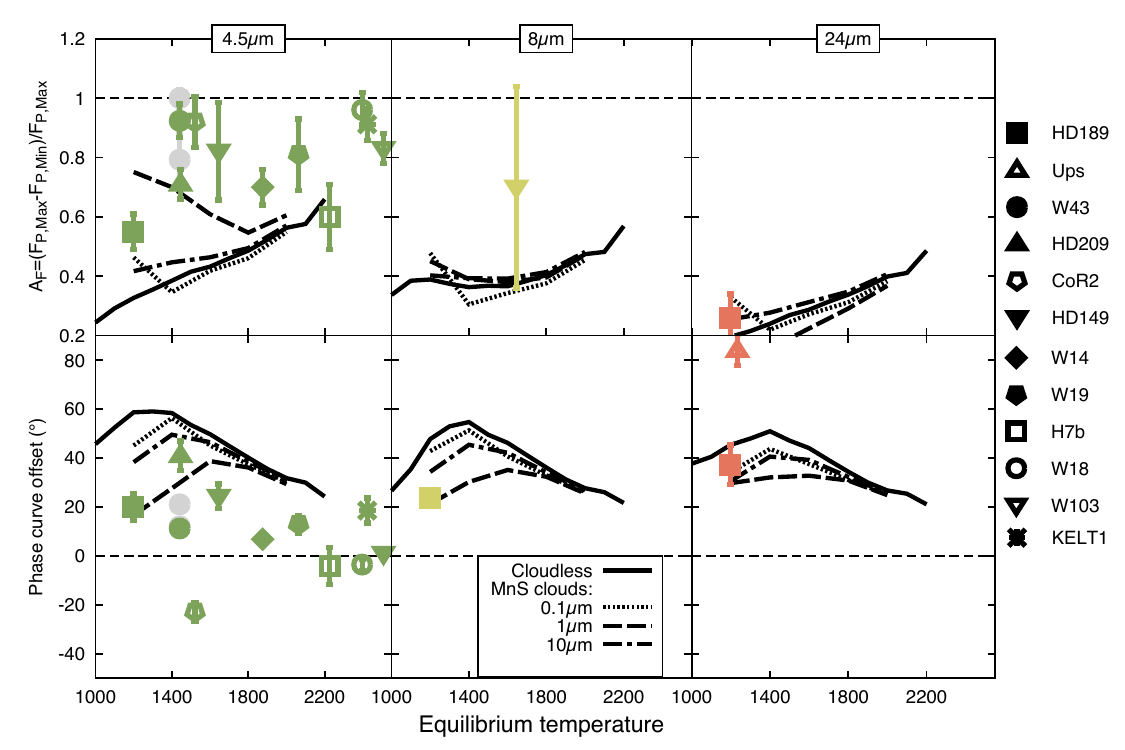}
\caption{Phase curve amplitude (top) and offset (bottom) for 6 different bandpasses. The observation are compared to predictions from cloudless model (plain line) and models with temperature-dependent MnS clouds for three different assumed particle sizes.}
\label{fig::MnS-Self-consistent}
\end{figure*}

We turn towards the set of simulations with temperature-dependent cloud distributions. The cloud maps of MnS are shown in Fig.~\ref{fig::TempMapMnSTeq} and depend strongly on the equilibrium temperature. At low equilibrium temperature, clouds cover the planet homogeneously. At high enough equilibrium temperature clouds almost completely disappear. In the intermediate temperature range, the cloud map is inhomogeneous and tracks the temperature map. As a consequence, as we show in Fig.~\ref{fig::MnS-Self-consistent}, the phase curves and offsets of our temperature-dependent MnS clouds are only affected for $T_{\rm eq}<1800\rm K$. 

In the Kepler bandpass, the temperature-dependent MnS clouds can reproduce the negative phase curve offset seen in the Kepler bandpass, particularly for the cooler planet. This is due to the presence of a strongly reflective crescent of clouds on the west side of the dayside atmosphere~\citep[see][for more details]{Parmentier2016}.

At longer wavelengths, and for the cooler planets, the $1\mu m$ temperature-dependent MnS cloud model usually increases the phase curve amplitude as strongly as in the nightside cloud case but does not reduce the phase curve offset as much. This is because in the temperature-dependent cloud model the eastern part of the dayside, east of the hot spot, is not fully covered by clouds (see section~\ref{sec::Effect}). Therefore, the temperature-dependent MnS cloud model provides a better simultaneous match to both the phase offset and the the phase curve amplitude of HD189733b and HD209458b at all thermal wavelengths. It cannot, however, reproduce the large phase curve amplitude of WASP-43b and CoRoT-2b and is no different than the cloudless model for higher temperature planets. Additionally, the negative offset of CoRoT-2b cannot be represented by our models.{ We note that \citet{Carone2020} provides an alternative explanation for these planets, where the momentum exchange between the deep atmospheric layers and the photosphere can lead to a westward hot spot offset and a cool nightside}. 

\section{Diverse clouds form diverse phase curves.}
To date the spectral signature of the cloud composition of hot Jupiters have never been detected. All clouds currently observed are well fit with a grey cloud or a Rayleigh slope. Therefore, our inferences into the chemical composition of these clouds are mainly based on the different expected physical properties of the different cloud species. For example clouds of different compositions are expected to form at different pressure and temperature levels, leading to different signatures in the transit spectrum~\citep{Barstow2017,Lee2015,Powell2018,Powell2019,Gao2020a}. Clouds of different compositions are also expected to form at different longitudes and latitudes in the planet, leading to different signatures in the optical phase curves~\citep{Oreshenko2016,Parmentier2016,Lee2016,Lee2017}. In the future, JWST and other platforms may be able to place constraints on the cloud composition, either through direct detection of solid-state spectral features of cloud condensates~\citep{Wakeford2015}, as has been done for brown dwarfs ~\citep[e.g.][]{Cushing2006} or through the differing effects that different cloud compositions will have on the phase curves, including the light curve amplitude and offset. Given the upcoming observations from JWST, ARIEL, and other platforms, it is therefore important to understand how differing cloud condensate species will influence infrared lightcurves and spectra over a large range of wavelengths.  We explore this topic in this section.

Given the large number of possible chemical and physical condensate properties~\citep{Burrows1999,Lodders2003a,Lee2015,Helling2016,Helling2020} we decided to use two specific cloud parameters -- particle size and composition -- to highlight the potential diversity of the effect of clouds on phase curves. To keep the study computationally feasible, we decided to use post-processed clouds:  we use the cloudless temperature map to determine the expected cloud map of each cloud species and include the corresponding cloud opacity only when calculating the spectra and phase curves (see Sect~\ref{Sec::Methods_clouds} for details). 

\begin{figure*}
\includegraphics[width=0.95\linewidth]{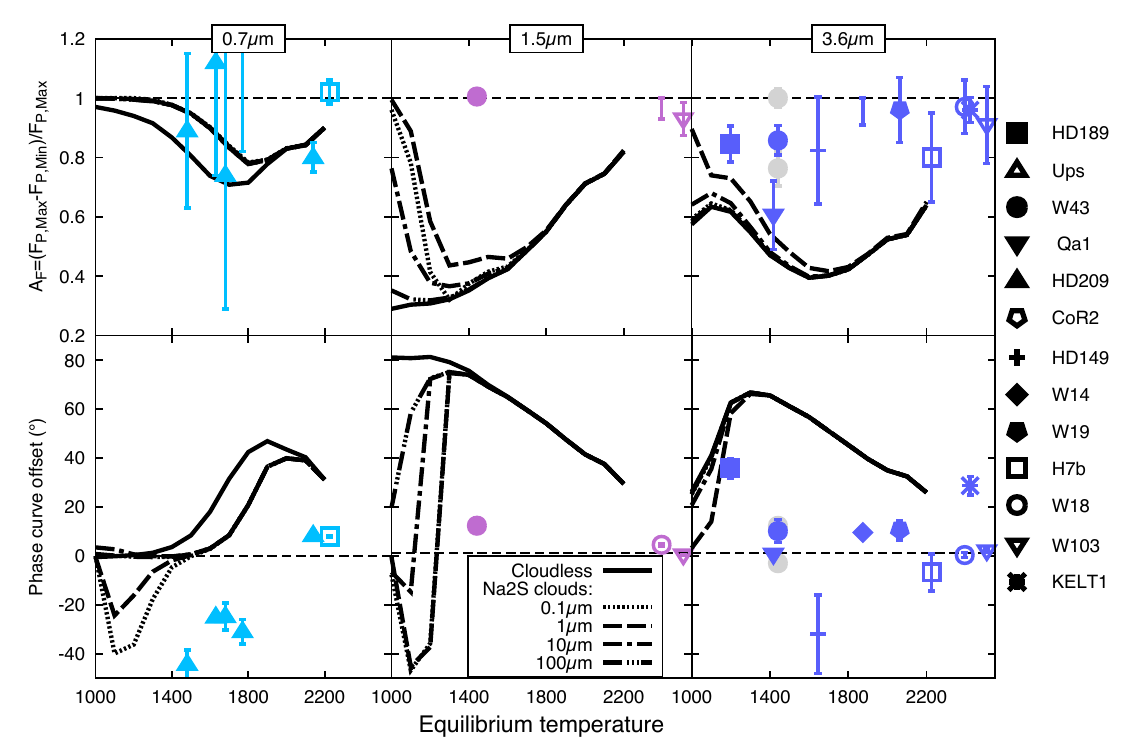}
\includegraphics[width=0.95\linewidth]{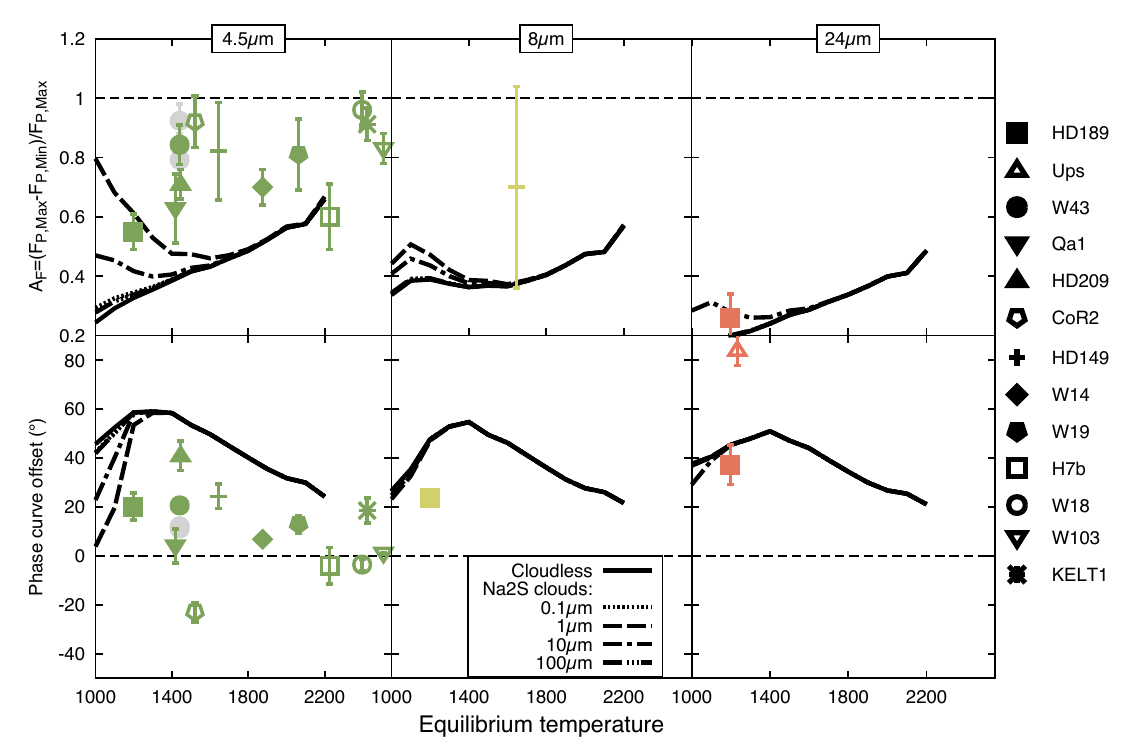}
\caption{Phase curve amplitude (top) and offset (bottom) for 6 different bandpasses. The observation are compared to predictions from cloudless model (plain line) and models with post-processed $\rm Na_2S$ clouds for three different assumed particle sizes.}
\label{fig::Na2S-NSC}
\end{figure*}

\begin{figure*}
\includegraphics[width=0.95\linewidth]{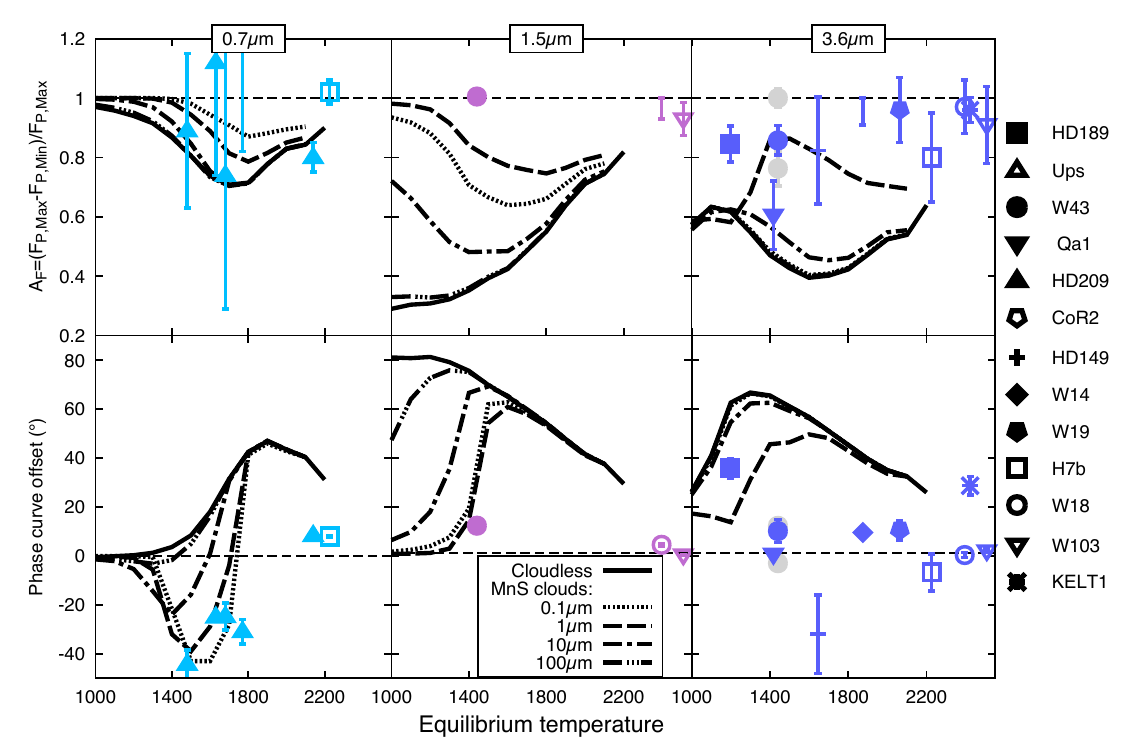}
\includegraphics[width=0.95\linewidth]{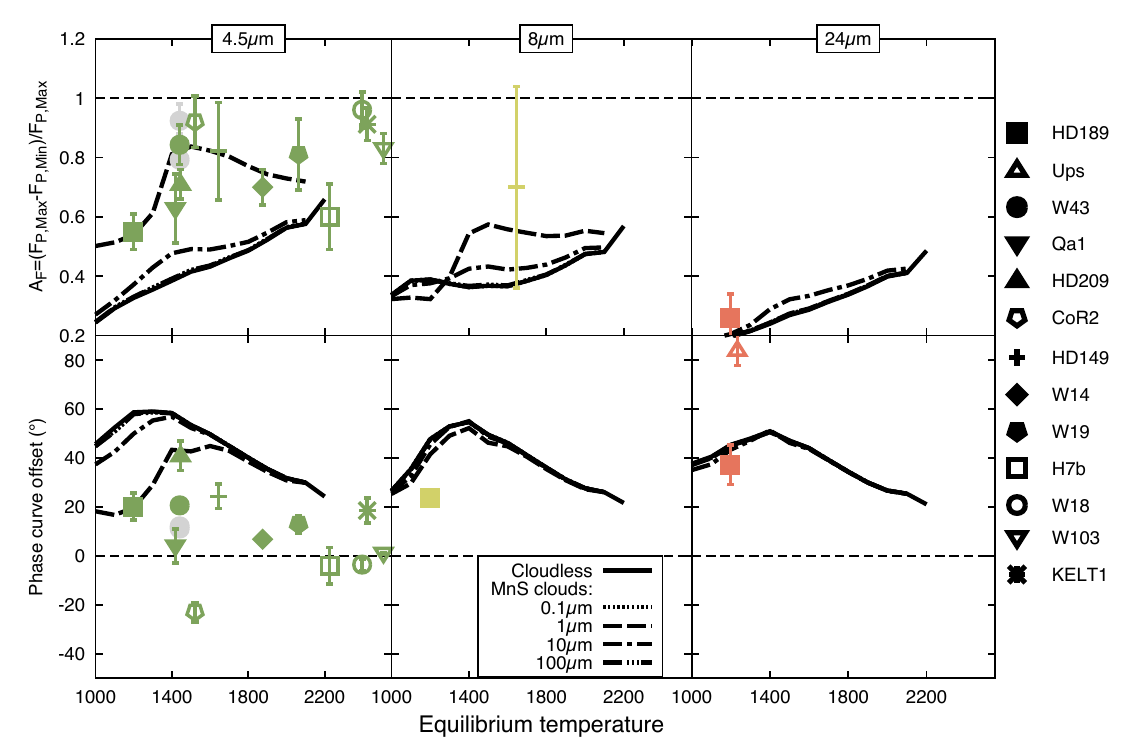}
\caption{Phase curve amplitude (top) and offset (bottom) for 6 different bandpasses. The observation are compared to predictions from cloudless model (plain line) and models with post-processed $\rm MnS$ clouds for three different assumed particle sizes.}
\label{fig::MnSCloudsNSC}
\end{figure*}

\begin{figure*}
\includegraphics[width=0.95\linewidth]{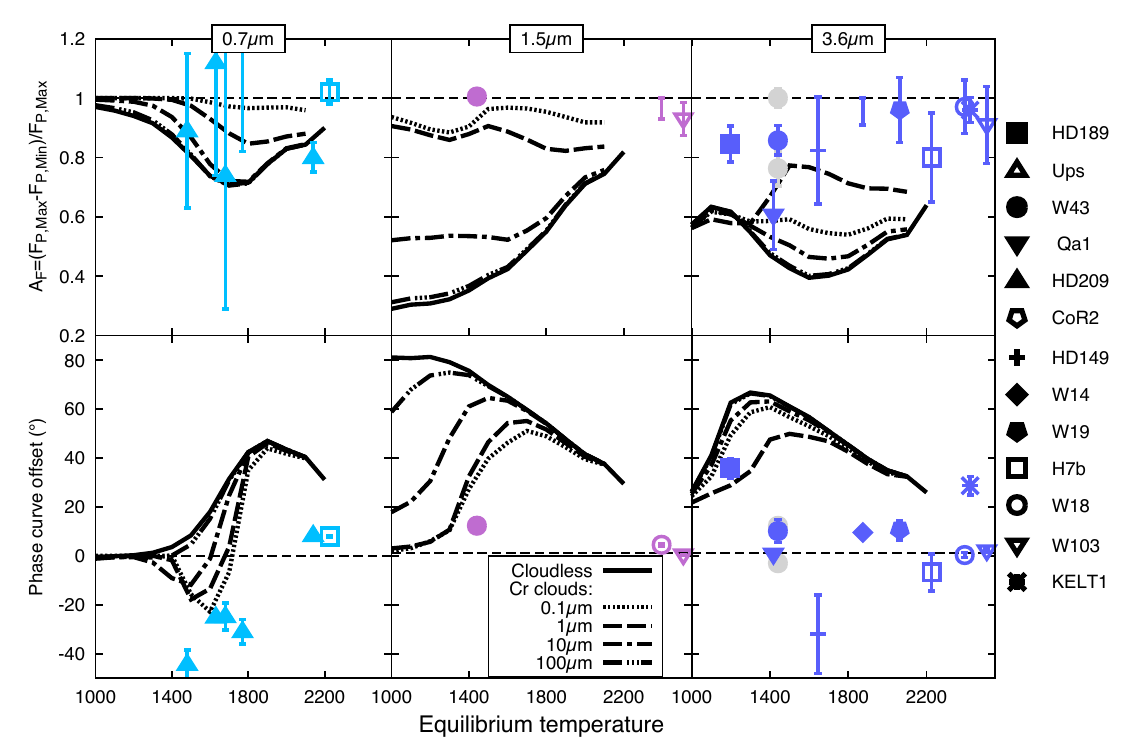}
\includegraphics[width=0.95\linewidth]{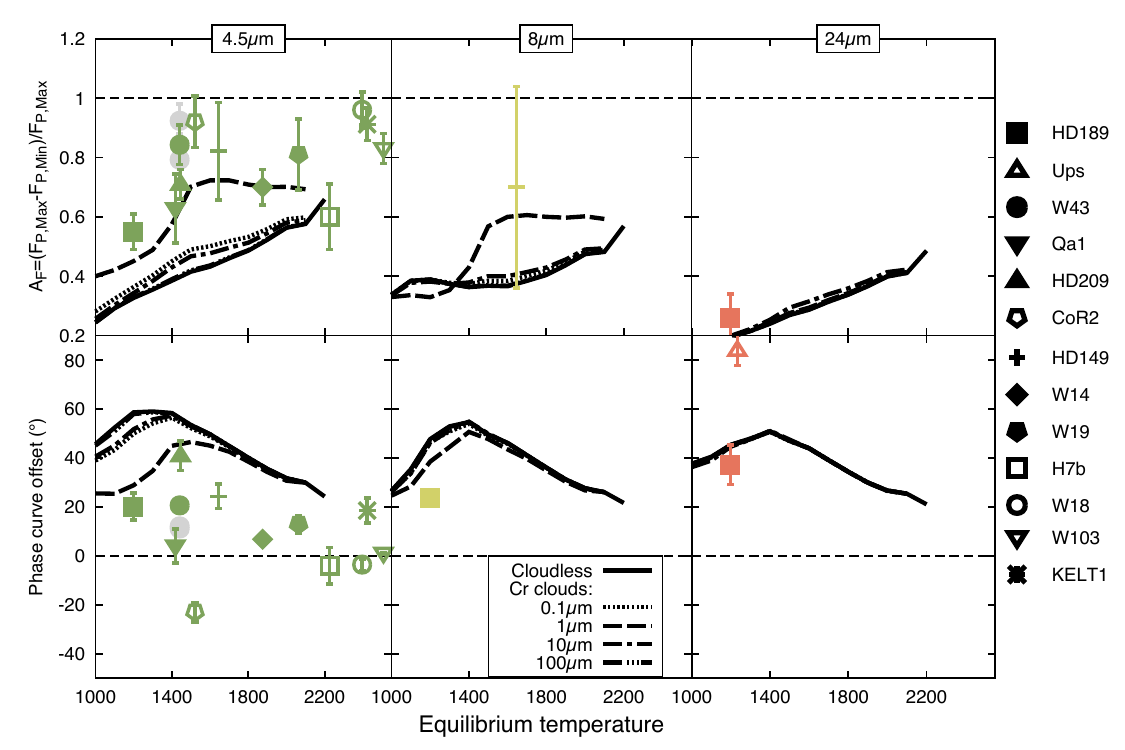}
\caption{Phase curve amplitude (top) and offset (bottom) for 6 different bandpasses. The observation are compared to predictions from cloudless model (plain line) and models with post-processed chromium clouds for three different assumed particle sizes.}
\label{fig::Cr-NSC}
\end{figure*}

\begin{figure*}
\includegraphics[width=0.95\linewidth]{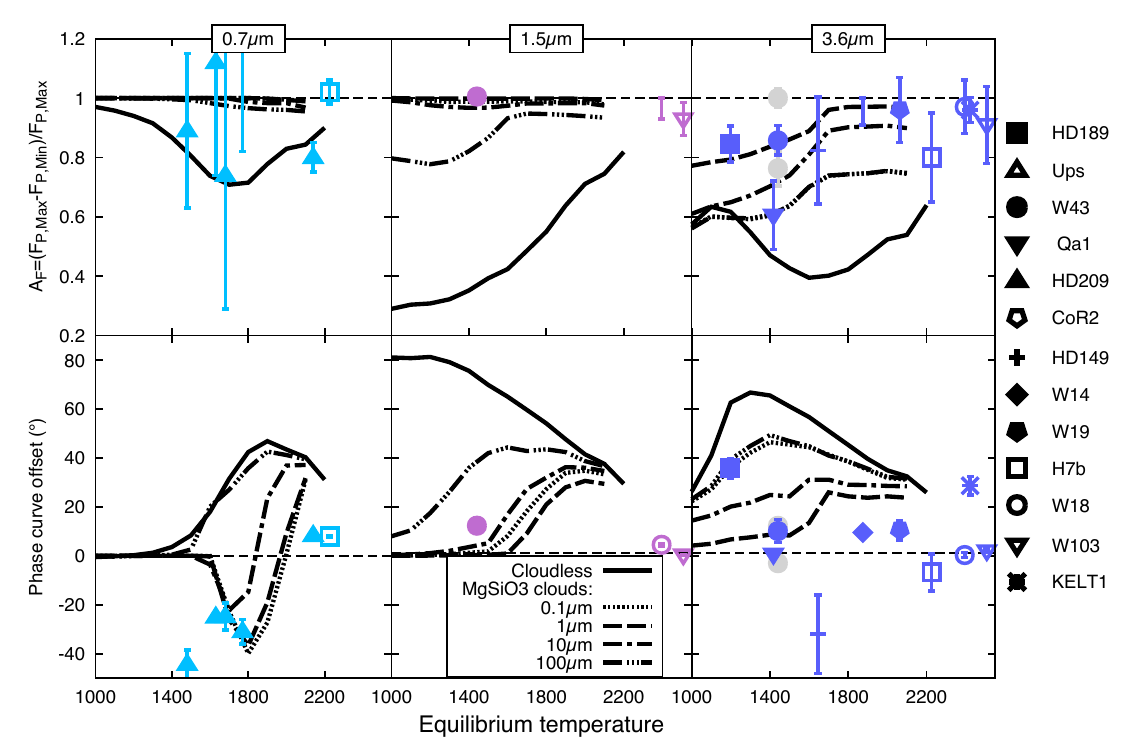}
\includegraphics[width=0.95\linewidth]{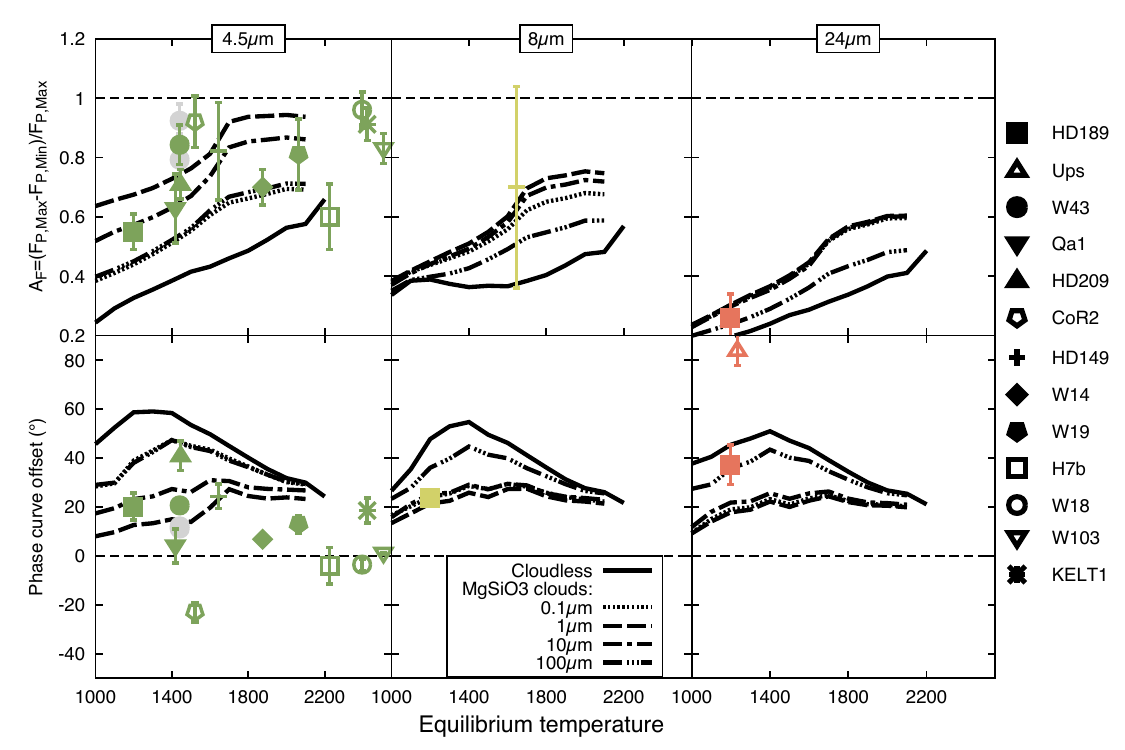}
\caption{Phase curve amplitude (top) and offset (bottom) for 6 different bandpasses. The observation are compared to predictions from cloudless model (plain line) and models with post-processed $\rm MgSiO_3$ clouds for three different assumed particle sizes.}
\label{fig::MgSiO3-NSC}
\end{figure*}

\begin{figure*}
\includegraphics[width=0.95\linewidth]{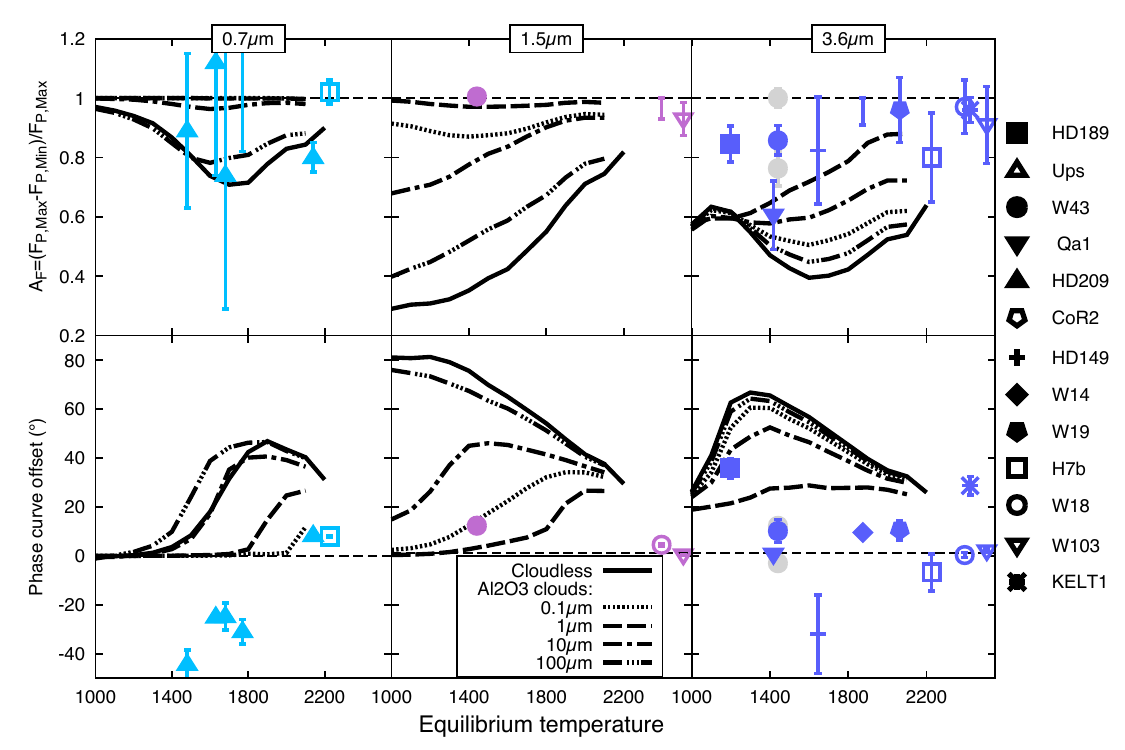}
\includegraphics[width=0.95\linewidth]{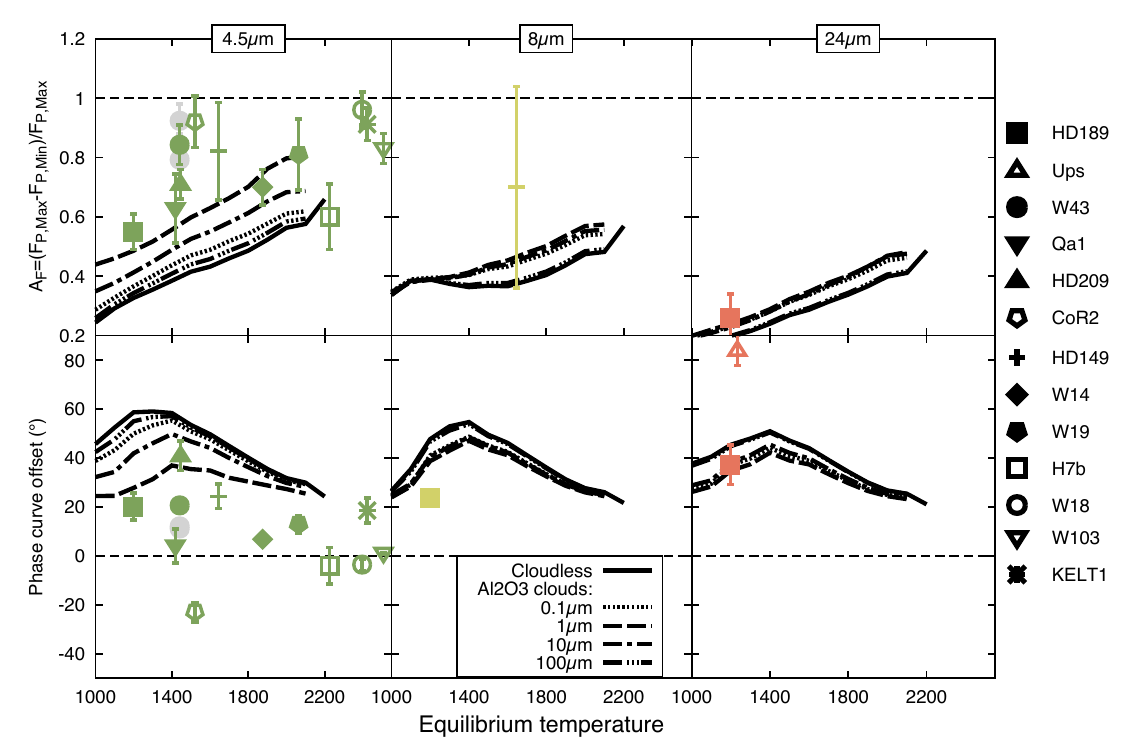}
\caption{Phase curve amplitude (top) and offset (bottom) for 6 different bandpasses. The observation are compared to predictions from cloudless model (plain line) and models with post-processed $\rm Al_2O_3$ clouds for three different assumed particle sizes.}
\label{fig::Al2O3-NSC}
\end{figure*}

\begin{figure*}
\includegraphics[width=0.95\linewidth]{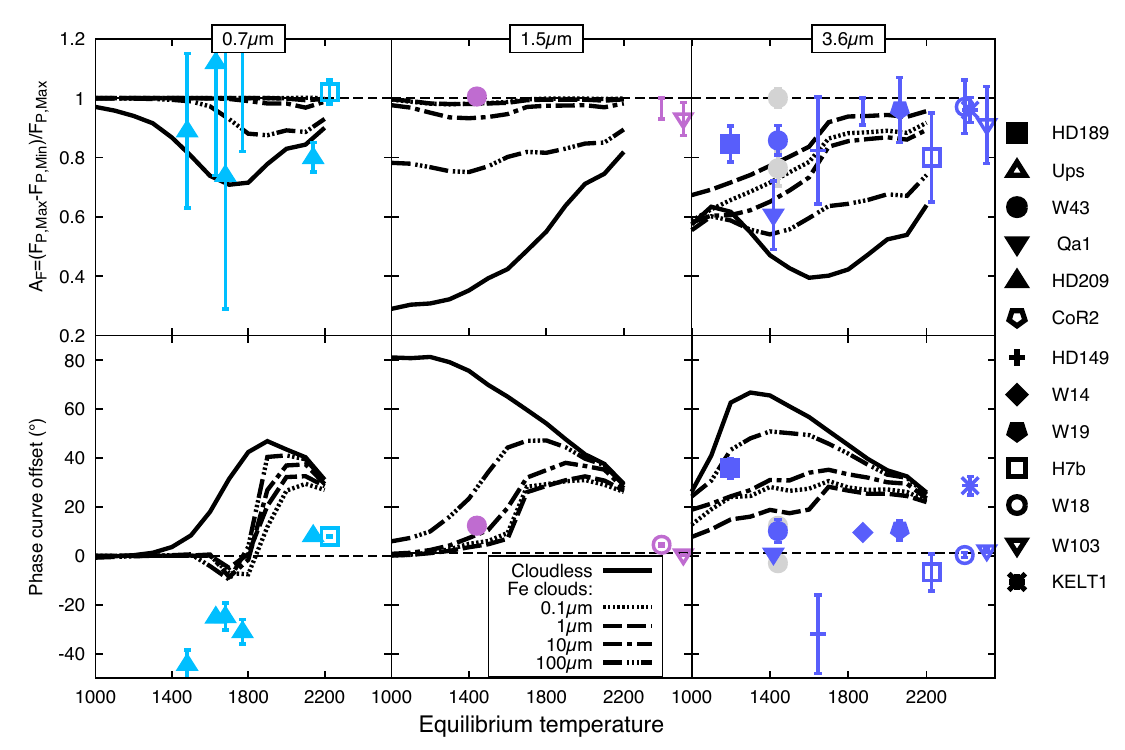}
\includegraphics[width=0.95\linewidth]{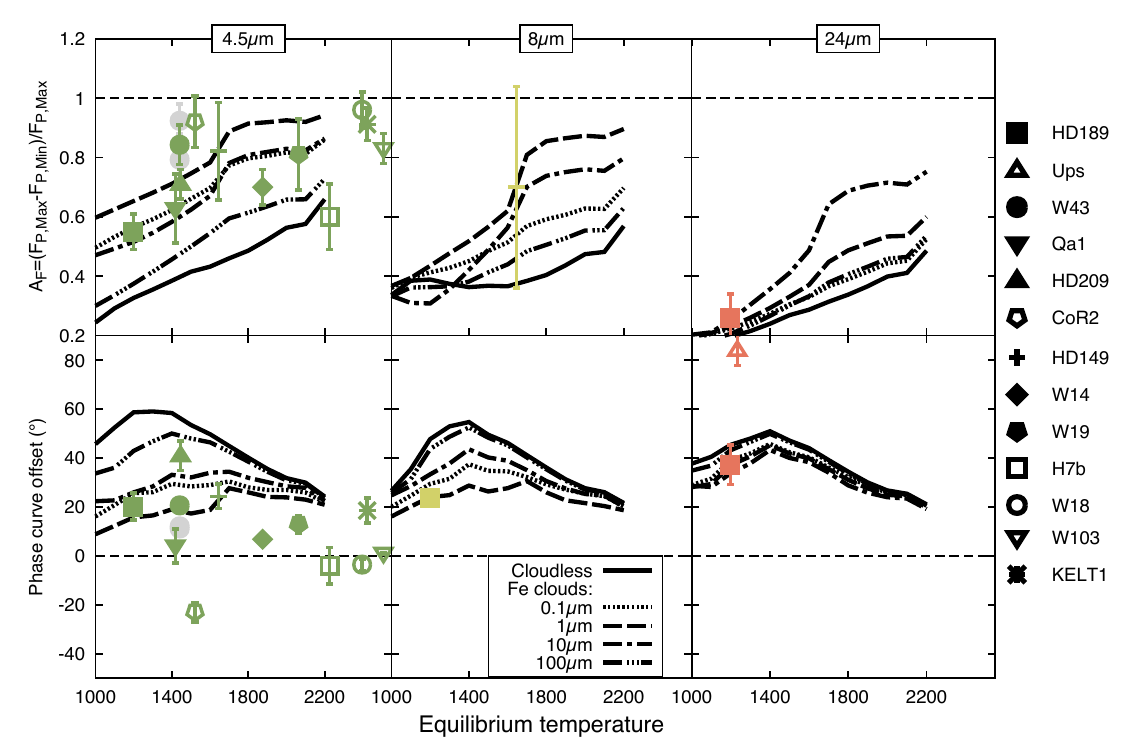}
\caption{Phase curve amplitude (top) and offset (bottom) for 6 different bandpasses. The observation are compared to predictions from cloudless model (plain line) and models with post-processed iron clouds for three different assumed particle sizes.}
\label{fig::Fe-NSC}
\end{figure*}

\begin{figure*}
\includegraphics[width=0.95\linewidth]{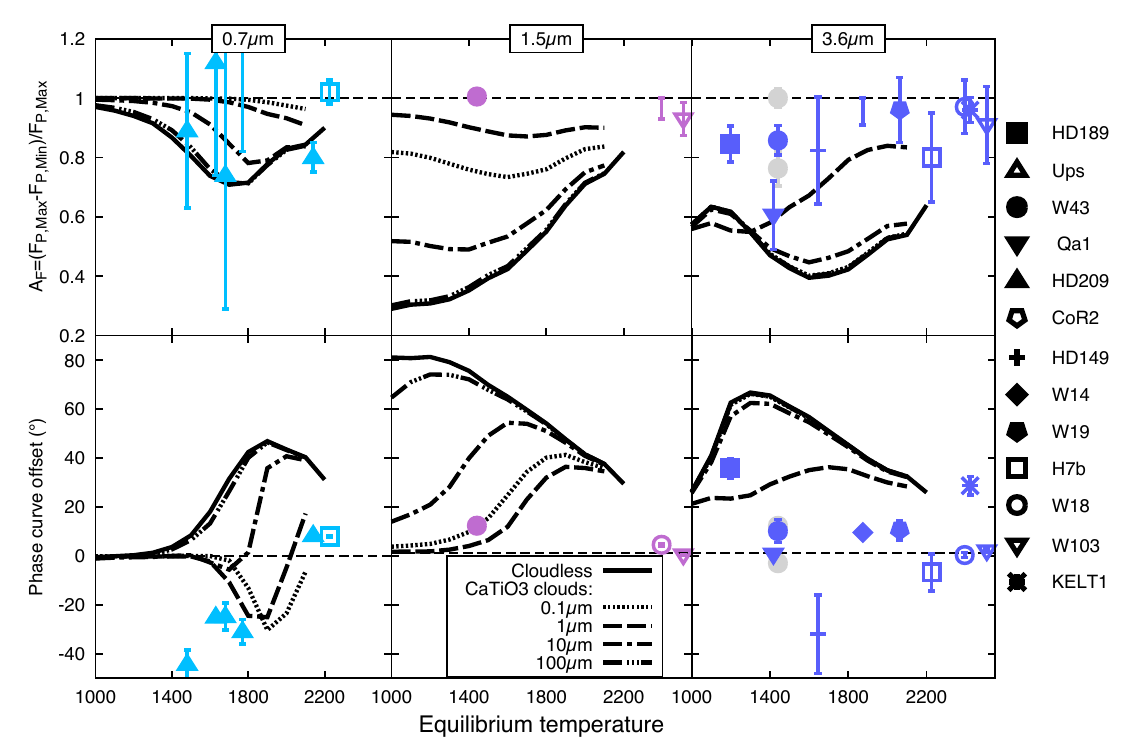}
\includegraphics[width=0.95\linewidth]{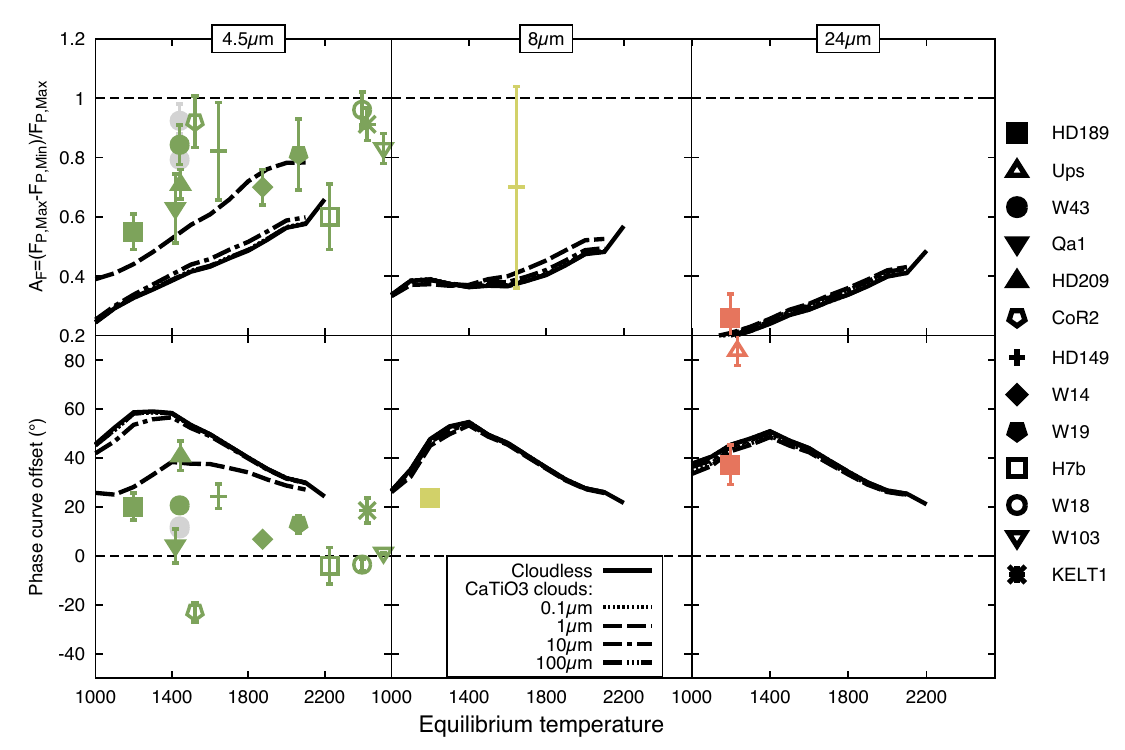}
\caption{Phase curve amplitude (top) and offset (bottom) for 6 different bandpasses. The observation are compared to predictions from cloudless model (plain line) and models with post-processed $\rm CaTiO_3$ clouds for three different assumed particle sizes.}
\label{fig::CaTiO3-NSC}
\end{figure*}

\begin{figure*}
\includegraphics[width=0.95\linewidth]{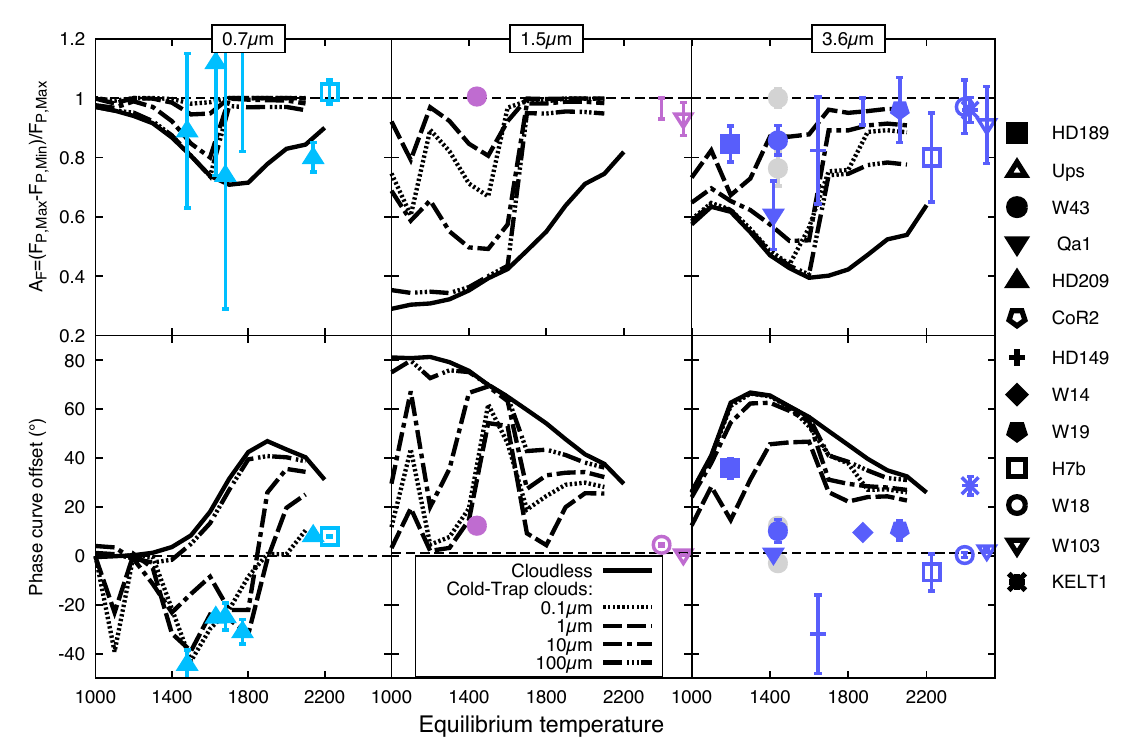}
\includegraphics[width=0.95\linewidth]{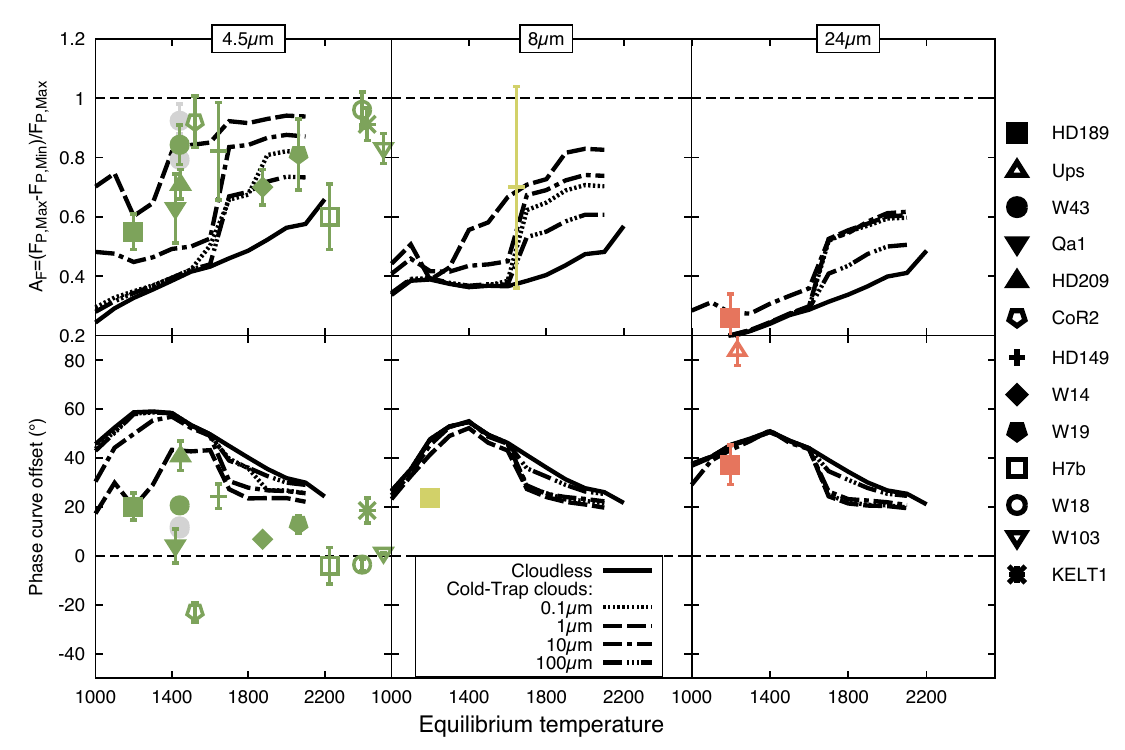}
\caption{Phase curve amplitude (top) and offset (bottom) for 6 different bandpasses. The observations are compared to predictions from cloudless model (plain line) and models with post-processed, temperature dependent clouds where the cloud chemical composition varies with the equilibrium temperature of the planet following the cold-trap model of~\citet{Parmentier2016}.}
\label{fig::ColdTrap}
\end{figure*}

\subsection{Individual cloud species}
\label{sec::PostProcess}

We choose to study the same set of clouds as in~\citet{Parmentier2016}: $\rm Na_2S$, MnS, Cr, $\rm MgSiO_3$, Fe, $\rm Al_2O_3$ and $\rm CaTiO_3$. Although not exhaustive, this list provides a first glimpse at the expected diversity of clouds in hot Jupiter atmospheres. Each cloud is determined by a specific condensation curve, cloud material abundance and optical properties, all taken similar to~\citet{Parmentier2016}. 

We now turn to figures \ref{fig::Na2S-NSC} to \ref{fig::CaTiO3-NSC} showing the phase curve amplitude and phase curve offset for all our models with radiatively passive clouds. As expected each cloud affects the phase curve over a specific range of equilibrium temperatures. In Fig.~\ref{fig::Na2S-NSC}, $\rm Na_{\rm 2}S$ significantly affect the phase curves of planets cooler than $\approx 1300\,\rm K$. For these planets the thermal emission is small enough in the HST/WFC3 bandpass that reflected light from the clouds can produce a large negative offset in the HST/WFC3 bandpass~\citep{Keating2017} . Additionally, the apparent match between the $1\mu m$ $\rm Na_{\rm 2}S$ clouds and HD189733b phase curve observations would deserve a separate, more specific study with radiatively active Na$_{\rm 2}$S clouds. 

The effect of neglecting radiative feedback on the phase curve amplitude and offset can be seen by comparing the MnS models without radiative feedback from Fig.~\ref{fig::MnSCloudsNSC} and the ones with radiative feedback previously shown in Fig.~\ref{fig::MnS-Self-consistent}. The main effect of the radiative feedback is to globally warm the atmosphere of the planet~\citep[e.g.][]{Charnay2015a, Oreshenko2016,Lines2019,Roman2019}. As a consequence, the curves representing the phase curve amplitude and phase curve offset are shifted by $\approx 100$ to $200\,\rm K$ towards higher temperatures when the cloud radiative feedback is not taken into account. e.g. the spatial distribution of the clouds of a model at a given equilibrium temperature is similar to the one of a cooler model where the radiative feedback of the cloud is taken into account. For a given planet at a given temperature, this translates to an overestimate of the phase curve amplitude and an underestimate of the phase curve offset. However, we expect the general trends seen in the post-processed case to be conserved if radiatively active clouds were considered. 

The chromium clouds shown in Fig.~\ref{fig::Cr-NSC} have a lower abundance than most of the clouds considered here. However, they can provide enough opacity to increase the phase curve amplitude to very high values in the HST/WFC3 bandpass and should be considered in future studies. It can also be a significant cloud opacity to consider when interpreting the Spitzer phase curve observations. 

As seen in Fig.~\ref{fig::MgSiO3-NSC}, silicate (MgSiO$_{\rm 3}$) clouds considered here can easily produce large negative offset in the Kepler bandpass. They can also drastically reduce the phase curve offset and increase the phase curve amplitude at all wavelengths shorter than $8\mu m$ for all the equilibrium temperatures considered here. Particularly, the silicate clouds could be a good match to the extremely high phase curve amplitude observed for WASP-43b at both 3.6 and 4.5$\mu m$~\citep[see][for a detailed discussion of this model applied to WASP-43b]{Venot2019}.

$\rm Al_2O_3$ (Fig.~\ref{fig::Al2O3-NSC}), iron (Fig.~\ref{fig::Fe-NSC}) and  $\rm CaTiO_3$ (Fig.~\ref{fig::CaTiO3-NSC}) clouds are all good candidates to explain the large phase curve amplitudes observed in the high equilibrium temperature planets (i.e. $T_{\rm eq}\approx1800-2200\,\rm K$) at the transition between hot and ultra hot Jupiters. They cannot, however, reproduce the small offsets observed. Mechanisms such as MHD~\citep{Rogers2014,Rogers2014a} could vary the dayside distribution of temperature and hence the phase curve offset without changing much the day/night contrast~\citep[e.g.][]{Komacek2017,Arcangeli2018}. Future studies coupling MHD models and non-grey radiative transfer should investigate this question further. 

Overall, if all clouds in hot Jupiters have similar chemical composition and physical properties we would expect he phase curve offset to increase or be constant with equilibrium temperature rather than decrease with equilibrium temperature as in the cloudless case. Particularly, if all hot Jupiter nightside clouds are silicate clouds with similar particle size distributions, we would expect a rise in the phase curve offset in the HST bandpass between equilibrium temperatures from 1200 to 2000 K, a prediction that could be tested with currently available telescopes.

\subsection{Cold-trap model}
As discussed in~\citet{Parmentier2016}, it is unlikely that the most refractory clouds will be present at the photosphere of the cooler hot Jupiters. Exactly at which equilibrium temperatures the clouds will be cold trapped depends strongly on the deep thermal profile and the strength of the vertical mixing in the deep atmospheric layers~\citep[e.g.][]{Spiegel2009,Powell2018}. Although recent insights from planetary interior modelling by~\citet{Thorngren2019} showed that the deep temperature is likely larger than in non-irradiated planets~\citep{Guillot2002,Spiegel2013,Fromang2016,Tremblin2017,Menou2019,Sainsbury-Martinez2019}, { the deep thermal profile inferred from the observation of the mass, radius and age of the planet is often degenerate with the planet metallicity, and large uncertainty remain.} In~\citet{Parmentier2016} we considered a cloud sequence driven by the possibility of clouds being cold trapped in the deep layer of a planet assuming a low entropy deep interior. The sequence happens to reasonably match the Kepler phase curve offset and apparent albedos that were observed. We show the variation of phase offset and phase curve amplitude of this model at different wavelengths in Fig.~\ref{fig::ColdTrap}. We see that the models matching the Kepler observations can produce very large variations of phase curve offset and amplitudes with equilibrium temperatures. Observing such a sequence would lead to direct evidence of cold-trap processes and better insight on the planet internal temperature. However, given the uncertainties in the cold-trap model, combined with the possible variations in particle sizes due to microphysics, it is likely hot Jupiters phase curve parameters will show mainly a scattered behaviour with any potential trend hidden behind the complexity of nightside clouds properties. 

\section{Consequences for interpreting observations}
{ We showed that the presence of nightside clouds on exoplanets significantly alter the observations by changing the height of the photosphere as a function of longitude and wavelength. In the following sections we discuss how these changes makes retrievals of dynamical quantities from the observations more complex and what mitigation strategies can be used to compare simplified models to observations. }

\label{sec::Consequences}
\subsection{Energy balance}
\label{sec::BrightnessTemp}
\begin{figure}
\includegraphics[width=\linewidth]{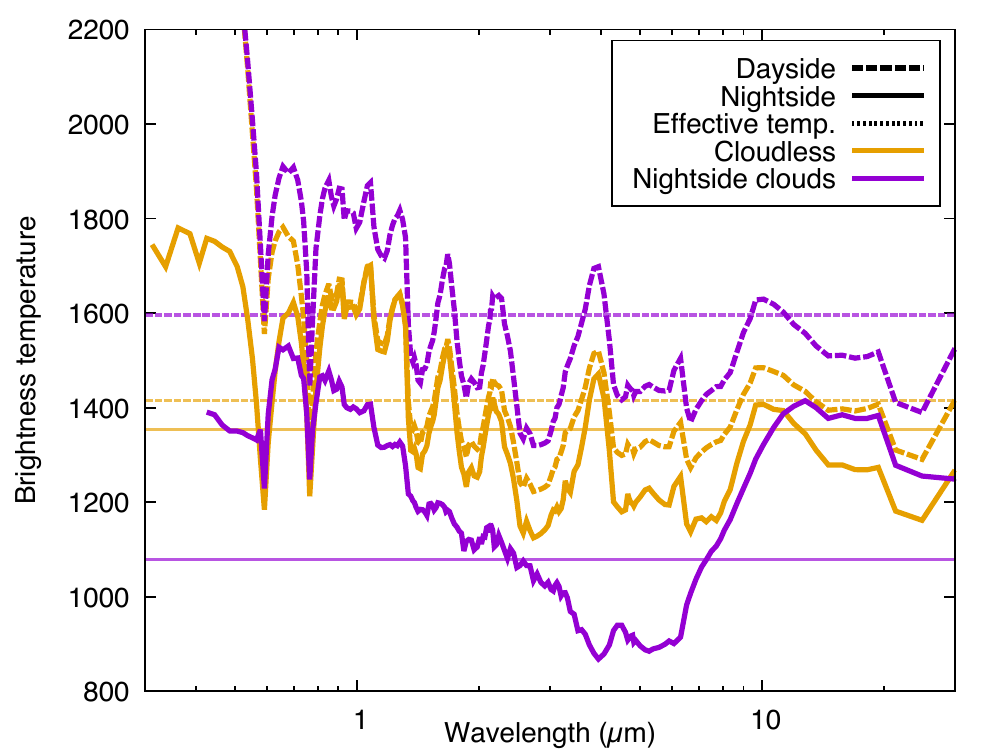}
\caption{Brightness temperature on the dayside and the nightside of our cloudless and nightside cloud model with $T_{\rm eq}=1400\,\rm K$ and particle size of $1\mu m$. The horizontal lines are the corresponding effective temperatures. The brightness temperature varies with wavelengths by $30\%$ on the clear case and by up to $60\%$ on the cloudy nightside case.}
\label{fig::Tbright}
\end{figure}

Measurements of the phase curve amplitude of exoplanets in specific bandpasses have often been used to infer atmospheric dynamical properties such as day/night heat transport and bond albedo through energy balance arguments~\citep[e.g.][]{Cowan2011,Schwartz2015,Schwartz2017}. One of the important approximations needed within this framework is related to the link between the brightness temperature observed at one wavelength at a given phase and the total flux emitted by the planet at a given hemisphere. ~\citet{Cowan2011} verified that one could link the broadband brightness temperature observed during secondary eclipse to the dayside effective temperature with approximately $10\%$ uncertainty based on a set of 1D radiative/convective models. We show in Figure~\ref{fig::Tbright} that the brightness temperature can vary by up to $\approx 30\%$ in the clear case and up to $\approx 60\%$ when nightside clouds are present. Importantly, the brightness temperature at $4.5\mu m$ is systematically lower than the effective temperature, leading to a potential bias towards lower temperatures when interpreting nightside temperatures. In the nightside cloud case, the nightside temperature is lower than the effective temperature at both $3.6\mu m$ and $4.5\mu m$. As a consequence, when nightside clouds are present averaging the $3.6\mu m$ and $4.5\mu m$ brightness temperatures such as in~\citet{Keating2019} would not lead to a better estimate of the effective temperature. {Estimating Bond albedos and heat redistribution efficiency based on Spitzer phase curves can therefore bias the results towards large Bond albedos, since the brightness temperature can be systematically lower than the effective temperature. }

\begin{figure}
\includegraphics[width=\linewidth]{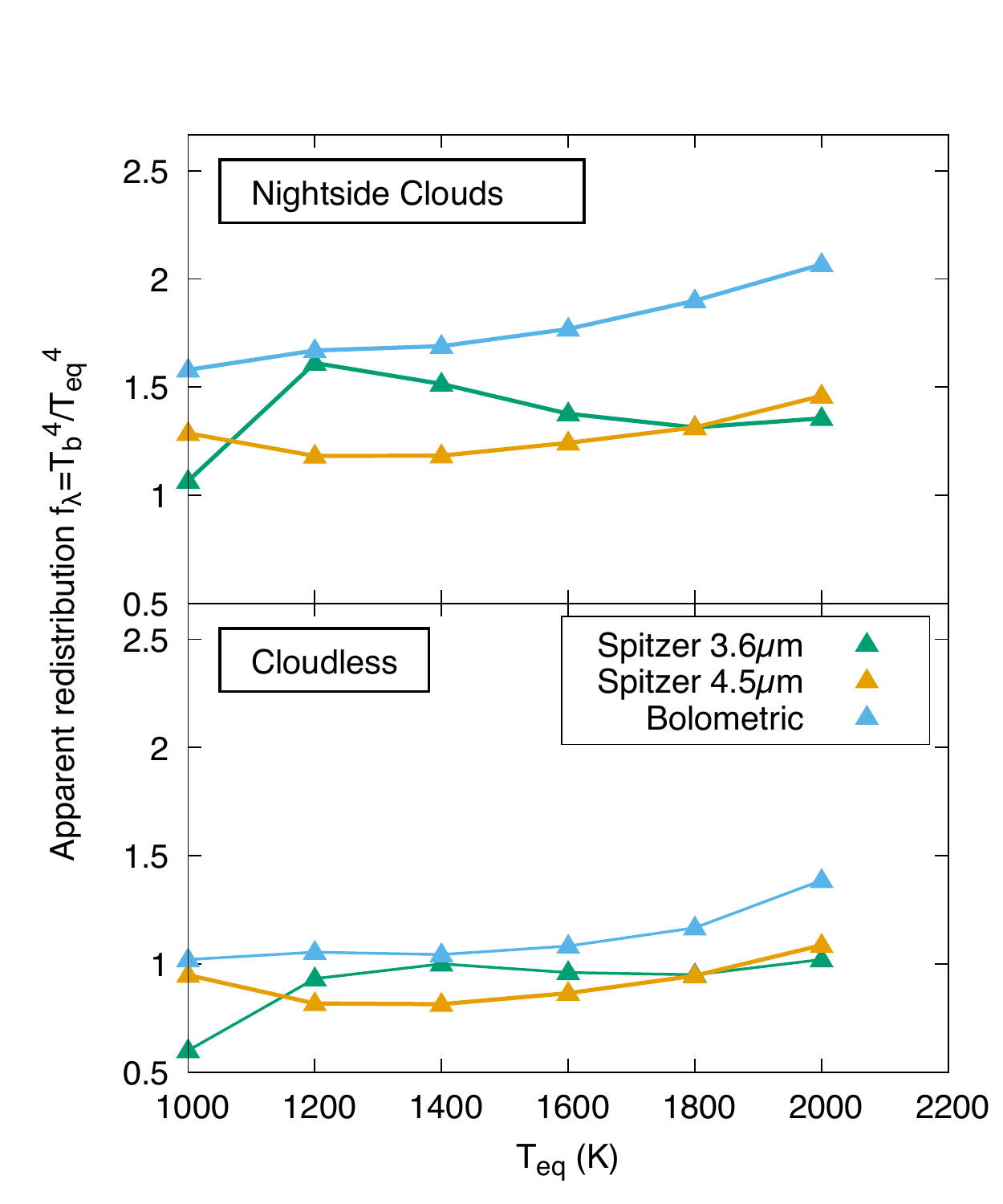}
\caption{Redistribution parameter inferred from only the Spitzer $3.6\mu m$ or $4.5\mu m$ dayside brightness temperatures compared to the real, bolometric redistribution for the cloudless case (bottom) and the nightside cloud case (top). The inferred redistribution based on Spitzer observations alone often overestimate the heat transport and show a different trend with equilibrium temperature than the what the real redistribution is.}
\label{fig::TbrightTrend}
\end{figure}

{  The discrepancy between brightness and effective temperature can also affect the trends in the planet population. As shown in Figure~\ref{fig::TbrightTrend} the trend of redistribution vs. equilibrium temperature that one would infer from observations in the Spitzer bandpasses can be very different from the trend shown by the actual heat redistribution.} Phase curves covering a wider range of wavelengths, obtained by JWST or Ariel are necessary to correctly estimate the effective temperatures and hence precisely estimate the bulk heat transport from the dayside to nightside of the planet.

\subsection{How to link observables and theoretical predictions ?}
Phase curves have also been used to decipher dynamical mechanisms. The phase curve amplitude and the phase curve offset at a given wavelength has been compared to analytical predictions~\citep{Cowan2011,Komacek2017,Zhang2017}, shallow-water models~\citep{Perez-Becker2013a,Hammond2018} and semi-grey global circulation models~\citep{Heng2011a,Rauscher2012b,Komacek2017,Roman2017,Roman2019} in order to gain insights in recirculation efficiency, potential drag mechanisms or cloud behaviour. All these models assume that any given phase curve is probing either isobars (for the semi-grey models) or a layer following a streamline~\citep{Cowan2011a,Hu2015}. However, we have seen in Fig.~\ref{fig::Photosphere} that phase curves at a given wavelength can probe very different pressure levels on the dayside and the nightside. The phenomenon, already explored by~\citet{Dobbs-dixon2017} for the cloudless chemical equilibrium case, is highly amplified when nightside clouds are present. It therefore becomes impossible to simply link the phase curve offset and amplitude at a given wavelength to the hot spot offset and the day/night temperature contrast on the planet itself. 

A more robust procedure would be to use a retrieval, taking into account the opacity variation with longitude, such as the 2.5D retrieval of~\citet{Irwin2019} that can provide a more robust estimates of the uncertainties on the hot spot position and the day/night contrast on isobars and use these corrected estimate to compare with theoretical model. Alternatively, dayside hot spot offsets could be more accurately through eclipse mapping~\citep[e.g.][]{DeWit2012,Majeau2012} with JWST as eclipse mapping measurements should not be strongly affected by the presence of nightside clouds. { Ultimately, parameter exploration with global circulation models taking into account the full feedbacks between atmospheric circulation, cloud, non-grey radiative transfer will likely be needed to fully understand the complexity of these objects. Although current models are too slow to efficiently explore the parameter space needed to fully interpret observations, new GPU based models~\citep{Mendonca2016,Deitrick2020} could significantly speed up the calculations and allow a more systematic 3D retrieval approach.}

\subsection{Identifying the nightside cloud composition}

Many concerns above would be solved if the cloud map were available for a given planet. This can be done in at least two ways. First, the use of reflected lightcurves observed with HST/UVIS or JWST/NIRSPEC would strongly improve our priors on the photospheric temperature at which clouds start to be present and could provide a good guess at the main chemical composition of the clouds~\citep[e.g.][]{Parmentier2016}. By combining this information with a thermal phase curve one could determine which longitudes are affected by the presence of clouds, both on the dayside and the nightside. 

\begin{figure}
\includegraphics[width=\linewidth]{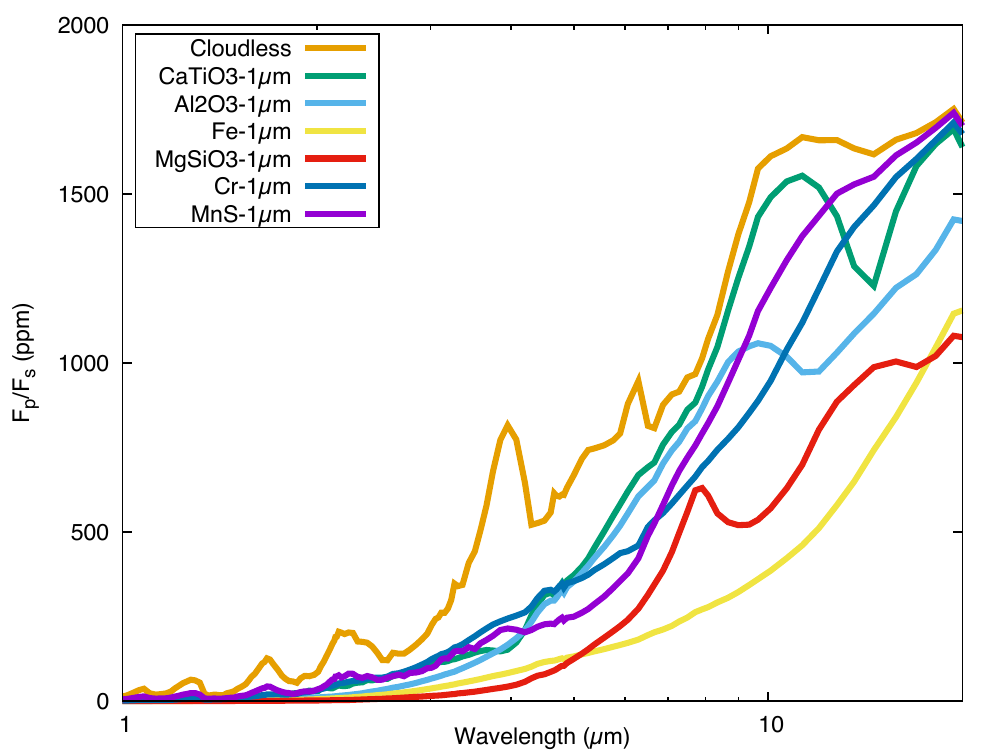}
\caption{Nightside spectrum of $T_{\rm eq}=1400\,K$ models with with different types of post-processed clouds. Every cloud shows its own specific spectral signature at long wavelengths due to its composition specific real and imaginary refractive indexes.}
\label{fig::Miri}
\end{figure}

A second possibility to determine the cloud composition on the planet's nightside would be to use phase curve observations with the MIRI instrument on the JWST. As shown in Figure~\ref{fig::Miri}, the nightside of hot Jupiters should be bright at wavelengths larger than $5\mu m$ even when no nightside flux is observed at shorter wavelengths. Additionally, this wavelength range exhibits the spectral signature of specific cloud compositions~\citep[see also][]{Wakeford2015,Kitzmann2018a,Taylor2020a}. We see that MgSiO$_{\rm 3}$,Al$_{\rm 2}$O$_{\rm 3}$ and CaTiO$_{\rm 3}$ show specific absorption features in the $8-15\mu m$ range. Iron, chromium and MnS clouds, however, show a rather featureless spectrum on the same spectral range. Interestingly, MgSiO$_{\rm 3}$ has a non-grey effect on the spectrum not only at $8-10\mu m$ through the well known resonance absorption band, but also in the $5-8\mu m$ range due to changes in its single scattering albedo~\citep[see][for more details]{Venot2019,Taylor2020a}.  

\section{Conclusion}
\label{sec::Conclusion}

The atmospheric circulation of tidally locked hot Jupiters transports heat from the hot dayside to the cold nightside of the planet. The efficiency of this heat transport determines the day and nightside temperatures and thus the observed secondary eclipse spectra and phase curves. Exoplanet phase curves in the infrared have been puzzling the community for a decade. Several solution have been proposed to explain the larger than expected phase curve amplitude, the smaller than expected nightside temperatures and the smaller than expected phase curve offset. 

Here we perform the first estimate of the variation of the day to night heat transport in hot Jupiters with equilibrium temperature using a global circulation models including non-grey radiative transfer and realistic gas and cloud opacities. { We show that the presence of nightside clouds leads to larger phase curve amplitudes, smaller phase curve offsets and cooler nightside brightness temperatures.} Our main conclusions are:
 
\begin{itemize}
\item  Heat redistribution for cloudless exoplanets should be efficient for $T_{\rm eq}<1600\,K$ then decrease with increasing equilibrium temperature, which confirms previous findings.When clouds are present on the nightside, the heat transport becomes much less efficient, even for low equilibrium temperatures. 
\item Nightside clouds significantly increase the phase curve amplitude and decrease the phase curve offset at all wavelengths, providing a good explanation for most observed hot Jupiter phase curves. Both quantities are extremely sensitive to the physical properties of the clouds such as mean particle size. Variation of cloud physical and chemical properties can easily wash out any trends expected in the cloudless case. 
\item The nearly constant nightside temperature observed in hot Jupiters can be explained by the strong dependence of the radiative timescale with temperature. Nightside clouds are needed to explain why this temperature is lower than expected. 
\item The phase curve offset does not necessarily track the planetary hot spot offset, particularly when clouds are present on the nightside. Secondary eclipse mapping could be a more robust way to determine the longitude of the hottest point on the planet.  
\end{itemize}

We additionally use our models to discuss more minor points: 
\begin{itemize}
\item Phase curve offsets and phase curve amplitude do not necessarily track each other, even in the cloudless case, pointing out that they are set by different dynamical mechanisms. 
\item The presence of nightside clouds homogenise the temperature on isobars by warming up the nightside. However, they shift the nightside photosphere to lower pressures, leading to a much lower brightness temperature than in the cloudless case.
\item The brightness temperature of a planet can significantly vary with wavelengths and care should be taken when deriving bulk heat transport properties from band averaged observations. Particularly, when clouds are present both the 3.6 and 4.5 Spitzer channels brightness temperatures can underestimate the effective temperature of the observed hemisphere and thus overestimate the Bond albedo of the planet.
\item Nightside cloud composition could be detectable with JWST phase curve observations at wavelengths longer than 5$\mu m$ through both the absorption and scattering properties of the clouds.
\item If all hot Jupiter nightside clouds are silicate clouds with similar particle size distributions, we would expect a rise in the phase curve offset in the HST bandpass between equilibrium temperatures from 1200 to 2000 K, a prediction that could be tested with currently available telescopes. 
\end{itemize}

{ Finally, although we showed that the diversity of current phase curve observations of hot Jupiters is possibly caused by a diversity of nightside cloud properties, work remains to be done to understand the specificities of each planet. A more through exploration of cloud properties, including the coupling between cloud, chemistry and radiative transfer on a planet to planet will likely be necessary to interpret the coming decade of exoplanets observations.}

\section*{Data availability}
The data underlying this article will be shared on reasonable request to the corresponding author.

\section*{Acknowledgement}
We thank Claire Baxter and Thomas Beatty for sharing the electronic version of the datasets used in Fig~\ref{fig::DayNightTemp}. We thanks Tad Komacek and Xianyu Tan for long discussions about atmospheric timescales. We thank Rosalba Perna and Mike Roman for sharing the necessary data for figure A.1. We thank the referee whose thoughtful comments improved the manuscript in 86 occasions. 
Adam Showman passed away during the last stages of this manuscript. He was both a friend and an inspirational scientist. His legacy will survive us all.

\bibliography{Phase-curves-V4-Final}

\appendix
\renewcommand\thefigure{\thesection\arabic{figure}}    
\section{Comparison with previous work.}
\setcounter{figure}{0}    
\label{AppA}
We compare our results with the other two studies that looked at the heat redistribution as a function of planet temperature using different sorts of global circulation models. For this we use the ratio of the nightside to the dayside bolometric flux, as presented in figure 2 of ~\citet{Perna2012}. A planet with no redistribution would have $F_{\rm night}/F_{\rm day}=0$ whereas a planet with full redistribution would have $F_{\rm night}/F_{\rm day}=1$. The main differences between our models and the two others are listed below.

~\citet{Perna2012} uses a semi-grey radiative transfer model with a constant infrared opacity of $\kappa=0.01$  cm$^2$/g, and the same relationship as us between orbital period and equilibrium temperature. They use two different values for the optical opacities, 0.5 and 2 times the infrared one and found no big changes in the heat redistribution between the two sets of models. Given that our models neglect the presence of TiO/VO, we compare our models with their $\gamma=0.5$ case.

\citet{Komacek2017} uses an opacity that vary with pressure (see their eq. 9), leading to a photosphere that is a larger pressures (around 0.2 bars) and more compact than~\citet{Perna2012}. The thermal opacity at the photosphere is $\kappa=4.5\times10^{-3}$ cm$^2$/g. Their optical opacity is constant $\kappa_{\rm v}=4\times10^{-3}$ cm$^2$/g. They assume a larger radius than us ($1.3R_{\rm jup}$), which should also lead to a reduced day/night heat transport efficiency.  We highlight three of their models. The first one is a clear atmosphere without drag and with a fixed rotation rate of 3.5 days. The second one has additional drag applied to the momentum and energy equations. Finally we also show a model with a varying rotation rate, which assumes that the planet is tidally locked around a star cooler and smaller than the sun. Although this last model is closest to our setup, for a given equilibrium temperature their modelled planet spins faster than ours, which should lead to a larger day/night contrast.

~\citet{Roman2020} uses a semi-grey radiative transfer model with a constant infrared opacity of $\kappa=0.018$  cm$^2$/g and an optical opacity 0.22 times smaller, smaller than the one used by~\citet{Perna2012}, a gravity of 10m/s and a planet radius of $1.3R_{\rm jup}$, similar to the one of~\citet{Komacek2017}. They use a fixed rotation period of 1.8 days, which is smaller than used by \citet{Komacek2017}  and should therefore lead to a larger day/night contrast.

\begin{figure}
\includegraphics[width=0.9\linewidth]{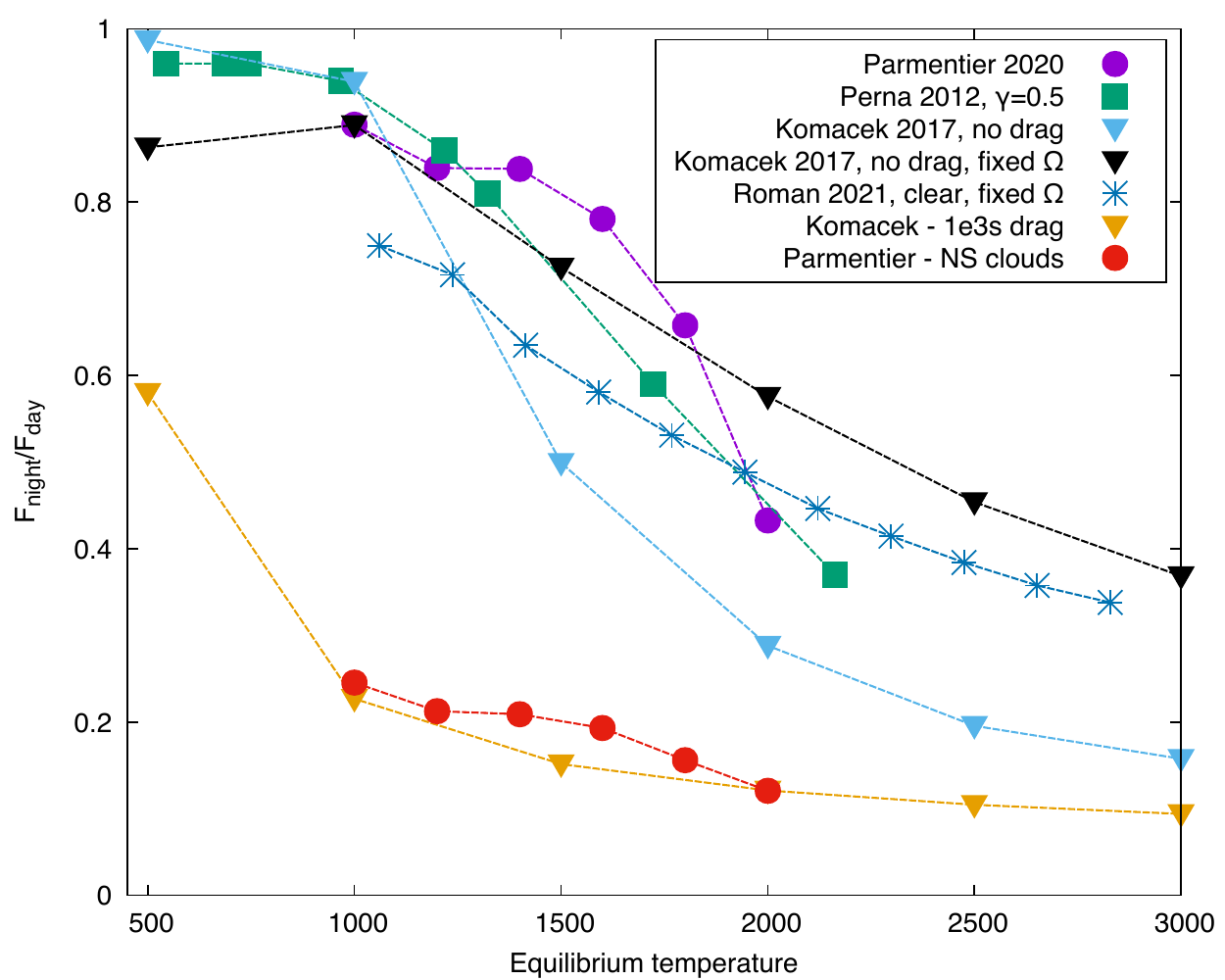}
\caption{Ratio of the nightside to the dayside bolometric flux as a function of equilibrium temperature for our models and the ones of~\citet{Perna2012} and ~\citet{Komacek2017} and~\citet{Roman2020}. Our cloudless model (purple points) agrees well with the ones of~\citet{Perna2012}. Models with nightside clouds (red circles) are as inefficient to transport heat as the models from~\citet{Komacek2017} including an extremely strong drag. }
\label{fig::Perna}
\end{figure}

As seen in Figure~\ref{fig::Perna}, all four models predict a redistribution efficiency that decreases with increasing temperature. The values from~\citet{Perna2012} are very similar to ours, pointing out that the non-greyness of the opacities does not play a major role in setting the bulk heat transport (although it does play a major role in shaping the observed, band averaged day/night temperature contrast as shown in this paper). Our model however, predicts a steeper variation of the heat redistribution with temperature. We attribute this to the change of opacity and hence photospheric pressure with equilibrium temperature in our model. As seen in Fig.~\ref{fig::Timescales}, the change of photospheric pressure contributes significantly to the variation of the radiative timescale for the cloudless case. 

The redistribution is less efficient in the models from~\citet{Komacek2017}. The larger opacities and the narrower extent of the photosphere used in~\citet{Komacek2017} would naturally lead to an increase of the heat transport. However, the faster rotation rate they use and larger radius are expected to decrease the efficiency of the heat transport.

Models of ~\citet{Komacek2017} and ~\citet{Roman2020} with a constant rotation rate follow a different slope. This is a consequence of the heat redistribution variation with rotation rate. At high temperature these models would have a smaller rotation rate than the ones with varying rotation rate, leading to a smaller day/night contrast. The opposite being true at small temperatures. The models from~\citet{Roman2020} have a larger day/night contrast than the ones of~\citet{Komacek2017}, which can be explained by their smaller rotation rate and larger infrared opacities.

Finally, the bottom two curves of Fig.~\ref{fig::Perna} compare the heat redistribution efficiency of our nightside cloud model to the~\citet{Komacek2017} model that includes a strong drag. We see that adding nightside clouds has an effect on the heat transport that is as large as having a very strong, $10^3s$ drag time constant. 

{ We note that the metric shown here is different than the $f$ and $\epsilon$ parameters presented above in the paper. The two metrics are different projections of the 2D planetary flux but there is no one-to-one comparison. For example, a planet with no nightside flux would have a  $F_{\rm night}/F_{\rm day}=0$ but that could correspond either to a dayside only redistribution ($f=2$) or to the no redistribution case ($f=2.666$) since both cases could have a nightside flux of zero~\citep[see also appendix of][]{Schwartz2017}. However, the metric was more available to perform our comparison.}

\end{document}